\documentclass[twocolumn,showpacs,rmp,aps]{revtex4}
\usepackage{graphicx}
\usepackage{amssymb}

\newcommand{\be}{\begin{equation}}
\newcommand{\ee}{\end{equation}}
\newcommand{\bea}{\begin{eqnarray}}
\newcommand{\eea}{\end{eqnarray}}
\begin{document}
\ifx\href\undefined\else\hypersetup{linktocpage=true}\fi 

\title{
       Microscopic dynamics in liquid metals: the experimental
       point of view.
      }
\author{
        Tullio~Scopigno$^{1}$}
\email{tullio.scopigno@phys.uniroma1.it}
\homepage{http://glass.phys.uniroma1.it}
\affiliation{
    $^{1}$INFM CRS-SOFT and Dipartimento di Fisica, Universit\'a di Roma "LaSapienza",~I-00185, Roma, Italy}
\author{
        Giancarlo~Ruocco$^{1}$}
\affiliation{
    $^{1}$INFM CRS-SOFT and Dipartimento di Fisica, Universit\'a di Roma "LaSapienza",~I-00185, Roma, Italy}
\author{Francesco~Sette$^{2}$}
\affiliation{
        $^{2}$European Synchrotron Radiation Facility, B.P.
220 F-38043 Grenoble, Cedex France.
}

\date{\today}

\begin{abstract}
The experimental results relevant for the understanding of the
microscopic dynamics in liquid metals are reviewed, with special
regards to the ones achieved in the last two decades. Inelastic
Neutron Scattering played a major role since the development of
neutron facilities in the sixties. The last ten years, however,
saw the development of third generation radiation sources, which
opened the possibility of performing Inelastic Scattering with X
rays, thus disclosing previously unaccessible energy-momentum
regions. The purely coherent response of X rays, moreover,
combined with the mixed coherent/incoherent response typical of
neutron scattering, provides enormous potentialities to
disentangle aspects related to the collectivity of motion from the
single particle dynamics.

If the last twenty years saw major experimental developments, on
the theoretical side fresh ideas came up to the side of the most
traditional and established theories. Beside the raw experimental
results, therefore, we review models and theoretical approaches
for the description of microscopic dynamics over different
length-scales, from the hydrodynamic region down to the single
particle regime, walking the perilous and sometimes uncharted path
of the generalized hydrodynamics extension. Approaches peculiar of
conductive systems, based on the ionic plasma theory, are also
considered, as well as kinetic and mode coupling theory applied to
hard sphere systems, which turn out to mimic with remarkable
detail the atomic dynamics of liquid metals. Finally, cutting
edges issues and open problems, such as the ultimate origin of the
anomalous acoustic dispersion or the relevance of transport
properties of a conductive systems in ruling the ionic dynamic
structure factor are discussed.
\end{abstract}


\maketitle
\tableofcontents 
\section{Introduction}

Liquid metals are an outstanding example of systems combining
great relevance in both industrial applications and basic science.
On the one hand they find broad technological application ranging
from the production of industrial coatings (walls of refinery
coker, drill pipe for oil search) to medical equipments
(reconstructive devices, surgical blades) or high performance
sporting goods. Most metallic materials, indeed, need to be
refined in the molten state before being manufactured.

On the other hand liquid metals, in particular the monoatomic
ones, have been recognized since long to be the prototype of
simple liquids, in the sense that they encompass most of the
physical properties of real fluids without the complications which
may be present in a particular system \cite{BALUCANI}.

In addition to that, metallic fluids such as molten sodium, having
similar density and viscosity as water, find application as
coolant in nuclear reactors.

The thermodynamic description of liquid metals can be simplified
by assuming a few parameters. Usually, if compound formation is
weak physical theory alone can be used while, if there is strong
compound formation, chemical theory alone is used. The
lowest-melting liquid metals are those that contain heavier
elements, and this may be due to an increase in ease of creating a
free-electron solution. Alkali metals are characterized by low
melting points, and they tend to follow trends. Binary associating
liquids show a sharp melting point, with the most noticeable
example being mercury ($T_m=234$ K). Melting points can be lowered
by introducing impurities into the metal. Often, to this purpose,
another metal with a low melting point is used. Mixing different
metals may often result in a solution that is eutectic. In other
words, from Henry's law it is understood that a melting point
depression occurs, and the system becomes more disordered as a
result of the perturbation to the lattice. This is the case, for
instance, of the well known eutectic Pb-Tin alloy, widely used in
soldering applications ($T_m=453$ K).


Until the sixties the understanding of the physical properties of
metals proceeded rather slowly. It was John Ziman, indeed, who
made the theory of liquid metals respectable for the first time
\cite{ZIMAN}, and the Faber-Ziman theory, developed in 1961-63 and
dealing with electronic and transport properties, is attractively
introduced in Faber's book, which is an excellent treatise of the
physical properties of liquid metals \cite{FABER}.

The other text which can be considered a classic is March's book
\cite{MARCHLM}, along with the more recent \cite{MARCHLMCT}, which
provides a comprehensive overview over liquid metals. It is from
these texts that a first clear definition of liquid metal can be
outlined. At first glance, indeed, the words "liquid metal" are
self-explanatory: by definition any metal heated to its melting
point can be cast in this category. Liquid metals, however, are
implicitly understood to be less general than the above
definition, and no literature clearly states an exact definition.
Although no precise agreement has been made, there are certain
characteristics shared by liquid metals, descending from a close
interplay between ionic structure, electronic states and transport
properties.

The book of Shimoji \cite{SHIMOJI} deals with the fundamentals of
liquid metals in an elementary way, covering the developments
achieved after the first book by March. It does not address,
however, the dynamical properties in great detail.

Addison's book \cite{ADDISON} is much like March's general book,
but is more focused on applications of alkali metals, especially
on their use in organic chemistry. In addition, Addison discusses
many methods for purifying and working with liquid alkali metals.
March is more theoretical whereas Addison is practical, but both
authors focus on a thermodynamic explanation of liquid metals.

For an appealing general introduction to the physics and chemistry
of the liquid-vapor phase transition (beyond the scope of this
review) the reader should certainly make reference to
\cite{HENSEL}, which also provides a bird eye view of the
practical applications of fluid metals, such as high-temperature
working fluid or key ingredients for semiconductor manufacturing.

There are, then, a number of books which are more general and more
specific at the same time, in the sense that they deal with with
the wider class of simple liquids (including noble fluids, hard
sphere fluids etc.), but they are mainly concerned with structural
and dynamical properties only
\cite{BALUCANI,HANSEN,BY,MARCH,EGELSTAFF}. They are practically
ineludible for those aiming at a rigorous approach to the
statistical mechanics description of the liquid state.

It can be difficult to find an exhaustive updated database of the
physical properties of liquid metals, especially as far as
dynamics is concerned. But the handbooks of \cite{ida} and
\cite{OSE} are remarkable exceptions, with the second one
specifically addressing liquid alkali metals.

\subsection{Historical background}

Early phenomenological approaches to the study of relaxation
dynamic in fluids can be dated back to the end of the nineteen
century \cite{max_visco,kel_visco}.

Only in the mid twentieth century, however, it was realized that a
deeper understanding of the physical properties of liquids could
have been reached only through a microscopic description of the
atomic dynamics. This became possible through the achievements of
statistical mechanics which provided the necessary tools, such as
correlation functions, integral equations etc. The mathematical
difficulties related to the treatment of real liquids brought to
the general attention the importance of simple liquids, as systems
endowed with the rich basic phenomenology of liquids but without
the complications arising, for instance, by orientational and
vibrational degrees of freedom.

As a consequence, the end of the fifties saw major experimental
efforts related to the development of Inelastic Neutron Scattering
(INS) facilities which, as we shall see, constitutes a privileged
probe to access the microscopic dynamics in condensed matter and,
in particular, in the liquid state \cite{egel_pio}. A sizable
library of experimental data on liquid metals has been constituted
since then, realizing the prototypical structural and dynamical
properties of these systems, representative of the whole class of
liquids.

In the sixties, the advent and the broad diffusion of
computational facilities brought a new era for two main reasons:
on the one side, realistic computer simulation experiments become
possible \cite{sch_sim}, on the other side the new computation
capabilities greatly facilitated the interpretation of INS
experiments. For instance, new protocols for accurate estimates of
the multiple scattering contribution affecting neutron scattering
were proposed \cite{cop_multiplo}. The theoretical framework of
the Inelastic Neutron Scattering, and the guidelines to interpret
the results, have been reviewed in the classical textbooks
\cite{LOVESEY,MARSHALL}.

The dynamics of liquid metals has been extensively investigated by
INS and computer simulations with the main purpose of ascertaining
the role of the mechanisms underlying both collective and
single-particle motions at the microscopic level. In the special
case of collective density fluctuations, after the seminal
inelastic neutron scattering study by Copley and Rowe
\cite{cop_rb} and the famous molecular dynamics simulation of
Rahman \cite{rahman_sim} in liquid rubidium, the interest in
performing more and more accurate experiments is continuously
renewed: it was soon realized, indeed, that well-defined
oscillatory modes could be supported even outside the strict
hydrodynamic region. In molten alkali metals, moreover, this
feature is found to persist down to wavelengths of one or two
interparticle distances, making these systems excellent candidates
to test the various theoretical approaches developed so far for
the microdynamics of the liquid state.

Up to ten years ago the only experimental probe appropriate to
access the atomic dynamics over the interparticle distance region
were thermal neutrons, and using this probe fundamental results
have been gained. There are, however, certain limitations of this
technique which can restrict its applicability: First, the
presence of an incoherent contribution to the total neutron
scattering cross section. If on one hand this allow to gather a
richer information, being simultaneously sensitive to collective
and single particle dynamics, on the other hand poses the problem
of decoupling the two contributions, when aiming at the
determination of collective properties only (i.e. of the coherent
dynamic structure factor $S(Q,\omega)$). In liquid sodium, for
instance, the incoherent cross section dominates; even in more
favorable cases (Li, K) at small $Q$ the intensity of the
collective contribution is low, and its extraction requires a
detailed knowledge of the single particle dynamics.

The second reason is dictated by the need of satisfying both the
energy and momentum conservation laws which define the $(Q-E)$
region accessible to the probe \cite{BALUCANI}. Roughly speaking,
when the sound speed of the system exceeds the velocity of the
probing neutrons ($\sim 1500$ m/s for thermal neutrons) collective
excitations can hardly be detected for $Q$ values below, let's
say, $Q_m,$ the position of the main diffraction peak of the
sample, which is the region where collective properties show the
richer phenomenology. As we shall see in section \ref{sec_klim},
given a certain kinematic region accessible to neutrons (basically
ruled by their thermal energy), by virtue of the $m^{-1/2}$
dependence of the sound velocity of an atomic system, the higher
its atomic number, the wider is the accessible energy-momentum
region of the excitations which can be studied. Taking as an
example alkali metals, indeed, accurate INS data are available for
heavier elements such as rubidium ($v \sim 1260$ m/s)
\cite{cop_rb,chi_rb,pas_rb} and cesium ($v\sim 970$ m/s)
\cite{bod_cs}, while more difficulties are met in the case of
lighter atoms. In particular, lithium represents the most critical
case due to its high sound speed ($v \sim 4500$ m/s) and to the
weak scattering cross section which, moreover, results from
comparable values of the coherent and incoherent contributions:
for this reason INS aiming to the study of collective properties
of Li represented a very hard challenge \cite{dej_li,dej_phd}.
From a general point of view the main outcome of most of these
early INS experiments, as far as collective properties are
concerned, is the evidence of inelastic excitations in
$S(Q,\omega)$ which have been necessarily analyzed within simple
models such as the damped harmonic oscillator (DHO)
\cite{fak_dho}, suitable to extract reliable and
resolution-corrected information on the peak positions but not
about the detail of the whole lineshape. Some additional
information have been achieved, for instance, in the case of
cesium \cite{bod_cs}, where information about an average
relaxation time have been extracted utilizing Lovesey's
viscoelastic model \cite{lov_visco} or, more recently, in molten
potassium, where a generalized hydrodynamic treatment as the one
described in sec. \ref{sec_collnonhydro} is undertaken
\cite{cab_k} and electronic screening effects have been explicitly
taken into account \cite{bov_k}.

Paralleling the development of INS facilities, new ideas arose on
both theoretical and numerical fields from 1975 and in the
intervening decades, driven by kinetic theory applied to Enskog's
fluid \cite{desh_hyd,desh_hyd0,kag_hs,all_hs,all_hs1}, allowing to
describe the hydrodynamic region in terms of three pole
approximation, or to reproduce the dynamic structure factor at
wavelengths comparable to the inverse mean particle distance in
terms of extended heat mode \cite{coh_hs}. Kinetic approaches
where eventually complemented by memory function formalism and by
Mode Coupling theory
\cite{Wah_kin,sjog_kin,sjog_kin1,sjog_sw,desh_mc,got_na,sjog_coll,beng_kinhs}.

Turning the attention to numerical advances in the liquid metals
field, the major achievement are probably related to the
introduction of the pseudo-potential concept
\cite{HARRISON,aus_psp,heine_psp,stroud_psp} which, beside
offering a deeper comprehension of physical properties such as
electrical resistivity, provided a clue for realistic numerical
simulations. In Molecular Dynamics, indeed, the choice of a
realistic interatomic potential - i.e. a potential model able to
reproduce structural properties - is crucial to allow the
determination of the dynamics of the system via the integration of
the classical Newton equations. Exploiting the pseudo-potential
theory, it has been possible to express the atomic interaction as
a sum of pairwise interactions, ruled by an effective density
dependent interaction. In this respect, one of the most successful
expressions is the effective potential proposed for alkali metals
\cite{pri_psp}.

The numerical simulation framework is particularly useful since
the single particle and the collective dynamics can easily be
investigated within technical restrictions due to the finite box
size (defining the minimum accessible wavevector) and computation
time (related to the statistical quality and to the energy
resolution of the calculated spectra). Broadly speaking the
features of the atomic collective motion i.e. the details of
$S(Q,\omega)$ lineshape, as outcome of MD run, turns out to be
less noisy and more straightforward than the correspondent INS
results: no absolute normalization is required, no mixing between
coherent/incoherent dynamics occurs and, above all, basically no
resolution corrections are needed.

The major experimental breakthrough, however, happened in the last
ten years when x-rays came up by the side of neutrons to study the
collective dynamics in a similar frequency and wavelength region.
The intriguing theoretical possibility of performing Inelastic X
ray scattering \cite{BURKEL} became real thanks to the advent of
the third generation sources \cite{mas_strum}, disclosing
previously unaccessible tasks in the physics of disordered systems
\cite{ruo_nat,set_sci,sin_al2o3,sco_sci}. In this case, the cross
section is mainly coherent, and the combination of the two
techniques can in principle serve to disentangle the two
contributions. Unfortunately, such complementarity has not yet
been exploited at full.

\subsection{Dynamical aspects}

There are mainly two routes to approach the dynamics of a viscous
melt. The first one stems from a quasi-crystalline picture, and
relies on the observation that, often, the diffusion coefficient
$D$ linearly depends on $1/T$. The same dependence, in fact, is
induced in crystalline solids by vacancies and defects. This
analogy suggests that diffusion in liquids is an activated
process, and many attempts have been done to relate the activation
energy in the liquid to the thermodynamics of its solid.

The other point of view is the kinetic theory, a gas-like picture
where the correlation functions are different in view of the
density which is typically much higher than in the gas state.
Within this framework an expression for the behavior of diffusion
coefficient and viscosity can be gained in terms of the friction
coefficient $\xi$. When a particle of the melt is moving with
constant velocity $v$ a net retarding force results from the
different rate between front and back collisions of the form
$F=-\xi v$. For an hard sphere gas it turns out from Fick's law of
diffusion that $\xi=\frac{8ng(\sigma)\sigma ^2 }{3}\sqrt{\pi mk_B
T}$, where $\rho$ is the atomic density, $m$ the atom mass, and
$4\pi \sigma^2 g(\sigma)$ the density probability of finding two
units at distance $\sigma$. The diffusion coefficient is therefore
$D=\frac{k_B T}{\xi}$, while the viscosity $\eta=\frac{\xi}{3\pi
\sigma}$ \cite{long_diff}. These simple expressions turn out to
describe remarkably well the dynamics of liquid metals as long as
the one of other simple fluids. However, better quantitative
agreement can be obtained introducing the velocity autocorrelation
function $\psi(\Delta t)=\frac{\langle \mathbf{v}(t+\Delta t)
\cdot \mathbf{v}(t)\rangle}{\langle v^2 \rangle}$, whose time
integral determines the diffusion coefficient. The $\psi(t)$ plays
a central role in liquid dynamics, not only for providing a
rigorous way to calculate $D$, but also because through its
Fourier transform one can grasp an insight into the detail of the
interatomic interactions. It can be accessed either by molecular
simulations (the calculation trivially follows from its
definition) or experimentally, mainly by Inelastic Neutron
Scattering (INS) with the methods detailed in section
\ref{sec_exp}.

Broadly speaking, $\psi(\Delta t)$ is related to the knowledge of
$G_s(r,t)$ i.e. the probability that a given particle travels a
distance $r$ in the time interval $t$. INS is always sensitive to
a combination of $G_s(r,t)$ and $G_d(r,t)$, being this latter
quantity the probability of finding two distinct particles at a
space distance $r$ and time distance $t$. The way to separate
these two contributions mixed in the instrumental response is one
of the major conundrum of the neutron scattering technique. The
direct knowledge of the coherent response, gained by means of IXS,
opens the possibility to the understanding of the high frequency
modes that have been seen to survive since the famous INS
experiments on the highly coherent scatterer liquid Rubidium
\cite{cop_rb}. In particular, from a solid point of view, the main
issue is to ascertain the relation between these modes and the
phonon excitations in the corresponding polycrystal just below the
melting. In a perfect crystal (i.e. a periodic assembly of atoms
or ions), indeed, atomic dynamics is mainly vibrational,
characterized by normal modes which are plane waves, due to the
harmonic nature of the interatomic forces. Consequently, a well
defined ratio (the sound velocity) exists between the frequency
and the wavevector of the density fluctuations, named in this case
\textit{phonon}. In real crystals, therefore, the energy spectrum
of a scattered probe results in sharp peaks, whose linewidth is
related to the presence of anharmonicities or lattice defects. In
a liquid things are more involved: other effects arise besides
anharmonicity, both structural (the average atomic positions are
randomly distributed) and dynamical (mass diffusion and activated
processes join the purely vibrational motion). But even in this
complex scenario distinct peaks survive and one can extract
dispersion relations.

Interestingly, the relation between crystals and liquids involves
more than the mere similarity between the sound velocity of the
crystalline acoustic branches and of the mode observed in the
liquid (i.e. the low $Q$ limit of the dispersion relation).
Despite the lack of periodicity intrinsic to the inherent liquid
structure, indeed, the presence of some residual correlation
(testified by the oscillation in static structure factor) seems to
warrant a support for the existence of umklapp modes similar to
the one existing in crystals. Such processes, i.e. the presence of
inelastic modes characterized by wavevectors which differ by
multiples of the reciprocal lattice spacing, have been early
reported by means of INS in liquid lead
\cite{rand_umk,coc_umk,dor_umk} and more recently in liquid
lithium by means of IXS \cite{scop_prbumk}. As correctly pointed
out by Faber, the presence of these excitations does not imply the
existence of genuine high $Q$ modes, it rather indicates that
umklapp processes may occur in liquid as much as in solid
\cite{FABER}.

From the liquid point of view the interest in this phenomenology
lies in the challenging extension of the simple hydrodynamics,
describing the density fluctuations in the long wavelenght limit,
down to the lenghtscale of the mean interparticle distances. As it
will be shown in the following, such an extension relies on
serious and sometimes not fully justified assumptions necessary to
walk in the uncharted and perilous territory between hydrodynamics
and single particle regimes.

\subsection{Peculiarities of liquid metals}

Apparently at odds with the previously mentioned classification of
liquid metals as prototype of simple liquids, even in the simplest
monoatomic case, metallic fluids are actually two component
systems. The interplay between electron and ions, indeed, is an
intrinsic aspect of liquid metals, and a rigorous approach should
therefore mimic the formalism utilized for binary mixtures. For
many aspect, however, one might be interested in ionic properties
only, and as far as atomic dynamics is concerned, this seems to be
the case. In such circumstance, one can look at a liquid metal as
an ionic assembly whose interaction is mediated by the conduction
electron gas. The treatment is in such way reduced to a one
component system, like for noble fluids one can introduce a
pairwise interaction, but this latter will be ruled by a density
dependent pseudopotential.

Although within the pseudopotential approach many results for
liquid metals are qualitatively similar to those for ordinary non
conductive fluids, some remarkable differences exist. One of the
most relevant of these differences concerns peculiar structural
properties involving short range order: in several liquid metals
the static structure factor exhibits an asymmetry or even a
shoulder just above the main peak. The origin of this anomaly has
been highly debated, and ascribed to the peculiar shape of the
interaction potential in those metals in which the hard sphere
description fails \cite{tsay_ga}. More specifically, it can be
interpreted in terms of a repulsive interaction composed by an
hard sphere part plus an adjacent ledge induced by electronic
effects, by a curvature change occurring at the nearest neighbor
distance, or by the interplay with Friedl oscillations.

The reported peculiarities extend also to the dynamics: while in
the long wavelength limit they are expected to behave similarly to
non conductive fluids, at finite wavevectors their departure from
ordinary hydrodynamics can be in principle influenced by the high
values of the thermal conductivity
\cite{FABER,sing_pre,scop_comm,sing_rep}. One of the most striking
quantitative differences with ordinary fluids concerns the
"visibility" of the inelastic features, i.e. the inelastic to
elastic ratio, which seems to be related to the softness of the
interaction potential \cite{cana_ljlm,baluc1_sim}. This latter is
responsible, for example, for the very favorable inelastic to
elastic ratio which makes alkali metals ideal systems to study
collective properties.

Though the whole dynamics of liquid metals seems to be
conveniently rationalized treating them as ordinary fluids
interacting via an effective, density dependent, pairwise
interaction potential, there is an alternative route which
explicitly takes into account electronic screening effect on the
ionic dynamics, relying on the introduction of a suitable model
for the wavelength dependent dielectric function. In this way one
his able to test different approximations comparing the
predictions for the mechanical compressibility (or, equivalently,
for the sound velocity) with the experimental values
\cite{burns_prb,bov_k}.

\subsection{Why this review}

As previously pointed out, some books offer a broad coverage of
the physics and chemistry of liquid metals, but none of them is
focused on the dynamical aspects. On the other side, books dealing
with liquid dynamics do not spot on liquid metals in great detail.
Egelstaff's review \cite{egel_rev} offers an exhaustive coverage
of transport phenomena in liquids but dynamical properties of
liquid metals are marginally addressed. The Copley and Lovesey
review \cite{coplov_rev} provides an invaluable sight on the
dynamics of simple liquids, and it presents many results obtained
for liquid metals, but again does not emphasize the peculiarity of
these systems. Moreover, it also addresses in detail numerical
results and last, but not least, it is necessarily not up to date
given the large amount of experiments recently performed. The last
decade, indeed, has seen important advances in the dynamics of
liquid metals, especially on the experimental side, driven by the
advent of new X-ray facilities and by the upgrades of the neutron
ones. On the theoretical side, in the eighthes, impressive results
have been achieved working on hard sphere models and/or mode
coupling approaches which, again, remain uncovered in the Copley
and Lovesey review. To conclude, it is certainly worth mentioning
the more recent review article by P. Verkerk \cite{verk_rev},
which offers a clear overview of the theoretical models developed
sofar for liquid dynamics, presenting a selection of experimental
results for liquified rare gases, molten metals and binary
mixtures.

Given this background, it seemed to us helpful to focus on the
experiments on liquid metals, and to discuss and summarize the
results and their interpretations in terms of the existing
theories, trying to emphasizes advantages and weakness of each
approach. This turned out to be a difficult task, given the
broadness of the matter, and we necessarily had to make some
choices. We left out, for instance, mixtures and alloys, and we
tried to focus on the most recent experimental achievements, say
of the last ten years: in most cases we quickly reference to older
results, unless they are particularly relevant in view of the most
recent ones.

The review is organized as follows. In section \ref{sec_bg} we
present different theoretical approaches to the dynamics of liquid
metals. We develop in parallel subsections the treatment of the
self and collective properties which are, in turn, organized
according to the different wavelength domains: hydrodynamic,
non-hydrodynamic, single particle. We also include a subsection
dealing with hard sphere treatment and one presenting the ionic
plasma approach, which is peculiar of conductive systems. In
section \ref{sec_exp} we describe the experimental approach to the
investigation of microscopic dynamics in liquids, outlining the
basics of the inelastic scattering problem. Since the case of X
rays is relatively newer, we decided to treat it in detail, but
continuous reference is made to neutron scattering in an effort to
emphasize merit, drawbacks and complementarities of the two
methods. Section \ref{sec_res} is the bulk part of this paper, in
which the experimental results are reviewed and ordered element by
element. Here, we make constant reference to section \ref{sec_bg}
to recall the different approaches utilized by different authors
to describe the experimental result. In section \ref{sec_sum},
finally, we try to summarize the arising scenario, pointing out
the issues which, in our opinion, deserves further investigations
and trying to draw, when possible, some conclusive pictures.

\section{Theoretical background \label{sec_bg}}

\subsection{General overview}

\subsubsection{Some basic definitions}

The investigation of microscopic dynamics of an ensemble of $N$
identical atomic or molecular units usually proceeds through the
study of correlation functions of dynamical variables, i.e. of
functions of the phase space variables, defined as the $6N$
positions ${\bf r}_i(t)$ and momenta ${\bf p}_i(t)=m{\bf v}_i(t)$
of the particles. Relevant dynamical variables are those stemming
from the microscopic density $\rho({\bf r},t)$ (whose average is
related to the number density $\rho=\langle \rho({\bf r},t)
\rangle$), momentum density ${\bf J}({\bf r},t)$ and kinetic
energy density E({\bf r},t):

\begin{eqnarray}
&& \rho({\bf r},t)\doteq \frac{1}{\sqrt N} \sum_{i} \delta({\bf r}-{\bf r}_i(t)) \nonumber \\
&& {\bf J}({\bf r},t) \doteq \frac{1}{\sqrt N} \sum_{i} {\bf v}_i(t) \delta({\bf r}-{\bf r}_i(t)) \nonumber \\
&& E({\bf r},t) \doteq \frac{1}{\sqrt N} \sum_{i} \frac{1}{2}m
v^2_i(t) \delta({\bf r}-{\bf r}_i(t)) \label{3dv}
\end{eqnarray}

In many cases, the study of the dynamics of a tagged particle $i$
can be of interest, and it relies on similar definitions for the
single particle dynamical variables:

\begin{eqnarray}
&& \rho_s({\bf r},t)\doteq \delta({\bf r}-{\bf r}_i(t)) \nonumber \\
&& {\bf J}_s({\bf r},t)\doteq {\bf v}_i(t) \delta({\bf r}-{\bf r}_i(t)) \nonumber \\
&& E_s({\bf r},t)\doteq \frac{1}{2}m v^2_i(t) \delta({\bf r}-{\bf
r}_i(t)) \nonumber
\end{eqnarray}

The well known van Hove distribution functions $G_d(r,t)$ and
$G_s(r,t)$ are related in the classic (not quantum) case to the
microscopic self and collective densities through:

\begin{eqnarray}
G({\bf r},t)=G_s({\bf r},t)+G_d({\bf r},t) \nonumber
\end{eqnarray}

\noindent with

\begin{eqnarray}
G_s({\bf r},t)&=&\frac{1}{N}\left \langle \sum_{i} \delta({\bf
r}+{\bf r}_i(0)-{\bf r}_i(t))\right \rangle \nonumber \\
G_d({\bf r},t)&=&\frac{1}{N}\left \langle \sum_{i}\sum_{j \neq i}
\delta({\bf r}+{\bf r}_j(0)-{\bf r}_i(t))\right \rangle \nonumber
\end{eqnarray}

As we shall see, experiments usually give information on the
correlation functions in the reciprocal, $Q$, space. Therefore, it
can be useful to define the space Fourier transform of the
microscopic quantities previously introduced:

\begin{eqnarray}
&&\rho({\bf Q},t)=\frac{1}{\sqrt N} \sum_{i} e^{-i{\bf Q} \cdot {\bf r}_i(t)} \nonumber \\
&& {\bf J}({\bf Q},t)=\frac{1}{\sqrt N} \sum_{i} {\bf v}_i(t) e^{-i{\bf Q} \cdot {\bf r}_i(t)} \nonumber \\
&& E({\bf Q},t)=\frac{1}{\sqrt N} \sum_{i} \frac{1}{2}m v^2_i(t)
e^{-i{\bf Q} \cdot {\bf r}_i(t)} \nonumber
\end{eqnarray}

\noindent and similarly for the single particle variables.

The time evolution of these microscopic variables is cross-linked
and can be studied by replacing them with their average over small
but statistically significant volumes. A closed set of equations,
basically conservation laws and constitutive relations, can easily
be written for these averages \footnote{From here on we will
implicitly assume that we are not dealing with microscopic
quantities but rather with the hydrodynamic quantities resulting
from their averages}. Once the expression for $a(\mathbf{Q},t)$ is
known ($a=\rho,\mathbf J, E$) one can calculate

\begin{equation}
\Phi_{\alpha \beta}(Q,t)=\left \langle a_\alpha^*({\bf Q})
e^{\mathcal{L}t} a_\beta({\bf Q}) \right \rangle_N \label{cf}
\end{equation}

In which $a_\alpha({\bf Q})$ are the appropriate dynamical
variables, and $\mathcal{L}$ is the Liouville operator ruling the
time evolution in the configurational space. The $\langle ...
\rangle_N$ indicates thermal averages evaluated over the
$N$-particle ensemble (from now on we will omit the subscript
$N$).

In particular, the autocorrelation functions of the microscopic
density (both self and collective) play a privileged role, and we
define, therefore:

\begin{eqnarray}
&& F(Q,t)=\langle \rho (Q,t)\rho(-Q,0) \rangle \label{fqt}\\
&& S(Q)=\langle |\rho(Q,0)|^2 \rangle \nonumber \\
&& \Phi(Q,t)=\Phi_{11}(Q,t)=\frac{F(Q,t)}{S(Q)} \nonumber \\
&& F_s(Q,t)=\langle \rho_s (Q,t)\rho_s(-Q,0) \rangle \nonumber \\
&&\Phi_s(Q,t)=F_s(Q,t) \nonumber
\end{eqnarray}

\noindent Consequently

\begin{eqnarray}
\Phi(Q,t)&=&\frac{1}{N S(Q)} \sum_{i,j} \left \langle e^{i{\bf Q} \cdot {\bf r}_i(0)}e^{-i{\bf Q} \cdot {\bf r}_j(t)} \right \rangle \nonumber \\
&=&\frac{1}{N S(Q)} \sum_{i,j} \left \langle e^{i{\bf Q} \cdot {\bf r}_i(0)}e^{\mathcal{L}t}e^{-i{\bf Q} \cdot {\bf r}_j(0)} \right \rangle \label{daf} \\
\Phi_s(Q,t)&=&\frac{1}{N} \sum_{i} \left \langle e^{i{\bf Q} \cdot
{\bf r}_i(0)}e^{-i{\bf Q} \cdot {\bf r}_i(t)} \right \rangle
\label{daf_s}
\end{eqnarray}

\noindent The above autocorrelation functions $F(Q,t)$ and
$F_s(Q,t)$ are, in fact, connected through their time-Fourier
transform to the self and collective dynamic structure factor
$S_s(Q,\omega)$ and $S(Q,\omega)$, respectively. These latter, in
turn, are the experimentally accessible quantities in neutrons and
X-rays inelastic scattering experiments and therefore, in the
following, we will mainly refer to the density-density correlation
functions.

\subsubsection{Spectral moments}

Before illustrating some models for the evolution of the density
autocorrelation function, it is worth to recall here some basic
relations involving the frequency moments of the dynamic structure
factor. These can be very useful to the experimentalist, as a way
to normalize the data (an example is given in section
\ref{sec_fetdq}, as well as to any theoretical approach, as a
direct test of sum rules - for example, we shall see how the
hydrodynamic expression for the density-density time correlation
function is valid up to the second frequency moment, because in
hydrodynamics the liquid is treated as continuum without atomic
structure, and such an information on structure and interatomic
potentials appears only within the fourth and higher frequency
moments of $S(Q,\omega)$. By expanding the density autocorrelation
function in a Taylor series, one can easily find a connection with
the frequency moments of the dynamic structure factor as:

\begin{equation}
\frac{d^n F(Q,t)}{dt^n} \Bigg |_{t=0} = (-i)^n
\int_{-\infty}^{+\infty} \omega^n S(Q,\omega)
d\omega=(-i)^n\langle \omega^n \rangle
\end{equation}

This relation holds for the frequency moments of both the
collective ($\langle \omega^n \rangle_S$) and the self ($\langle
\omega^n \rangle_{S_s}$) dynamic structure factor. From the
previous definitions, it easily follows that $\langle \omega^0
\rangle_S=S(Q)$ and $\langle \omega^0 \rangle_{S_s}=1$. The first
frequency momentum are, on the other side, $\langle \omega^1
\rangle_S=\langle \omega^1 \rangle_{S_s}=\frac{\hbar Q^2}{2m}$ and
 therefore are zero for any classical theory, characterized by
 symmetric spectral functions. The second frequency moments are $\langle
 \omega^2 \rangle_S=\langle \omega^2 \rangle_{S_s}=\frac{k_B T Q^2}{m}+O(\hbar
 ^2)$. Higher order spectral moments depend on the details of the
microscopic interactions, and can be analytically derived for
additive pairwise interatomic potential \cite{dej_ema}. Some
examples will be given in sections \ref{sec_mfs} and \ref{sec_mf},
while pratical usage of sum rules for normalization purposes will
be outlined in section \ref{sec_fetdq}.

\subsubsection{Quantum aspects}

The models that we will illustrate in the next sections have been
developed for a classical system. The main effect of
quantum-mechanical corrections stems from the well-known
inequality of the positive and negative-frequency parts of the
spectra, connected by the detailed balance factor $e^{\hbar \omega
/ k_B T}$. Additional sources of non-classical behavior, such as
those associated with a finite value of the de Broglie wavelength
$\Lambda =\left( 2\pi \hbar
^2/mk_BT\right) ^{1/2}$, are small (for lithium at melting $\Lambda $%
\ is only $0.11$ times the average interparticle distance, and
this ratio decreases for heavier metals) and can safely be
neglected. Since the effects of the detailed balance are clearly
visible in the experimentally measured dynamic structure factors,
we briefly discuss a possible procedure to account for this
quantum feature in a consistent way, while preserving the inherent
advantages of the classical description. In doing this, for the
sake of clarity we shall denote all the previous classical
quantities with the subscript {\it cl}, while the notation {\it q}
will refer to the quantum case.

The natural theoretical counterpart of the classical density
correlation function is the so called Kubo canonical relaxation
function \cite{kubo_quantum}

\begin{equation}
K_q(Q,t)=\frac 1{\beta N}\sum_{i,j}\int_0^\beta d\lambda \left\langle e^{-i%
{\bf Q\cdot }\widehat{{\bf r}}_i(0)}e^{-\lambda \widehat{H}}e^{i{\bf Q\cdot }%
\widehat{{\bf r}}_j(t)}e^{\lambda \widehat{H}}\right\rangle
\label{kubo}
\end{equation}

where $\beta =1/k_BT$ and the angular brackets denote now a
quantum statistical
average. In the classical limit ($\beta \rightarrow 0,$ $\hbar \rightarrow 0$%
) the quantum operators become classical commuting dynamical
variables
and $K_q(Q,t)\rightarrow F_{cl}(Q,t).$ It can be shown \cite{lov_visco} that $%
K_q(Q,t)$ is a real even function of time, so that its spectrum $%
K_q(Q,\omega )$ is an even function of frequency. On the other
hand, the
experimental scattering cross section involves the Fourier transform $%
S_q(Q,\omega )$ of the quantum density correlator $F_q(Q,t)=(1/N)\sum_{i,j}%
\left\langle e^{-i{\bf Q\cdot }\widehat{{\bf r}}_i(0)}e^{i{\bf Q\cdot }%
\widehat{{\bf r}}_j(t)}\right\rangle .$ The relation between
$S_q(Q,\omega )$ and $K_q(Q,\omega )$ reads \cite{LOVESEY}

\[
S_{q}(Q,\omega )=\frac{\beta \hbar \omega }{1-e^{-\beta \hbar \omega }}%
K_{q}(Q,\omega )
\]

and, as can be easily checked, satisfies the detailed balance
condition. Moreover, it can be seen that the relation

\begin{equation}
\langle \omega^{2n}\rangle _K=\frac 2{\beta \hbar }\langle
\omega^{2n-1}\rangle _S \label{dispari}
\end{equation}

connects the even frequency moments of $K_q$ with the odd ones of
$S_q.$ In addition to that, the same memory function framework
which will be outlined in sections \ref{sec_mfs} and \ref{sec_le}
can be phrased for the Kubo relaxation function and for its Laplace transform $\widetilde{K}_q(Q,s)$

By virtue of all these properties, in a situation where the
quantum aspects not associated with detailed balance are marginal,
it is reasonable (although not strictly rigorous) to identify the
spectrum $K_q(Q,\omega )$ with the classical quantity
$S_{cl}(Q,\omega )$ so that

\begin{equation}
S_{q}(Q,\omega )\simeq \frac{\beta \hbar \omega }{1-e^{-\beta \hbar \omega }}%
S_{cl}(Q,\omega )  \label{squantclass}
\end{equation}

Having assumed such a correspondence, from now on we will drop out
the subscript {\it cl} and refer to the classical quantities as in
fact done at the beginning of this section.

The transformation (\ref{squantclass}) allows one to test
classical models against experimental data. It is worth to point
out, however, that it alters the frequency moments: as shown in
the previous section, for instance, it introduces a $\hbar ^2$
correction to the second frequency moment, though this effect has
been shown to be hardly noticeable in liquid metals
\cite{scop_jpc}.

\subsection{Single particle dynamics in the hydrodynamic regime} \label{sec_selfhydro}

The time evolution of the single particle density can be easily
obtained through the continuity equation and the constitutive
relation (Fick's law) relating density and current variables:

\begin{eqnarray}
&&\dot{\rho}_s({\bf r},t)+ \nabla \cdot {\bf J}_s({\bf r},t)=0 \nonumber \\
&&{\bf J}_s({\bf r},t)=-D\nabla \rho_s({\bf r},t) \nonumber
\end{eqnarray}

\noindent It is worth to stress that while the first equation is
exact, the second is a phenomenological "closure". Combining the
two equations, one gets the diffusion equation straightforward:

\begin{equation}
\dot{\rho}_s({\bf r},t)=D\nabla ^2 \rho_s({\bf r},t) \nonumber
\end{equation}

which, in the reciprocal space, has the solution

\begin{equation}
\rho_s(Q,t)=\rho_s(Q)e^{-DQ^2t} \nonumber
\end{equation}

The normalized autocorrelation function of the single particle
density is now obtained as

\begin{equation}
\Phi_s(Q,t)=\langle \rho_s(Q,t)\rho_s(-Q) \rangle=e^{-DQ^2t}
\nonumber
\end{equation}

while its fourier transform, the self dynamic structure factor,
will read

\begin{equation}
S_s(Q,\omega)=\frac{1}{\pi} \frac{DQ^2}{\omega^2+(DQ^2)^2}
\label{fick}
\end{equation}

i.e. a Lorenzian function centered at $\omega=0$ with FWHM equal
to $2DQ^2$. It is worth to point out how, in the hydrodynamic
limit, the diffusion coefficient is related to the dynamic
structure factor as $D=\lim_{Q\to 0}\frac{\omega^2}{Q^2}\pi
S_s(Q,\omega)$.

Finally, for completeness, the corresponding Van-Hove self
correlation function is:

\begin{equation}
G_s(r,t)=\frac{1}{(4\pi Dt)^{3/2}} e^{-r^2/4Dt} \label{Gs_hydro}
\end{equation}

\subsection{Collective dynamics in the hydrodynamic regime} \label{sec_collhydro}

In a similar manner as in the previous section, one can build
again a set of closed equations but, in this case, Fick's law does
not apply, and the situation is more involved. The constitutive
relations, indeed, couple together the three conservation laws for
the microscopic variables density, momentum and energy, which, in
terms of the correspondent fluxes reads:

\begin{eqnarray}
&&\dot{\rho}({\bf r},t)+ \nabla \cdot {\bf J}({\bf r},t)=0 \label{3idro} \\
&&\dot{\bf J}({\bf r},t)+ \nabla \cdot {\bf \sigma}({\bf r},t)=0 \nonumber \\
&&\dot{E}({\bf r},t)+ \nabla \cdot {\bf H}({\bf r},t)=0 \nonumber
\end{eqnarray}

\noindent were we have defined the momentum flux $\mathbf \sigma
(\mathbf r ,t)$ and the energy flux $\mathbf H(\mathbf r ,t)$

\begin{eqnarray}
\sigma_{\alpha, \beta}(\mathbf r ,t)&=&\delta_{\alpha,
\beta}P(\mathbf r ,t)-\eta \left ( \frac{\partial u_\alpha
(\mathbf r ,t)}{\partial r_\beta} + \frac{\partial u_\beta
(\mathbf r ,t)}{\partial r_\alpha} \right ) \nonumber \\
&+& \delta_{\alpha, \beta}(\frac{2}{3}\eta-\xi) \mathbf \nabla
\cdot \mathbf u(\mathbf r ,t) \nonumber \\
\mathbf H(\mathbf r ,t)&=& h \mathbf u(\mathbf r ,t)-\kappa
\mathbf \nabla T(\mathbf r ,t) \nonumber
\end{eqnarray}

\noindent here $\eta$ and $\xi$ are the shear and bulk
viscosities, $P$ and $T$ are the local pressure and temperature
fields, $h$ is the enthalpy density and $\kappa$ is the thermal
conductivity.

Two additional constitutive relations (the Navier Stokes equation
and the Fourier law) call into play the two additional
thermodynamic variables pressure $P$ and temperature $T$. Invoking
the thermal equilibrium and the equations of state one obtains a
closed set of equations which can be solved to get the density
density correlation function. The detailed derivation of
$\rho(Q,t)$ is beyond the purpose of the present review, and can
be easily retrieved in classical textbooks \cite{BERNE,HANSEN}.
Here we will only recall the final \textit{approximate} results
\footnote{The exact hydrodynamics expression contains a small
additional contribution which makes the Brillouin components
asymmetric, as emphasized in \cite{nic_hyd,verk_rev}. The
resulting lineshape can be easily recognized as the already
mentioned Damped Harmonic Oscillator, originally proposed within a
solid-like picture \cite{fak_dho}, which can be actually retrieved
by a liquid-like point of view within the memory function
formalism, shown in section \ref{sec_mf}.} which are:

\begin{eqnarray}
&&\frac{\rho(Q,t)}{\rho(Q)}=\left[ \left( \frac{\gamma-1}{\gamma}
\right ) e^{-D_T Q^2t} + \frac{1}{\gamma}e^{-\Gamma Q^2 t} \cos
c_sQt \right] \nonumber \\
&&\frac{S(Q,\omega)}{S(Q)}=\frac{1}{2\pi}\left[ \left(
\frac{\gamma-1}{\gamma} \right )
\frac{2D_TQ^2}{\omega^2+(D_TQ^2)^2}\right] \nonumber \\ && +
\frac{1}{\gamma}\left[ \frac{\Gamma Q^2}{(\omega+c_sQ)^2+(\Gamma
Q^2)^2} + \frac{\Gamma Q^2}{(\omega-c_sQ)^2+(\Gamma Q^2)^2}
\right] \nonumber \\ \label{S_hidro}
\end{eqnarray}

having defined

\begin{eqnarray}
&&\gamma=\frac{c_P}{c_V} \nonumber \\
&&D_{T}=\frac{\kappa}{\rho m C_{P}} \nonumber \\
&&\Gamma=\frac{1}{2\rho m}\left[ \frac{4}{3}\eta_s + \eta_B
+\frac{(\gamma-1)\kappa}{c_P} \right] \label{gidro}
\end{eqnarray}

In the above expressions $c_P$ and $c_V$ are the specific heat
ratios at constant pressure and volume, $\eta_s$ and $\eta_B$ are
the shear and bulk viscosities, $\kappa$ is the thermal
conductivity and $c_s=\sqrt{\frac{\gamma}{m}\left (\frac{\partial
P}{\partial \rho}\right )}_T$ is the adiabatic sound velocity.

The classical hydrodynamics, therefore, predicts in the long
wavelength limit ($Q \rightarrow 0$) a frequency spectrum for the
density fluctuations constituted by two main features. The central
part of the spectrum is dominated by a quasi-elastic, non
propagating mode related to entropy fluctuations (Rayleigh
component) of linewidth $\Gamma_{qe}=2D_T Q^2$, which reflects the
fact that thermal fluctuations decay over a finite lifetime
$\tau=2/\Gamma_{qe}$. Beside, two symmetrically shifted inelastic
components peaked at frequency $\omega_s=\pm c_s Q$ are the
signature of propagating pressure waves (Brillouin doublet), which
are damped by a combination of viscous and thermal effects. The
ratio between the Rayleigh and the Brillouin component is given by
the Landau-Placzeck ratio

\begin{equation}
\frac{I_R}{2 I_B}=\gamma -1 \label{lp}
\end{equation}

Usually, the hydrodynamic regime is investigated by visible light
scattering spectroscopy (BLS, Brillouin Light Scattering). In the
case of liquid metals, the light scattering study of density
fluctuations, is prevented by the non-transparent nature of these
systems \footnote{In the case of copper, for instance, the
distance to the Fermi surface is 2.3 eV. Thus, electrons are
promoted by energies associated with the blue-green end of the
spectrum. As a result, red and orange light at the opposite end of
the spectrum is reflected back and gives copper its characteristic
color. With the alkali metals, the \textit{s} electron is involved
in promotion to the Fermi level. There is little overlap to the
empty \textit{3p} and \textit{3d} orbitals that contribute to the
conduction band. Therefore, only radiation close to the
ultraviolet region is absorbed and visible light is reflected,
hence the silver-like appearance of the alkali liquid-metals.}.

Inelastic scattering experiment can, in fact, only be performed by
means of higher energy photons (X-rays) or by neutrons, but in
both cases the probed wavevectors are fairly outside the strict
hydrodynamic region.

The way the Brillouin triplet evolves at finite $Q$ is far from
being fully understood, though some simplified phenomenological
models have been proposed in the past \cite{mcg_hyd}. Primarily,
one should account for the frequency dependence of the transport
coefficients which, as we shall see, corresponds to abandon the
hypothesis of a Markovian dynamics. Second, once the wavevector
approaches the inverse interparticle distances, structural effects
are expected in the form of a $Q$ dependence of all the
thermodynamic quantities. Last, but not least, it is highly
questionable whether the role of the thermal and viscous processes
remains well separated at high $Q$. In particular, specifically in
the case of liquid metals, due to the high thermal conductivity
one expects $D_TQ^2$ to become soon of the order of the brillouin
frequency, so that entropy and density fluctuations become closely
interwove. Strangely enough, this aspect, with a few exceptions
\cite{FABER} did not receive much attention in the past, although
lately it has been the matter of some debate
\cite{sing_pre,scop_comm,sing_rep}.

Before coming in the discussion of the evolution of $S(Q,\omega)$
at increasing $Q$ values, above the hydrodynamic limit, it is
worth to discuss another analytically solvable case: the high $Q$
limit.

\subsection{The short wavelength limit \label{sec_swl}}

In the previous sections we examined the self and collective
motion at very small wavevectors and frequencies. In the opposite
regime, i.e. at short distances and timescales, the particles of a
fluid are expected to move as they were free, as it happens in an
ideal gas. Since in this case the behaviors of different particles
are uncorrelated ($G_d(r,t)=0$), the self and collective dynamic
structure factors coincides in this limit.

The correlation function of Eq. (\ref{fqt}) can be easily
calculated for a free particle, i.e. for ${\bf r}_i(t)={\bf
r}_i(0)+{\bf v}_i t$. The classical free particle correlation
function is:

\begin{equation}
F(Q,t)=\frac{1}{N} \sum_i \left\langle e^{-i{\bf Q} \cdot
\frac{{\bf p}_i}{m_i} t} \right\rangle=\left\langle e^{-i{\bf Q}
\cdot \frac{\bf p}{m} t} \right\rangle \label{fqt_free_class}
\end{equation}

\noindent evaluating the thermal average one easily gets

\begin{eqnarray}
F(Q,t)=\int e^{-\frac{p^2}{2mk_BT}} e^{-i Q p m t} dp =
e^{-\frac{k_B T Q^2 t^2}{2m}}
\end{eqnarray}

\noindent and the corresponding dynamic structure factor is

\begin{equation}
S(Q,\omega)=\sqrt{\frac{m}{2\pi k_B T Q^2}}e^{-\frac{m\omega^2}{2
k_B T Q^2}} \label{S_fp_cl}
\end{equation}

By virtue of the previously mentioned considerations, the Van-Hove
self correlation function reads:

\begin{equation}
G_s(r,t)=\left ( \frac{m}{2\pi k_B T t^2} \right )^{3/2}
e^{-\frac{mr^2}{2 k_B T t^2}} \label{Gs_single}
\end{equation}

In the quantum case the correlation function
(\ref{fqt_free_class}) can be evaluated treating ${\bf r}_i$ and
${\bf p}_i$ as operators, and paying attention to the fact that,
in this case, the product of the exponential in Eq.(\ref{fqt}) can
not be reduced to a single exponential, as in the classical
treatment. Invoking the identity $e^{\hat{A}} e^{\hat
{B}}=e^{\hat{A} + \hat{B} + \frac{1}{2} \left [ \hat{A},\hat{B}
\right ]}$, holding when ,as in the present case, $\left [
\hat{A},\hat{B} \right ]$ is a number, one can write:

\hspace{-2cm}\begin{eqnarray}
&&\left \langle e^{-i{\bf Q}\cdot {\bf r}_i(0)}e^{i{\bf Q}%
\cdot {\bf r}_i(t)}\right\rangle =  \nonumber \\
&&\;\;\;\left\langle e^{i{\bf Q}\cdot \left ({\bf r}_i(t)-{\bf
r}_i(0) \right )+\frac{{\bf Q}^2}{2} \left [{\bf r}_i(0),{\bf
r}_i(t)
\right ]}\right\rangle = \;\;\;\;\;\;\;\; \nonumber \\
&&\;\;\;\left\langle e^{i{\bf Q} \cdot \frac{{\bf p}_i}{m}
t-\frac{{\bf Q}^2}{2} \left [\frac{{\bf p}_i}{m} t,{\bf r}_i(t)
\right ]} \right\rangle = \nonumber \\
&&\;\;\; e^{i\hbar\frac{Q^2t}{2m}}\left\langle e^{-i{\bf Q} \cdot
\frac{\bf p}{m} t} \right\rangle \nonumber
\end{eqnarray}

\noindent and, using the result of Eq.(\ref{fqt_free_class}), one
obtains:

\begin{eqnarray}
F(Q,t)=e^{-\frac{Q^2}{2m} \left(k_B T t^2-i \hbar t \right ) }
\end{eqnarray}

\and the correspondent dynamic structure factor:

\begin{equation}
S(Q,\omega)=\sqrt{\frac{m}{2\pi k_BTQ^2}}e^{-\frac{m}{2k_BTQ^2}
\left(\omega-\frac{\hbar Q^2}{2m} \right )^2 }
\label{sqw_free_quant}
\end{equation}

Summing up, the quantum dynamic structure factor for a free moving
particle is a gaussian with recoil energy $\omega_R(Q)=\frac{\hbar
Q^2}{2m}$ and linewidth $\sigma=\sqrt {\frac{k_BT}{M}}Q$. It is
worth to point out that Eq.(\ref{sqw_free_quant}) satisfies the
detailed balance condition $S(Q,\omega)=e^{\frac{\hbar \omega}{k_B
T}}S(Q,-\omega)$ and coincides with the classical case for $\hbar
\rightarrow 0$ or, equivalently, for $T \rightarrow \infty$. As
far as the sound velocity is concerned, one can still define the
apparent frequencies $\omega_l(Q)=\frac{1}{2}(\omega_R \pm
\sqrt{8\sigma ^2 + \omega_R^2})$ as the positive and negative
maxima of the longitudinal current $C_L(Q,\omega)=\frac{\omega^2
S(Q,\omega)}{Q^2}$. In the classical case, the two values
coincides and are $\omega_l(Q)=\sqrt{2}\sigma$:

\subsection{The non-hydrodynamic region: single particle} \label{sec_selfnonhydro}

\subsubsection{The gaussian approximation \label{sec_gaussapp}}

As can be easily noticed looking at the expressions
(\ref{Gs_hydro}) and (\ref{Gs_single}), both the hydrodynamic and
the ideal gas limit end up with the Van Hove correlation functions
that are gaussian in $r$. On the basis of this observation seems
natural to assume the gaussian dependence as valid in the whole
dynamical range. In terms of the second moment of $G_s(r,t)$ one
can write the following expression:

\begin{equation}
G_s(r,t)=\sqrt{\frac{3}{2\langle r^2(t) \rangle}}
e^{-\frac{3r^2}{2\langle r^2(t) \rangle}} \label{Gs_gauss}
\end{equation}

where $\langle r^2(t) \rangle$ is the mean square displacement
which, in the hydrodynamic and single particle approximations
reads $\langle r^2(t) \rangle=6Dt$ and $\langle r^2(t)
\rangle=\frac{3KT}{m}t^2$, respectively.

In the gaussian approximation, therefore, the self scattering
function is related to the mean square displacement which can be,
for instance, inferred by molecular dynamics simulations.

\subsubsection{The jump diffusion model \label{sec_jd}}

The jump diffusion model was firstly introduced by Chudley and
Elliot \cite{chud_jd}. The particle is thought to live for a
residence time $\tau_0$ in the cage of its neighbors, and at some
point to change cage. In some sense, therefore, it is the opposite
of collisional models, where the free diffusion of a particle is
sometimes interrupted by collisional events. The jump diffusion
model sets a rate equation for the Van-Hove self scattering
function of the kind:

\begin{eqnarray}
\frac{\partial G_s(\mathbf r,t)}{\partial t}=\frac{1}{\mathcal{N}
\tau_0} \sum_{\mathbf l} G_s(\mathbf {r+l},t)-G_s(\mathbf r,t)
\nonumber
\end{eqnarray}

\noindent with $\mathcal{N}$ the number of available residence
sites.

By fourier transform in space and time one immediately gets:

\begin{equation}
S_s(Q,\omega)=\frac{1}{\pi}\frac{f(Q)}{\omega^2+f(Q)^2} \label{jd}
\end{equation}

\noindent which is a lorenzian function with $Q-$ dependent
damping

\begin{equation}
f(Q)=-\frac{1}{n\tau_0}\sum_{\mathbf l}(e^{i\mathbf Q \cdot
\mathbf l}-1) \label{jd_width}
\end{equation}

\noindent This latter can be conveniently estimated supposing that
the vectors $\mathbf {l}$ have random and continuous orientations
and distributions. In this case one can average
Eq.(\ref{jd_width}):

\begin{eqnarray}
f(Q)=-\frac{1}{\tau_0}\left [ 1-\frac{1}{1+Q^2l_0^2} \right ]
\nonumber
\end{eqnarray}

It can be easily shown that Eq.(\ref{jd}) tends to the Fick's free
diffusion expression (\ref{fick}) \cite{EGELSTAFF}

\subsubsection{The mode coupling theory \label{sec_mct}}

The Laplace transform of the intermediate scattering function can
be generally written as:

\begin{eqnarray}
\tilde{S}_s(Q,s)=\frac{1}{s+Q^2\tilde{U}(Q,s)} \nonumber
\end{eqnarray}

\noindent being $\tilde{U}(Q,s)$ a generalized frequency and
wavevector dependent diffusion coefficient. The mode coupling
theory provides a self consistent expression for
$\tilde{U}_s(Q,s)$ \cite{desh_mc}, and the resulting self dynamic
structure factor reads:

\begin{equation}
S_s(Q,\omega)\approx \frac{1}{\pi} \frac{DQ^2}{\omega^2+(DQ^2)^2}
+\frac{1}{\pi D Q Q^*}\mathrm{Re}G\left(
\frac{i\omega+DQ^2}{\delta DQ^2} \right) \label{mct1}
\end{equation}

\noindent with

\begin{equation}
G(s)=arctan \left(\frac{1}{\sqrt{s-1}}  \right) -
\frac{(s-2)\sqrt{s-1}}{s^2} \label{mct2}
\end{equation}

\noindent where $Q^*=16\pi m \rho D^2 / k_BT$ and
$\delta=D/(D+\nu)$, being $D$ the diffusion coefficient and $\nu$
the kinematic viscosity.

An estimate of the FWHM can be numerically evaluated
\cite{dej_phd}, yielding:

\begin{equation}
\Delta \omega \approx \left
[1-\frac{Q}{Q^*}H(\delta)+O(Q^{3/2})\right ]DQ^2 \label{mct3}
\end{equation}

\noindent with

\begin{equation}
H(\delta) \approx 1.45\delta^{3/2}\left
[1-0.73\delta-0.15\delta^2-O(\delta ^3)\right ] \label{mct4}
\end{equation}

\subsubsection{The Nelkin-Ghatak model \label{sec_ng}}

Nelkin and Ghatak have considered a dilute gas in which the atomic
motion is dominated by binary collisions, with a distribution
function obeying a linearized Boltzman's equation, valid in a
small disturbance limit, i.e. for arbitrary large fluctuations
compared to the mean collision time. In terms of the reduced
variables $x=-\omega / Qv_0$ and $y=\alpha / Qv_0$, with
$v_0=\sqrt{2KT/m}$ and $\alpha$ an adjustable parameter, and by
introducing the real ($u(x,y)$) and imaginary ($v(x,y)$) parts of
the probability integral for complex argument
$z(x+iy)=\int_{-\infty}^{+\infty} e^{-t^2}(z-t)^{-1}dt$

\begin{eqnarray}
U(x,y)=\sqrt \pi y u(x,y) \\
V(x,y)=\sqrt \pi y v(x,y)
\end{eqnarray}

one gets the following expression for the incoherent scattering
function \cite{nelkin_inco}:

\begin{equation}
S^{NG}_s(Q,\omega)=\frac{1}{\pi
\alpha}\frac{U(1-U)-V^2}{(1-U^2)+V^2} \label{NG}
\end{equation}

\noindent it can be easily noticed that Eq.(\ref{NG}) has the
correct low $Q$ (lorenzian) and high Q (gaussian) limits: in the
first case it is sufficient to pose $\alpha=v_0^2/2D$, while at
high $Q$'s one has that $y\rightarrow 0$ and the familiar gaussian
shape is recovered.

\subsubsection{The memory function formalism \label{sec_mfs}}

The easiest way to abandon the hydrodynamic region is to assume
the frequency dependence of the transport coefficients, entering
the so called generalized hydrodynamics. The natural playground
for performing such step is the memory function framework
\cite{mori_mf}: we will recall here the basic formalism while a
detailed treatment can be found in specialized books
\cite{BALUCANI,HANSEN}. Let ${\bf M}^{(0)}$ be the correlation
matrix of a given set of dynamical variables $\bf A$
($M^{(0)}_{\nu \sigma}=\langle A_\nu ^* A_\sigma (t) \rangle $).
The equation of motion of ${\bf M}^{(0)}(t)$ can be conveniently
expressed in terms of a chain of arbitrary order $n$ of
integro-differential equation involving appropriate memory
functions ${\bf M}^{(i)}(t)$ for $i=1...n$:

\begin{equation}
\frac{d{\bf M}^{(i-1)}}{dt}-i {\bf \Omega}^{(i-1)} {\bf M}^{(i-1)}
+ \int_0^t {\bf M}^{(i)}(t-t') {\bf M}^{(i-1)}(t') dt'=0 \nonumber
\label{memory}
\end{equation}

\noindent with

\begin{equation}
i{\bf \Omega}^{(i-1)}={\bf \dot M}^{(i-1)}(0) \cdot \left [{\bf
M}^{(i-1)}(0)\right ]^{-1} \label{frequency}
\end{equation}

here ${\bf \Omega}^{(i-1)}$ is a set of generalized frequency
matrixes, while the memory kernels ${\bf M}^i(t)$ rule the
dynamical evolution of the observables correlation matrix ${\bf
M}^{(0)}$.

In the specific case of self dynamics, as we have seen in section
(\ref{sec_selfhydro}) the relevant set of variable is given by the
self density only, therefore: $M^{(0)}(t)=\phi_s (t)$. The
equation of motion for the density correlation function will be:

\begin{equation}
\frac{d\phi_s(Q,t)}{dt} + \int_0^t M^{(1)}(Q,t-t') \phi_s(Q,t')
dt'=0 \label{self_memory} \nonumber
\end{equation}

\noindent being $\Omega^{(0)}=0$ due to the orthogonality of
$\rho_s$ and $\dot{\rho_s}$.

A good description of the evolution of $\phi_s(Q,t)$ is normally
gained utilizing the first two equations of the chain
(\ref{memory}). In terms of Laplace trasform it holds:

\begin{equation}
{\tilde \phi}_s(Q,t)=\left [s+{\tilde M}^{(1)}(Q,s) \right]^{-1}
=\left [s+\frac{M^{(1)}(Q,t=0)}{s+{\tilde M^{(2)}}(Q,s)} \right
]^{-1} \label{phis_laplace}
\end{equation}

The initial values ${\bf M}^{(i)}(t=0)$, can be easily obtained
from the general relation:

\begin{equation}
{\bf M}^{(i)}(0)=-{\bf \ddot M}^{(i-1)}(0) \cdot \left [{\bf
M}^{(i-1)}(0)\right ]^{-1}-\left [{\bf \Omega}^{(i-1)} \right]^2
\end{equation}

\noindent which, in turns, is obtained deriving Eq.(\ref{memory})
and exploiting Eq.(\ref{frequency}). For the first two memory
function it holds therefore:

\begin{eqnarray}
&&M^{(1)}(Q,0)=\frac{k_BTQ^2}{m}=\langle \omega^2(Q) \rangle _{S_{s}} \nonumber \\
&&M^{(2)}(Q,0)=2\langle \omega^2(Q) \rangle
_{S_{s}}+\Omega_0^2=\frac{\langle \omega^4(Q) \rangle
_{S_{s}}}{\langle \omega^2(Q) \rangle _{S_{s}}}-\langle
\omega^2(Q) \rangle _{S_{s}} \nonumber \\
\label{initial_Ms}
\end{eqnarray}

\noindent Here $\langle \omega^n(Q) \rangle _{S_{s}}$ are the
frequency moments of $S_{s}(Q,\omega)$ and the quantity
$\Omega_0^2$ is related to the mean squared force $\langle |{\bf
F}|^2 \rangle$ acting on the diffusing particle and, for a system
of identical particle interacting via pairwise interactions
potential $V(r)$, it holds:

\begin{equation}
\Omega_0^2=\frac{\langle |{\bf F}|^2 \rangle }{3mk_BT}
=\frac{\rho}{3m}\int \nabla^2 V(r)g(r)d{\bf r} \label{mom4_s}
\end{equation}

\noindent where $g(r)$ is the pair distribution function.

It is worth to stress that Eq. (\ref{phis_laplace}) is the exact
solution of motion in which all the dynamics is detailed by the
shape of the second order memory function $M^{(2)}(Q,t)$. The most
common way of solving the equation is making a guess on the shape
of the memory function. A useful approximation is provided by the
exponential shape, which has the advantage of being easily Laplace
transformed:

\begin{equation}
M^{(2)}(Q,t)=\Delta^2_s(Q)e^{-t/\tau_s(Q)}=\left [ 2\langle
\omega^2(Q) \rangle _{S_{s}}+\Omega_0^2 \right] e^{-t/\tau_s(Q)}
\end{equation}

With such a choice it follows straightforward:

\begin{equation}
S_s(Q,\omega )=\frac{1}\pi \frac{\langle \omega^2 \rangle _{S_{s}}
( 2\langle \omega^2 \rangle _{S_{s}}+\Omega_0^2) \tau_s}
 {\omega^2 \tau_s (\omega ^2-3\langle
\omega^2 \rangle _{S_{s}}-\Omega_0^2) ^2+( \omega^2-\langle
\omega^2 \rangle _{S_{s}})^2} \label{sqw_s_1t}
\end{equation}

It is interesting now to look at the FWHM $\Gamma_s(Q)$ of
Eq.(\ref{sqw_s_1t}). In the small $Q$ limit it is easy to show
that

\begin{equation}
\frac{\Gamma_s(Q)}{DQ^2}=\frac{1}{\sqrt
{1+\frac{2k_BT}{m\Omega_0^2}Q^2}}
\end{equation}

The memory function approach with exponential kernel, therefore,
predicts a quasielastic lineshape narrower then the hydrodynamic
one, a result which is in agreement with several experimental
data. Contrarily, in the gaussian approximation the linewidth is
always larger than the hydrodynamic value.

Theoretical expressions have been proposed in the past for the
memory function in terms of kinetic theory, splitting the memory
function in a contribution due to uncorrelated binary collision,
obtained by a Fokker-Plank equation, and a long time contribution
representing the coupling of a tagged particle to the collective
motion of the surrounding particles. This approach have been
tested in hard spheres, Lennard-Jones and alkali metals
\cite{sjog_kin1,sjog_sw,beng_kinhs}

\subsection{The non-hydrodynamic region: collective motion \label{sec_collnonhydro}}

\subsubsection{The Langevin equation \label{sec_le}}

The most reliable approach to the study of collective dynamics at
finite wavevectors parallels the one adopted for the single
particle in section \ref{sec_selfnonhydro}, i.e. an extension of
the classical hydrodynamics assisted by the formalism of the
memory function ruling the Langevin equation of motion of the
density fluctuations.

We will deal, therefore, with a $3 \times 3$ correlation matrix
${\bf M}^{(0)}$, and a set of $n$ (with $n$ arbitrary large)
memory matrixes ${\bf M}^{(i)}$ with the same dimensionality and
coupled by the chain of $n$ equations (\ref{memory}). Actually, in
order to work with a orthogonal set of variables (and energy and
density are not), one normally prefers to replace the microscopic
energy with the microscopic temperature $T(Q)=\frac{1}{m \rho
c_V(Q)}\left [ E(Q)-\frac{\left \langle E ^*(Q) \rho(Q) \right
\rangle }{\left \langle \rho ^*(Q) \rho(Q) \right \rangle }\rho(Q)
\right ]$. With this choice, solving for $\phi(Q,t)=M^{(0)}_{\rho
\rho}(Q,t) / M^{(0)}_{\rho \rho}(Q,0)=F(Q,t)/S(Q)$ in terms of
Laplace transform one has:

\begin{equation}
{\tilde \phi}(Q,s)=\frac{1}{s+\frac{\omega_0^2(Q)}{s+{\tilde
M^{(1)}}_{JJ}(Q,s)-\frac{({\tilde M^{(1)}}_{JT}(Q,s)-i
\Omega^{(0)}_{JT})({\tilde M^{(1)}}_{TJ}(Q,s)-i
\Omega^{(0)}_{TJ})}{s+{\tilde M^{(1)}}_{TT}(Q,s)}}}
\label{coll_contfrac2}
\end{equation}

\noindent where

\begin{equation}
\omega _0^2(Q)=k_BTQ^2/mS(Q) \label{a2}
\end{equation}

Recalling the definition of the isothermal sound velocity
$c_t=\sqrt{1/\rho m \chi _T}$, and the expression for the low
$Q\rightarrow 0$ limit of the static structure factor $S(0)=\rho
\chi _T k_B T$ it seems natural to introduce a finite $Q$
generalization of the isothermal sound velocity as
$c_t(Q)=\omega_0(Q)/Q$.

It is easy to recognize in Eq. (\ref{coll_contfrac2}) the same
structure of Eq.(\ref{phis_laplace}), once the following
identification is made:

\begin{eqnarray}
\tilde{M}^{(eff)}_{\rho \rho}(Q,s)&=&{\tilde M^{(1)}}_{JJ}(Q,s)
\nonumber
\\ &-& \frac{({\tilde M^{(1)}}_{JT}(Q,s)-i
\Omega^{(0)}_{JT}(Q))({\tilde M^{(1)}}_{TJ}(Q,s)-i
\Omega^{(0)}_{TJ}(Q))}{s+{\tilde M^{(1)}}_{TT}(Q,s)} \nonumber
\end{eqnarray}

The effective memory function $\tilde{M}^{(eff)}_{\rho
\rho}(Q,t)$, therefore, can be formally considered as the second
order memory function of a chain of two equations for the single
variable $\rho(Q,t)$:

\begin{eqnarray}
&&\frac{dF(Q,t)}{dt} + \int_0^t M^{(1)}(Q,t-t') F(Q,t') dt'=0
\\
&&\frac{dM^{(1)}(Q,t)}{dt} + \int_0^t  M^{(2)}(Q,t-t')
M^{(1)}(Q,t') dt'=0 \label{collective_chain2} \nonumber
\end{eqnarray}

\noindent which, as can be easily verified, corresponds to the
single second order integro-differential equation:

\begin{eqnarray}
\stackrel{..}{\phi}(Q,t)&&+\omega _0^2(Q)\phi(Q,t) \\ &&+\int_0^tM(Q,t-t^{\prime })%
\stackrel{.}{\phi}(Q,t^{\prime })dt^{\prime }=0  \label{langevin}
\nonumber
\end{eqnarray}

\noindent From here on, to save writing, we define
$M(Q,t)=M^{(2)}(Q,t)=\tilde{M}^{(eff)}_{\rho \rho}(Q,t)$.

From the knowledge of $\widetilde{\phi}(Q,s)$ one straightforwardly obtains $%
S(Q,\omega )=[S(Q)/\pi] \Re \{ \widetilde{\phi}(Q,s=i\omega ) \}$ in terms of the real ($%
M^{\prime }$) and imaginary ($-M^{\prime \prime }$) parts of the
Fourier-Laplace transform of the memory function:

\begin{equation}
S(Q,\omega )=\frac{S(Q)}\pi \frac{\omega _0^2(Q)M^{\prime }(Q,\omega )}{%
\left[ \omega ^2-\omega _0^2-\omega M^{\prime \prime }(Q,\omega
)\right] ^2+\left[ \omega M^{\prime }(Q,\omega )\right] ^2}
\label{sqwgenerale}
\end{equation}

The spectral features of the dynamic structure factor can be
characterized by its frequency moments $\langle \omega^n (Q)
\rangle _S\equiv \int \omega ^nS(Q,\omega )d\omega $, where, for a
classical system, only the even frequency moments (such as
$\langle \omega^0 (Q) \rangle _S=S(Q)$ and $\langle \omega^2 (Q)
\rangle _S=\frac{k_B T}{m}Q^2 $ are different from zero.

It can be easily proven that the dynamic structure factor is
related to the longitudinal current spectrum through the relation
$C_{L}(Q,\omega )=\left( \omega ^{2}/Q^{2}\right) S(Q,\omega )$.
The presence of the factor $\omega ^{2}$ wipes out the low
frequency portion of the dynamic structure factor, and
consequently emphasizes the genuine inelastic features of
$S(Q,\omega )$. After its definition and Eq.(\ref{phis_laplace}),
it is readily seen that the Laplace transform
$\widetilde{C}_{L}(Q,s)$ satisfies

\begin{eqnarray}
\widetilde{C}_{L}(Q,s) &=&-s[s\widetilde{F}(Q,s)-S(Q)]  \label{cqz} \\
\ &=&\frac{k_BT}{m}\left\{ {s+[\omega _0^2(Q) / s]\ +\
\widetilde{M}(Q,s)}\right\} ^{-1} \nonumber
\end{eqnarray}

Again, the spectrum $C_L(Q,\omega )$ can be expressed as $(1/\pi
)\Re \{ \widetilde{C}_{L}(Q,s=i\omega ) \}$. Then the position and
the width of the
inelastic peaks in $C_L(Q,\omega )$ are determined by the poles of $%
\widetilde{C}_{L}(Q,s)$.

\subsubsection{Collective memory function and hydrodynamics\label{sec_mf}}

The effective, second order memory function $M(Q,t)$ accounts for
all the relaxation mechanisms affecting collective dynamics and,
consequently, is the central quantity in most theoretical
approaches. In analogy with the single particle case (Eq.
(\ref{initial_Ms})), the initial value of $M(Q,t)$ is related to
the spectral moments of $S(Q,\omega )$ by:

\begin{eqnarray}
M(Q,0)=\frac{\langle \omega^4(Q) \rangle _S}{\langle \omega^2(Q)
\rangle _S}-\langle \omega^2(Q) \rangle _S \doteq \Delta^2(Q)
\label{initial_M}
\end{eqnarray}

Along the same line, relations similar to (\ref{initial_Ms}) and
(\ref{mom4_s}) holds, and an explicit expression for $\langle
\omega^4(Q) \rangle _S$ can be given, involving both the
derivatives of the interparticle potential and the pair
distribution function:

\begin{equation}
\frac{\langle \omega^4(Q) \rangle _S}{\langle \omega^2(Q) \rangle
_S}=\frac{3K_{B}TQ^{2}}{m}+\frac{\rho}{m}\int
\frac{\partial^{2}V(r)}{\partial z^{2}} (1-e^{-iQz}) g(r) d^3 r
\label{winf}
\end{equation}

The second and fourth frequency moments are particularly
significant, as they rule the sound velocity in the whole
$Q-\omega$ domain. For sufficiently large $s$, indeed,
$\widetilde{M}(Q,s)\simeq M(Q,t=0)/s$ and Eq. (\ref{cqz}) is seen
to have poles at $s=\pm i\sqrt{\omega _0^2(Q)+\Delta ^2(Q)} \doteq
\pm i\omega_\infty(Q)$. This latter relation defines the frequency
$\omega_\infty(Q)$ which characterizes the instantaneous
collective response of the liquid at the wavevector $Q$ and, in
turns, defines the unrelaxed sound velocity as $c_\infty (Q)\equiv
\omega_\infty(Q)/Q$. In the opposite limit, i.e. for small $s$,
one easily verifies that $\widetilde{M}(Q,s)\simeq \int M(Q,t)
dt$. In this (relaxed) regime, therefore, the poles of the
longitudinal current are located at $s=\pm i \omega _0^2(Q)$, i.e.
the longitudinal modes propagate with the isothermal sound
velocity $c_t(Q)=\omega_0(Q)/Q$. Summing up, whatever the details
of the memory function are, in presence of a relaxation process
one observes a transition of the longitudinal sound velocity
between two different regime, associated to the evolution of the
longitudinal current correlation maxima. It is interesting to
study in a parallel way the evolution of $S(Q,\omega)$. In the low
frequency limit, Eq. \ref{sqwgenerale} reduces to the spectrum of
a damped harmonic oscillator (DHO) of characteristic frequency
$\omega_0$ (which, in general, does not coincides with the
position of the inelastic maximum), with damping $\Gamma = \int
M(Q,t) dt$.

\begin{equation}
\frac{S(Q,\omega)}{S(Q)}=\frac{1}{\pi}\frac{\omega_0^2(Q)\Gamma(Q)}{(\omega^2-\omega_0^2)+\omega^2\Gamma^2}
\end{equation}

In the opposite, high frequency, limit, again Eq.
(\ref{sqwgenerale}) tends to an harmonic oscillator with no
damping and with an higher characteristic frequency $\pm
\omega_\infty(Q)$. Additionally, it appears an elastic peak of
area $\frac{\Delta^2(Q)}{\omega_0^2(Q) + \Delta^2(Q)}$. The same
results can be directly retrieved in the time domain from Eq.
(\ref{langevin}), substituting the high frequency limit (constant)
and low frequency limit (delta shaped) of the memory function, in
turn, and taking the Fourier transform.

The memory function formalism leads, in the small wavevectors
limit, to the hydrodynamic prediction Of Eq.(\ref{S_hidro}). It
can be easily proven, indeed, that the diagonal terms of the
memory matrix appearing in Eq. (\ref{coll_contfrac2}) have a $Q^2$
dependence, while the cross terms follow a $Q^4$ dependence. For
$Q\rightarrow 0$, therefore, these latter can be neglected.
Moreover, from the continuity equations (\ref{3idro}) it follows
that in the same limit the time dependence of the conserved
quantities and their associated currents becomes increasingly
slow. Consequently, one can model the decay of the terms
$M^{1}_{JJ}(Q\rightarrow 0,t)$ and $M^{1}_{TT}(Q\rightarrow 0,t)$
as instantaneous, or equivalently, constant in the Laplace domain.
By making the identifications ${\tilde M}^{1}_{JJ}(Q\rightarrow
0,s)=D_V Q^2$ and ${\tilde M}^{1}_{TT}(Q\rightarrow 0,s)=\gamma
D_T Q^2$, and computing $\Omega^{(0)}_{JT}(Q)
\Omega^{(0)}_{TJ}(Q)=(\gamma-1)\omega_0^2(Q)$, from
Eq.(\ref{coll_contfrac2}) one gets for the memory function:

\begin{eqnarray}
\tilde{M}(Q\rightarrow
0,s)&=&D_VQ^2+\frac{(\gamma-1)\omega_0^2(Q)}{s+\gamma D_TQ^2} \nonumber \\
M(Q\rightarrow 0,t)&=&2D_VQ^2\delta
(t)+\omega_0^2(Q)(\gamma-1)e^{-\gamma D_TQ^2t} \nonumber \\
\label{mem_hydro}
\end{eqnarray}

The dynamic structure factor can be obtained by substituting Eq.
(\ref{mem_hydro}) in the general expression (\ref{sqwgenerale}).
In the $Q\rightarrow 0$ limit, moreover, one has
$\omega_0=c_tQ>>\gamma D_TQ^2$. In this limit, $S(Q,\omega)$ is:
i) a DHO function around the Brillouin peaks, which are located at
$\omega_s=\sqrt \gamma c_t Q$ (adiabatic sound propagation) and
have a linewidth $D_VQ^2+(\gamma -1) D_T Q^2$ ii) a lorenzian
function around $\omega=0$, whose linewidth is $2 D_TQ^2$. In the
small damping limit, the DHO is well approximated by two
symmetrically shifted lorenzian, and the hydrodynamic limit of
Eq.(\ref{S_hidro}) is finally recovered.

The advantage of the memory function approach is, however, in
providing a way to generalize the hydrodynamic result for
wavevector and frequency dependent transport coefficients. To this
purpose, from the very start it is convenient to separate in
$M(Q,t)$ the decay channels which explicitly involve couplings to
thermal fluctuations ( $M_{th}(Q,t)$ ) from those directly
associated with longitudinal density modes ( $M_{L}(Q,t)$ ).

\subsubsection{Finite wavelengths generalization: beyond hydrodynamics}

A straightforward generalization of ordinary hydrodynamics at
finite wavevectors suggests for the thermal contribution the
following form

\begin{equation}
M_{th}(Q,t)\approx (\gamma(Q)-1)\omega_0^2(Q)\exp[-\gamma
D_T(Q)Q^{2}t] \label{memoryth}
\end{equation}

where $D_T(Q)$ and $\gamma(Q)$ can be regarded as a finite $Q$
{\it generalization} of the quantities $D_{T}=\kappa /nC_{P},$
being $\kappa $ the thermal conductivity, and $\gamma=c_P / c_V$.

It must be stressed, however, how the extension to finite
wavevectors, in the special case of highly conductive systems,
requires attention to the physics behind this model. As pointed
out by Faber \cite{FABER}, in a liquid metal, owing to the high
thermal conductivity, the quantity $\gamma D_T(Q)Q^2$ may easily
become larger than the Brillouin frequency as soon as $Q\approx
\frac{c_t(Q)}{D_T(Q)}$ \footnote{More specifically, the
discrepancy observed in liquid lead between the sound velocity
measured with ultrasound (adiabatic) and with inelastic neutron
scattering has been tentatively assigned to the isothermal nature
of the sound propagation at the wavevectors probed with
neutrons.}. Actually, both $c_t(Q)$ and $D_T(Q)$ are expected to
decrease on approaching values comparable to the inverse mean
interparticle distance, i.e. in coincidence with the first sharp
rising edge of $S(Q)$. But assuming that the transition occurs
below this edge, one finds a crossover condition of
$Q=\frac{c_t}{D_T}$ which, considering typical values of sound
speed and thermal diffusivity of metals (a few thousands
meters/second and $\approx 50$ nm$^{-2}$/ps, respectively), lies
at wavevectors around $0.1\div 0.5$ nm$^{-1}$, which is indeed
consistent with the initial assumption. In other words, on
increasing $Q's$, the thermal peak broadens ultimately overlapping
with the Brillouin lines, the sound propagation turns from
adiabatic to isothermal, and independent thermal fluctuations
become impossible. By the memory function point of view, at
wavevectors $Q>>\frac{c}{D_T}$ (i.e. when the condition $\omega
\tau_{th} << 1$ holds), $M_{th}$ decays instantaneously in a
similar fashion to $M_L$. In this condition, the dynamic structure
factor is a DHO with isothermal characteristic frequency, while
the brillouin damping is given by the area under the memory
function which is $D_VQ^2 + \frac{(\gamma-1)c_t^2}{\gamma D_T}$.
The existence of this adiabatic to isothermal crossover, is
expected just below the lower accessible $Q's$ of an inelastic
scattering experiment ($\approx 1$ nm$^{-1}$ at present) and
therefore has never been observed directly. Moreover, the
magnitude of this effect is ruled by $\gamma-1$, which is normally
very small. A remarkable exception to this latter condition is
constituted by Nickel \cite{ber_ni} ($\gamma=1.88$), which has
been studied by INS, as it will be discussed in the next section.

Beside the thermal contribution, also the viscous term $D_V$ is
expected to exhibit a $Q$ dependence. For simplified Lennard-Jones
pairwise interaction it has been shown a decay of $\eta_l(Q)$ of a
factor ten up to the main peak of $S(Q)$, with $\eta_B$ going
negative over the same region \cite{tank_visco}.

\subsubsection{Finite frequencies generalization: viscoelasticity}

A second important generalization of transport coefficients
concerns their possible frequency dependence.

This latter case stems when the frequencies of the observed
density fluctuations are high enough that their timescale competes
with the one ruling the decay of $M_L$. In this condition, one has
to drop the hypothesis on the instantaneous (Markovian) nature of
the viscous term, and has to introduce a finite timescale ($\tau$)
for the decay of $M_L$. As discussed at the beginning of this
section, the timescale of $M_L$ sets a new crossover between two
different regimes, characterized by different sound velocities and
attenuations.

The simplest practical way to go beyond the hydrodynamic result
(\ref{mem_hydro}) is to allow for an exponential decay of
$M_L(Q,t)$:

\begin{equation}
M_L(Q,t)=\Delta _L^2(Q)e^{-t/\tau (Q)}\label{memoryL1}
\end{equation}

\noindent with $\Delta _L^2(Q)=\omega_L(Q)^2-\gamma
\omega_0(Q)^2$, in order to have the correct normalization of the
whole $M(Q,t=0)$.

Although this has the advantage of analytical simplicity when
dealing with Fourier transform, a drawback of this ansatz lies in
the violation of some basic short-time features of the memory
function (such as a zero derivative at $t=0$), causing the
divergency of $\langle \omega^n \rangle _S$ for $n\geq 6.$

Neglecting thermal effects, Eq. (\ref{memoryL1}) yields the
so-called {\it viscoelastic model} \footnote{Actually, the
viscoelastic model stems from the approximation (\ref{memoryL1}),
with the additional condition $\gamma=1$ ($M_{th}(Q,t)=0$). Within
the viscoelastic framework, indeed thermal effects are neglected,
in the sense that the hydrodynamic limit is isothermal (i.e.
$\Delta _L^2(Q)=\omega_L(Q)^2-\omega_0(Q)^2$)} for $S(Q,\omega )$
\cite{lov_visco}. Since as $Q\rightarrow 0$ $\widetilde{M}%
_L(Q,s=0)/Q^2$ can written as $[c_\infty ^2-c_0^2]\tau
(Q\rightarrow 0)$, the requirement that this coincides with $\eta
_L/nm$ shows that the time $\tau (Q)$ must be finite as
$Q\rightarrow 0$. Such a connection with viscous effects justifies
the physical interpretation of the rate $1/\tau (Q)$ as a
parameter giving an overall account of all relaxation processes by
which the longitudinal response of the liquid is affected by
time-dependent disturbances. In particular, for slow perturbations
developing over a timescale $t\gg \tau (Q)$ the system can adjust
itself to attain local equilibrium and the usual viscous behavior
holds. In contrast, for density fluctuations fast enough that
$t\ll \tau (Q)$ the liquid responds instantaneously, with a
solid-like (elastic) behavior. The crossover between these
limiting situations marked by frequencies $\omega $ such that $%
\omega \tau (Q)\approx 1$) is ultimately responsible for the
gradual changes often detected in the sound dispersion of several
liquids at increasing wavevectors.

Although appealing, the simplicity of the viscoelastic model can
be deceptive. First of all, the model itself provides no clue for
the physical origin of the decay mechanisms leading to the rate
$1/\tau (Q)$.

Actually, the situation is even more involved. In earlier MD
studies of Lennard Jones fluids, it was soon realized that the
viscous dynamics in the microscopic regime (i.e. at wavelength
comparable with the inverse mean inter-particle separation)
proceeds through two distinct processes, characterized by two well
separate time scales \cite{lev_2t}. More recently, the advent of
IXS provided the experimental evidence substantiating these
speculations
\cite{scop_prlli,scop_prena,scop_preal,scop_prlga,mon_k}.

In view of these result, the obvious remedy is to modify the
simple ansatz (\ref{memoryL1}) by allowing a two step decay of
$M_L(Q,t)$:

\begin{equation}
M_{L}(Q,t)=\Delta _{L}^{2}(Q){\ }\left[ {(1-\alpha (Q))e}^{{-\gamma }_{1}{%
(Q)t}}{+\alpha (Q)e}^{{-\gamma }_{2}{(Q)t}}\right] \label{x2tempi}
\end{equation}

where the rate ${\gamma }_{1}{(Q)}$ is chosen to be larger than ${\gamma }%
_{2}{(Q)}$, so that the dimensionless factor ${\alpha (Q)}$
measures the relative weight of the ''slow'' decay channel.
Besides being more flexible than the viscoelastic model, we shall
see that the ansatz (\ref{x2tempi}) has the much more important
merit that the presence of two different timescale does in fact
have a definite physical interpretation.

A simplified version of the previous ansatz has been implicitly
introduced in the viscoelastic analysis of Brillouin Light
Scattering (BLS) spectra of glass forming materials (see for
example \cite{cumm_visco}. In fact, in these works, the general
expression of $M_L(Q,t)$ for a two times decay is always expressed
as

\[
M_L(Q,t)={2}\Delta _\mu ^2(Q)\tau _\mu (Q){\delta (t)+}\Delta _\alpha ^2(Q){e}^{%
{-t/\tau }_\alpha (Q)}
\]

with explicit reference to the so-called $\alpha $- (structural)
relaxation process as responsible for the long lasting tail, and
to the $\mu -$ microscopic process as additional, faster,
relaxation dominant over a very short timescale (in the BLS window
the condition $\omega \tau _\mu <<1$ holds).

On the contrary, the above mentioned IXS works it have shown how
this approximation is no longer tenable in the case of liquid
metals at the IXS frequencies. In particular, it has been pointed
out that the slower (-$\alpha$) relaxation time satisfies the
condition $\omega_B(Q) \tau_\alpha(Q)>>1$, i.e. some part of the
viscous flow is frozen. As a consequence, at the wavevectors
typical of the IXS experiments ($Q=1\div 20$ nm$^{-1}$) the
quasielastic spectrum acquires a component arising from this
frozen structural relaxation.

The origin, at the atomic level, of this fast decay channel is
still an open issue: the rapidly decaying portion of $M_L(Q,t)$ is
customarily attributed to largely uncorrelated collisional events,
similar to those occurring in a dilute fluid. In addition, at the
high densities typical of the liquid state, non-negligible
correlations among the collisions can be expected, making no
longer valid an interpretation only in terms of ''binary''
collisions. Although the magnitude of the correlation effects is
relatively small and their buildup slow, once established their
decay is even slower, and for $t>1/{\gamma }_1{(Q)\equiv \tau
}_\mu (Q)$ this relaxation channel dominates the decay of
$M_L(Q,t)$, which consequently may exhibit a small but
long-lasting ''tail''. The ansatz (\ref{x2tempi}) can incorporate
most of this physics: on the basis of the latter, one may
reasonably anticipate that ${\alpha (Q)\ll }1$, and that the time $1/{\gamma }_2{%
(Q)\equiv \tau }_\alpha (Q)$ is distinctly longer than $1/{\gamma
}_1{(Q)\equiv \tau }_\mu (Q)$. In this picture, the best fitted
values of the viscoelastic rate $1/\tau (Q)$ clearly represent
some sort of ''weighted average'' between ${\gamma }_1{(Q)}$ and
${\gamma }_2{(Q)}$. Finally, we may argue that at increasing $Q$
(namely, over a shrinking length scale) the magnitude ${\alpha
(Q)}$ of correlation effects should decrease, and that at higher
temperatures the value of ${\alpha (Q)}$ at a given wavevector
should
equally decrease. On a general basis, the requirement that $%
\lim_{Q\rightarrow 0} \tilde{M}_L(Q,s=0)/Q^2\rightarrow \eta
_L/nm$ now takes the form

\begin{equation}
(c_\infty ^2-c_0^2)\left[ \frac{{(1-\alpha (Q\rightarrow 0))}}{{{\gamma }_1{%
(Q\rightarrow 0)}}}{\ +\ }\frac{{\alpha (Q\rightarrow 0)}}{{{\gamma }_2{%
(Q\rightarrow 0}\ )}}\right] \rightarrow \eta _L/nm
\label{limite}
\end{equation}

The refined model (\ref{x2tempi}) yields a dynamic structure
factor given by

\begin{eqnarray}
&&S(Q,\omega )=\frac{S(Q)}\pi \nonumber \\
&&\ \times Re\left\{ \frac{\omega _0^2(Q)}{i\omega +\frac{\Delta _\mu ^2(Q)}{%
i\omega +{\gamma }_1{(Q)}}+\frac{\Delta _\alpha ^2(Q)}{i\omega +{\gamma }_2{%
(Q)}}+\frac{\Delta _{th}^2(Q)}{i\omega +{\gamma (Q)
D_T(Q)Q}^2}}\right\} ^{-1} \label{x3tempi}
\end{eqnarray}

One of the major drawbacks of expression (\ref{x3tempi}), is that
one normally overestimates the relaxation strength of the faster
viscous process. Such problem is somehow expected as one is trying
to force an exponential dependence to reproduce the memory
function at short times, which, instead, has a zero derivative in
the $t \rightarrow 0$ limit. As a consequence, adjusting the
characteristic time of the exponential memory on the experimental
data one can reproduce the decay of the true memory function but
will overestimate the short time limit, due to the cusp behavior
of the exponential at $t=0$.

An obvious remedy it is a better choice of the memory function
model. An alternative possibility could be a gaussian shape. Also
the $sech$ function has been proposed as solution of the Mori
Equation \cite{tank_sech}, but this latter case is of limited
practical interest, as the fourier transform is related to the
digamma function and, therefore, the expression (\ref{x3tempi})
must be numerically evaluated.

\subsection{Kinetic theories: the hard sphere approximation \label{sec_kin}}

Special attention has been devoted in the past to the theoretical
and numerical study of the hard sphere model, as it conveniently
mimics the behavior of more realistic simple liquids
\cite{lebo_hs,furt_hs}. In the eighthes, transport coefficients
\cite{all_hs} and neutron scattering response \cite{all_hs1} have
been evaluated by means of molecular dynamics. On the theoretical
side, the dynamical properties have been investigated by a
revisiting the so called "Enskog fluid", i.e. by means of the
spectral decomposition of the Enskog operator
\cite{desh_hyd0,desh_hs,coh_hs1,kag_hs,mry_gcm}. This latter
approach turned out to be particularly useful to describe INS
experimental data \cite{coh_hs}, and we will briefly recall the
basics in this section.

The main idea beyond the Enskog's theory is to evaluate the
correlation functions (\ref{cf}) replacing the Liouville operator
$\mathcal{L}$ and the dynamical variables $a_\alpha(Q)$, defined
at the $N-$ particle ensemble level, with the one particle
Enskog's operator $L$, and appropriate one particle variables
$\phi_\alpha(Q)$. It can be easily recognized, indeed, that

\begin{eqnarray}
a_\alpha(Q)=\frac{1}{\sqrt N}\sum_j \phi_\alpha(\mathbf{v}_j)
e^{-i\mathbf{Q} \cdot \mathbf{r}_j} \nonumber
\end{eqnarray}

\noindent where, for the first three dynamical variables
(\ref{3dv}), it holds

\begin{eqnarray}
&&\phi_1(\mathbf{v})=\frac{1}{S(Q)} \nonumber \\
&&\phi_2(\mathbf{v})=\sqrt{\frac{m}{k_BT}}\frac{\mathbf{Q}\cdot
\mathbf{v}}{Q} \nonumber \\
&&\phi_3(\mathbf{v})=\frac{3-mv^2/k_BT}{\sqrt 6} \nonumber
\end{eqnarray}

For an hard sphere system, a possible asymmetric representation of
$L$ reads \cite{desh_hs}:

\begin{equation}
L(\mathbf{Q},\mathbf{v}_1)=-i\mathbf{Q}\cdot\mathbf{v}_1+\rho
g(\sigma) \Lambda(\mathbf{Q}) + \rho A(\mathbf{Q}) \label{enskog}
\end{equation}

\noindent where $g(\sigma)$ is the pair distribution function
evaluated at the contact point between two spheres. The first term
appearing in (\ref{enskog}) is a free streaming contribution. The
second term accounts for binary collisions through the operator
$\Lambda(\mathbf{Q})$, defined through its action over a generic
function of the velocity $f(\mathbf{v}_1)$

\begin{eqnarray}
&&\Lambda(\mathbf{Q})f(\mathbf{v}_1)=-\sigma^2 \int d\hat{\sigma}
\int d\mathbf{v}_2 \xi(v_2) \mathbf{\delta}
\theta(\mathbf{\delta}) \nonumber
 \\ && \{f(\mathbf{v}_1)-f(\mathbf{v}_1')+e^{-i\mathbf{Q} \cdot \hat
{\sigma} \sigma} [f(\mathbf{v}_2)-f(\mathbf{v}_2')] \}
\label{enskog_2}
\end{eqnarray}

\noindent with $\xi(v)$ the normalized Maxwell distribution
function, $\theta(x)$ the Heaviside step function, $\hat{\sigma}$
the unit vector, $\mathbf{\delta}=(\mathbf{v}_1-\mathbf{v}_2)\cdot
\hat{\sigma}$ and $\mathbf{v}_{1,2}'=\mathbf{v}_{1,2}\mp \delta
\hat{\sigma}$ the post-collision velocity. The third term,
finally, is a mean field operator defined through

\begin{eqnarray}
&&\rho A(\mathbf{Q})f(\mathbf{v}_1)=[C(Q)-g(\sigma)C_0(Q)]
\nonumber
 \\ &&  \int
d\mathbf{v}_2 \xi(v_2) i \mathbf{Q} \cdot \mathbf{v}_2
f(\mathbf{v}_2) \label{enskog_3}
\end{eqnarray}

\noindent where $C(Q)=[1-\frac{1}{S(Q)}]$ and
$C_0(Q)=\lim_{\rho\rightarrow 0} C(Q)$.

An explicit expression for the dynamic structure factor can be
easily retrieved thought the spectral decomposition of $L$:

\begin{eqnarray}
L(\mathbf{Q},\mathbf{v}_1)=-\sum_j |\Psi_j(\mathbf{Q},\mathbf{v}_1
\rangle z_j(Q) \langle \Phi_j(\mathbf{Q},\mathbf{v}_1) | \nonumber
\end{eqnarray}

\noindent in which $z_j$, $\Psi_j$ and $\Phi_j$ are eigenvalues,
left and right eigenfunctions of $-L$, respectively. $S(Q,\omega)$
then reads

\begin{eqnarray}
\frac{\pi S(Q,\omega)}{S(Q)}&=&\mathrm{Re}\left \langle
\frac{1}{i\omega-L(\mathbf{Q},\mathbf{v}_1)}\right \rangle_1
\nonumber
\\ &=&\mathrm{Re} \sum_j \frac{B_j(Q)}{i\omega+z_j(Q)} \label{ehm}
\end{eqnarray}

with

\begin{equation}
B_j(Q)=\langle \Psi_j(\mathbf{Q},\mathbf{v}_1)\rangle_1 \langle
\Phi_j^*(\mathbf{Q},\mathbf{v}_1) \rangle_1 \label{hscoeff}
\end{equation}

The subscript $\langle ... \rangle_1$ explicitly indicates that we
are now dealing with single particle averages over the Maxwell
Bolzmann velocity distribution function.

There are several approaches to determine the spectrum of $L$, but
the main point is that different approximations can be performed
according to the density and kinematic region of interest. These
regions are marked by the values of the reduced density
$V_0/V=\rho \sigma^3/\sqrt 2$, $V_0$ being the closed packed
volume for spheres or radius $\sigma$, and the Enskog mean free
path $l_E=l_0/\chi$ with $l_0$ the Bolzman mean free path
$l_0=1/\sqrt 2 \rho \pi \sigma^2$ and $\chi=g(\sigma)$ the pair
distribution function evaluated at the contact point between two
spheres.

The lower three eigenvalues of $L$ always go to zero with
$Q\rightarrow 0$. In the same limit it can be shown that these
latter eigenvalues are:

\begin{eqnarray}
&&z_1(Q)=z_h(Q)=D_{TE}Q^2 \nonumber \\
&&z_{2,3}(Q)=z_\pm (Q)=\pm i c_0Q+\Gamma_E Q^2 \label{coeff_hyd}
\end{eqnarray}

\noindent and only the first three coefficients (\ref{hscoeff})
are relevant, so that the eigenfunctions are linear combinations
of the density, current and energy variables. In other words the
hydrodynamic result of Eq. (\ref{S_hidro}) is recovered, with the
dynamic structure factor composed of three Lorenzian functions: a
diffusive heat mode and two propagating modes with the adiabatic
sound velocity, with dampings $D_{TE}$ and $\Gamma_E$ as given
within the Enskog's transport theory. The relative intensities of
the thermal and acoustic contributions are ruled by the Landau
Placzek ratio.

This limit is practically attained at low densities ($V_0/V <0.1$
and therefore $l_E\approx l_0$) and sufficiently small $Q$'s
($Q\sigma<<1$), when the contribution (\ref{enskog_3}) can be
safely neglected and the term (\ref{enskog_2}) can be replaced
with $\Lambda(Q\sigma=0)$. This normally occurs up to the case of
light scattering of dilute gases ($Ql_0\approx 1$), and one speaks
in terms of tree extended hydrodynamic modes. Still in the same
low density-momentum range, but at wavevectors $1\le Ql_0 \le 3$ a
description in terms of a few hydrodynamic modes is no longer
allowed, while for $Ql_0> 3$ one can evaluate the first order
corrections to the free particle result of Eq.(\ref{S_fp_cl})
which reads:

\begin{equation}
S(Q,\omega)=\sqrt{\frac{m}{2\pi k_B T
Q^2}}\left[e^{-\frac{m\omega^2}{2 k_B T
Q^2}}+\frac{S^B(\omega/Q)}{Ql_0}+O(1/Ql_0)^2\right ]
\label{S_fp_hs}
\end{equation}

\noindent with $S^B(\omega/Q)$ the leading correction to the free
streaming term $-i\mathbf{Q}\cdot\mathbf{v}_1$ of Eq.
(\ref{enskog}) due to a single binary collision event, which can
be numerically estimated \cite{kag_hs}.

Conversely, for dense fluids ($V_0/V>0.35$), the hydrodynamic
scheme breaks down at as soon as $Ql_E>0.05$. Above this value,
only the extended heat mode is well separated from all the other
modes. Still a description in terms of three \textit{effective}
hydrodynamic modes apply, and the low $Q$ limit of these modes is
again coincident with the hydrodynamic result. For $0<Ql_E<1$ the
free streaming and the mean field contributions of the Enskog
operator can be treated as perturbation to the binary collision
term (\ref{enskog_2}), and the following approximate expression
for the extended heat mode can be given:

\begin{equation}
z_h(Q)=\frac{D_EQ^2}{S(Q)}d(Q) \label{zh_hs}
\end{equation}

\noindent in which

\begin{eqnarray}
D_E=-\left \langle v_{1x}\frac{1}{\rho \chi \Lambda(Q\rightarrow
\infty)}v_{1x}\right \rangle \nonumber
\end{eqnarray}

\noindent is the Enskog diffusion coefficient and

\begin{eqnarray}
d(Q)&=&-\frac{\left \langle v_{1x}\frac{1}{\rho \chi
\Lambda(Q)}v_{1x}\right \rangle}{D_E}\approx \frac{\left \langle
v_{1x}\Lambda(Q\rightarrow \infty)v_{1x}\right \rangle}{\left
\langle
v_{1x}\Lambda(Q)v_{1x}\right \rangle} \nonumber \\
&=& \frac{1}{1-j_0(Q\sigma)+2j_2(Q\sigma)}
\end{eqnarray}

can be approximated in terms of the first two even spherical
Bessel functions. Enskog's diffusion coefficient is related to the
Bolzman diffusion coefficient

\begin{eqnarray}
D_0=\frac{3}{8\rho \sigma^2}\sqrt{\frac{k_B T}{\pi m}} \approx
\frac{0.216}{\rho \sigma^2}\sqrt{\frac{k_B T}{m}} \nonumber
\end{eqnarray}

\noindent by the collision enhancing term $g(\sigma)$ as
$D_E=D_0/g(\sigma)$. Using the analytic expression of $g(\sigma)$
for non attractive hard spheres one finally gets an expression for
$D_E$ in terms of the packing fraction $\varphi=\pi \rho \sigma^3
/ 6$:

\begin{eqnarray}
D_E=\frac{1}{16}\sqrt{\frac{\pi k_B T}{m}}\sqrt[3]{\frac{6}{\pi
\rho \varphi ^2}} {\frac{(1-\varphi)^3}{1-\varphi/2}} \nonumber
\end{eqnarray}

As in the low density case, for $Ql_E>3$ the the free streaming
limit is recovered along with the binary collisions corrections.
In this case, however, a different approximation holds, since the
binary collision term (\ref{enskog_2}) is now replaced by the
Lorentz-Boltzmann operator $\Lambda(Q\rightarrow \infty)$. One
ultimately gets an expression similar to (\ref{S_fp_hs}) in which
$S^B(\omega/Q)$ is replaced by the different Lorentz-Boltzmann
expression.

At intermediate densities, finally, again the hydrodynamic result
does not hold for $Ql_E\gtrsim 0.05$. Moreover, in this regime
three effective modes do not suffice and one has to extend the
description with two additional kinetic modes, i.e. including the
kinetic part of the z-z component of the stress tensor and the
z-component of the heat flux \cite{kag_hs}. Like in the high
density case, for $1<Ql_E<3$ all the modes are closely interwoven,
while at larger $Q$'s the free streaming limit with the
Lorentz-Boltzmann correction is retrieved.

According to kinetic theory, therefore, $S(Q,\omega)$ in simple
liquids not too far from the melting temperature (such as Argon),
should be described in terms of three lorenzians up to relatively
large wavevectors ($Q\approx 30$nm$^{-1}$). Consequently, sound
modes should exist even in a $Q$ region where side peaks are not
distinctly observable. In this respects, Lovesey argued that the
extended hydrodynamic picture should break up above $Q\approx 3$
nm$^{-1}$, while above a viscoelastic theory should be utilized
\cite{lov_nokin}.

Exploiting the previous results obtained for the coherent case,
one can describe the incoherent dynamics via the Lorenz-Enskog
operator

\begin{equation}
L_s(\mathbf{Q},\mathbf{v}_1)=-i\mathbf{Q}\cdot\mathbf{v}_1+\rho
g(\sigma) \Lambda(\mathbf{Q\rightarrow \infty}) \label{enskog_sp}
\end{equation}

As already mentioned, at large $Q$'s the self and the collective
dynamics coincide, and therefore the spectrum of $L$ will tend to
the one of $L_s$ ($A(Q\rightarrow \infty)=0$). In this limit the
extended heat mode $z_h$ will tend to the self-diffusion mode
$z_D$. In the opposite, $Q\rightarrow 0$ limit, they both approach
their hydrodynamic values: $z_h(Q\rightarrow 0)=D_{TE}Q^2$ and
$z_D(Q\rightarrow 0)=D_{E}Q^2$, with $D_{TE}$ and $D_E$ the
thermal diffusivity and the self diffusion coefficient of the
Enskog's theory. At intermediates $Q$ values, $z_h$ oscillates
around $z_D$, with a periodicity dictated by $S(Q)$ and $d(Q)$
according to Eq. (\ref{zh_hs}).

Sears has calculated the moments of the self part of the Van Hove
scattering function al large wavevectors \cite{sears_inco}.
Starting from the general case of a velocity independent central
force field, he specialized the result to the hard sphere case,
evaluating the leading correction to the impulse approximation due
to final state interactions, which turns out to be $Q^{-1}$:

\begin{equation}
\omega_H(Q)\approx \sqrt{\frac{2k_BTln2}{m}}Q\left(
1-\frac{0.27}{Ql_E}+O(Q^{-2})\right) \label{sears}
\end{equation}

The transition from the Fickian to the Gaussian regime occurs in
this case at $Ql_E\approx 1$.

Finally, it is worth to recall here one the most significant
achievement of the mode coupling theory applied to hard sphere
fluids, but which has been shown to apply to the wider class of
simple fluids. Ernst and Dorfman \cite{ern_pdis} have shown how
Eq.(\ref{coeff_hyd}), which retrieves the hydrodynamic expression
of the sound velocity and attenuation, is actually the leading
term of an expansion of the kind:

\begin{eqnarray}
z_h(Q)&=&\alpha_h Q^2 - \beta_h Q^{5/2} + O(Q^{11/4}) \nonumber \\
z_\pm (Q)&=&\pm i c_0Q+\alpha_{{_\pm}} Q^2 + (\pm
i-1)\beta_{{_\pm}} Q^{5/2} + O(Q^{11/4})  \nonumber \\
\label{coeff_hyd_MCT}
\end{eqnarray}

\subsection{The ionic plasma \label{plasma}}

The dynamical descriptions given up to this point are extensions
of models holding for ordinary fluids which, in some cases, are
modified \textit{ad hoc} to account for the high thermal
conductivity of liquid metals.

A totally different approach is to look at liquid metals, from the
very start, as a one component plasma, i.e. as an assembly of
identical, point-like charged particles (ions) embedded in a
uniform background (the electrons) which neutralizes the total
charge \cite{Baus_ocp}. The long-ranged, Coulomb interaction
active in this case give rises to peculiar phenomena such as
screening effects and plasma oscillations.

The ionic number density $\rho(\mathbf{r},t)$ and the current
density $\mathbf j(\mathbf{r},t)$ are related through the
continuity equation:

\begin{eqnarray}
\frac{\partial \rho(\mathbf{r},t)}{\partial t}=-\nabla \cdot
\mathbf j(\mathbf{r},t) \nonumber
\end{eqnarray}

The Poisson equation, for a fluid with $Ze$ charge, reads

\begin{eqnarray}
\nabla \cdot \mathbf E(\mathbf{r},t)= 4\pi Ze \delta \rho
(\mathbf{r},t) \nonumber
\end{eqnarray}

\noindent in which $\delta \rho (\mathbf{r},t)$ is the deviation
of the ionic density from its average value $\rho$ (neutralized by
the opposite uniform electronic density). Neglecting thermal
conductivity effect (collisionless regime), from the Newton's law
for the equation of motion one can find a third equation and close
the system. In the limit of large wavelengths fluctuations
(compared to the ionic size) one can write

\begin{eqnarray}
m\frac{\partial \mathbf j(\mathbf{r},t)}{\partial t}=\rho
Ze\mathbf E(\mathbf{r},t) \nonumber
\end{eqnarray}

being $m$ the ionic mass. Combining the previous equations one
easily gets:

\begin{eqnarray}
\frac{\partial^2 \rho(\mathbf{r},t)}{dt^2}=-\Omega^2_p \delta
 \rho (\mathbf{r},t) \nonumber
\end{eqnarray}

\noindent which describes ionic plasma oscillation of
characteristic $Q$ independent frequency

\begin{equation}
\Omega^2_p=\frac{4\pi \rho Z^2e^2}{m} \label{oplas}
\end{equation}

This simplified picture, therefore, contrasts with the
experimental evidence of long wavelength excitations whose
frequency vanishes in the $Q\rightarrow 0$ limit. The commonly
used remedy is to "dress" the plasma frequency accounting for the
electron screening effect, i.e. to take into account that the
background electrons have their own dynamics that can be described
introducing the dielectric response $\epsilon(Q)$. In such  a way
the coulomb interactions and, in turn, the plasma frequency are
renormalized leading to the expression:

\begin{equation}
\omega^2_p=\frac{\Omega^2_p}{\epsilon(Q)} \label{oplas_d}
\end{equation}

In the small wavevector limit the Thomas-Fermi expression within
the Random Phase Approximation (RPA, in with the system is assumed
to have a free electron gas compressibility) yelds the useful
expression \cite{MARCH}

\begin{equation}
\epsilon(Q)=1+\frac{Q^2_{TF}}{Q^2} \label{TF}
\end{equation}

\noindent in which $Q_{TF}=6\pi e^2\rho_e/E_F$, with $E_F=\hbar^2
(3\pi^2\rho_e)^{2/3} / 2m_e$, $\rho_e$ and $m_e$ the Fermi energy,
the electronic density and the electronic mass, respectively. At
small $Q$ values, Eq.(\ref{TF}), together with Eqs.(\ref{oplas})
and (\ref{oplas_d}), leads to the so called Bohm-Staver
expression, i.e. a dispersive excitation with sound velocity

\begin{equation}
c_{BS}=\frac{\omega_p}{Q}=\sqrt{\frac{m_eZ}{3m}}v_F \label{BS}
\end{equation}

\noindent in which $v_F$ the electron velocity at the Fermi level.

As we will show in the following, the experimental values for
sound velocities in conductive liquids often contrast with the
prediction of  Eq.(\ref{BS}), especially at small electron density
or, equivalently, at large values of the reduced ionic radius
$r_s=(\frac{3}{4\pi \rho_e a_0^3})^{1/3}$, where the Thomas-Fermi
approximation is no longer valid.

The interplay between plasma oscillations and sound waves can be
better accounted for within a two component plasma description
(TCP) in which nuclei and electrons are treated separately
\cite{chih_tcp}, eventually exploiting the framework of
Mori-Zwanzig \cite{hans_tcp}. The details of the TCP are, however,
beyond the scope of this review, the interested reader might want
to consult more specialized literature.

\section{Experimental study of the microscopic dynamics \label{sec_exp}}

The main experimental ways to study microscopic dynamics in liquid
metals are acoustic spectroscopy and inelastic scattering
experiments. These latter have to be necessarily performed with
probes which can penetrate enough into the sample to give boundary
free information, a requirement which restrict the choice to
neutrons and X-rays only. Actually, a few attempts have been
performed by means of visible light scattering \cite{dil_rev}, for
instance on liquid mercury and gallium \cite{dil_blsmet}, but
Brillouin scattering from opaque, liquid matter, presents several
difficulties, mainly associated with the ill definition of the
exchanged momentum in absorbing media.

The basics of the neutron interaction with matter have been
surveyed in many papers \cite{coplov_rev} and books
\cite{MARSHALL,LOVESEY,EGELSTAFF} which provide exhaustive survey
of this issue. Inelastic X-ray scattering is a relatively newer
technique, and therefore we will detail the basics of an IXS
experiment recalling from time to time INS features for
comparison.

\subsection{The scattering problem}

The measured signal in an inelastic scattering experiment is
determined by the double differential scattering cross-section.
Within the linear response theory, where it is assumed that the
coupling between the probe and the system is weak, this scattering
differential cross-section can be written quite generally as the
product of three terms: i)  One term describes the intensity of
the probe-sample coupling, and it is independent from the energy
of the incident particle. ii)  A second  one is a kinematic term
related to the phase-space volumes of the incident and scattered
particles. iii)  The third term is the space and time Fourier
transform of the correlation function of the observable in the
system that couples to the probe particle. This last quantity is
the one related to the elementary excitations characteristic of
the system.

\subsubsection{The photon-electron interaction Hamiltonian}

The actual expression for the scattering cross section can be
derived by a perturbation expansion from the probe-system
interaction Hamiltonian. In the case of the interaction of charges
with the electromagnetic field, in the weak relativistic limit
(i.e. to first order in $v^2/c^2$), neglecting the direct coupling
of the field with the nuclei (i.e. to zero order in the
electron-to-nuclei mass ratio $m_e /m$ ), and neglecting the
magnetic terms (i.e. to the zero order in the electron spin) one
gets:

\begin{eqnarray}
H_{INT}&=&\frac{e^2}{2m_ec^2} \sum_j \mathbf{A}(r_j)\cdot
\mathbf{A}^*(r_j) + \frac{e}{2m_ec} \sum_j \{ \mathbf{A}(r_j),
\mathbf{p}_j \} \nonumber \\ &\doteq& H^{(1)}_{INT} +
H^{(2)}_{INT} \nonumber
\end{eqnarray}

\noindent where the symbol $\{\}$ denotes the anti-commutator
operator. The two "electric" terms, $H^{(1)}_{INT}$ and
$H^{(2)}_{INT}$ contain, respectively, two and one field operators
$\mathbf{A}(r)$. It is clear, therefore, that - in a perturbation
expansion treatment of the interaction Hamiltonian - the term
$H^{(1)}_{INT}$ will give rise to two-photons processes at first
order while the term $H^{(2)}_{INT}$ will give rise to one-photon
processes at first order. To have the two-photon processes from
the latter (the so called $\mathbf{p}\cdot \mathbf{A}$
contribution), necessary to describe the scattering events, one
must consider the second order in the perturbation expansion,
which is consequently completely negligible in the off-resonance
case. In the following, therefore, we will consider only the first
charge scattering term.

\subsubsection{The X-ray scattering cross-section.}

The double differential cross-section,
$\frac{\partial^2\sigma}{\partial E \partial \Omega}$, is
proportional to the probability that an incident particle is
scattered with a given energy and momentum variation within an
energy range $\Delta E$ and a solid angle $\Delta \Omega$. In the
process, a photon of energy $E_i$, wavevector $\mathbf{k}_i$, and
polarization $\mathbf{\epsilon}_i$, is scattered into a final
state of energy $E_f$, wavevector $\mathbf{k}_f$, and polarization
$\mathbf{\epsilon}_f$, and the electron system goes from the
initial state $|I\rangle$ to the final state $|F\rangle$ (states
with energies $E_I$ and $E_F$, respectively). According to this
definition, the double differential cross section can be related
to the quantity $\frac{dP_{i\rightarrow f}}{dt}$ which is the
probability rate per sample and probe unit that a probe particle
makes the transition from the initial state to the final state:

\begin{eqnarray}
\frac{\partial^2\sigma}{\partial E \partial
\Omega}=\frac{dP_{i\rightarrow f}}{dt} \frac{1}{j}
\frac{\partial^2 n}{\partial E \partial \Omega} \nonumber
\end{eqnarray}

In this equation $j$ is the incident particle current density
($j=\rho v$, being $\rho$ the particle density and $v$ its
velocity) and $\frac{\partial^2 n}{\partial E \partial \Omega}$
the density of states of the scattered particle. For zero mass
particles, the latter two quantities can be written as:

\begin{eqnarray}
j&=&\frac{c}{V_0} \\
\frac{\partial^2 n}{\partial E \partial
\Omega}&=&\frac{V_0}{8\pi^3} \frac{k_f^2}{\hbar c}
\end{eqnarray}

Therefore, the double differential cross section becomes

\begin{equation}
\frac{\partial^2\sigma}{\partial E \partial
\Omega}=\frac{V_0^2}{8\pi^3} \frac{k_f^2}{\hbar c^2}
\frac{dP_{i\rightarrow f}}{dt} \label{csec}
\end{equation}

The transition of the incident particles between states $i$ and
$f$ involves, in general, different possible elementary
excitations in the sample. This implies that, indicating with
$\frac{dP_{i,I \rightarrow f,F}}{dt}$, the scattering probability
involving the transition in the sample from the state $|I\rangle$
to the final state $|F\rangle$, the total probability
$\frac{dP_{i\rightarrow f}}{dt}$ can be expressed as:

\begin{eqnarray}
\frac{dP_{i\rightarrow f}}{dt}=\sum_{F,I} \frac{dP_{i,I\rightarrow
f,F}}{dt} \nonumber
\end{eqnarray}

Equation (\ref{csec}) is particularly useful, as the transition
probability per unit time $\frac{dP_{i,I\rightarrow f,F}}{dt}$ can
be calculated from the perturbation theory. To first order this
quantity is written as (Fermi golden rule):

\begin{eqnarray}
\frac{dP_{i,I\rightarrow f,F}}{dt} &=& \frac{2\pi}{\hbar}  \left |
\langle i,I | H_{INT} |f,F \rangle \right |^2
\delta(E_i+E_I-E_f-E_F) \nonumber \\ \label{fgr}
\end{eqnarray}

Inserting the term $H^{(1)}_{INT}$ into Eq.(\ref{fgr}), using
Eq.(\ref{csec}), and considering the initial and final photon
states as plane waves one gets:

\begin{eqnarray}
\frac{\partial^2\sigma^{(1)}}{\partial E \partial \Omega}&=&\left
(\frac{e^2}{m_ec^2}\right )^2 \frac{k_f}{k_i} (\mathbf{\epsilon}_i
\cdot \mathbf{\epsilon}_f)^2 \nonumber \\ & \times &\sum_{F,I} P_I
\delta(E-(E_F-E_I)) \left | \langle F | \sum_j e^{i \mathbf{Q}
\cdot \mathbf{r_j}} |I \rangle \right |^2 \nonumber \\
\label{csec_1}
\end{eqnarray}

where $Q=\mathbf{k}_i-\mathbf{k}_f$ $(E=E_f-E_i)$ is the momentum
(energy) transferred from the photons to the system. The sum over
the initial and final states is the thermodynamic average, and
$P_I$ corresponds to the equilibrium population of the initial
state.

Apart from the sum over the phase factors of the photons scattered
from the different particles, whose interference give rise to a
truly $Q$ dependent scattering signal, the energy- and angle-
integrated cross section is of the order of the square of the
classical electron radius $r_0=\frac{e^2}{m_e c^2}$.

\subsubsection{The adiabatic approximation and the dynamic
structure factor}

From Eq. (\ref{csec_1}), which implicitly contains the correlation
function of the electron density, one arrives at the correlation
function of the atomic density on the basis of the following
considerations: i) One assumes the validity of the adiabatic
approximation, and this allows to separate the system quantum
state $|S\rangle$ into the product of an electronic part,
$|S_e\rangle$, which depends only parametrically from the nuclear
coordinates, and a nuclear part, $|S_n\rangle$: $|S\rangle
=|S_e\rangle |S_n\rangle$. This approximation is particularly good
for exchanged energies that are small with respect to the
excitations energies of electrons in bound core states: this is
indeed the case in basically any atomic species when considering
values in the range of phonon energies. In metals we neglect the
small portion of the total electron density in proximity of the
Fermi level. ii) One limits to consider the case in which the
electronic part of the total wavefunction is not changed by the
scattering process, and therefore the difference between the
initial state $|I\rangle =|I_e\rangle |I_n\rangle$ and the final
state $|F\rangle =|I_e\rangle |F_n\rangle$ is due only to
excitations associated with atomic density fluctuations. Using
these two hypothesis we then obtain:

\begin{eqnarray}
\frac{\partial^2\sigma}{\partial E \partial \Omega}&=&\left
(\frac{e^2}{m_e c^2}\right )^2 \frac{k_f}{k_i}
(\mathbf{\epsilon}_i \cdot \mathbf{\epsilon}_f)^2
\sum_{F_{n},I_{n}} P_{I_{n}}
\delta(E-(E_F-E_I)) \nonumber \\
&\times & \left | \langle F_n | \sum_j f_j(Q) e^{i \mathbf{Q}
\cdot \mathbf{R}_j} |I_n \rangle \right |^2
\end{eqnarray}

where $f_j(Q)$ is the atomic form factor of the $j$-th atom at
$R_j$ and the sum is now extended to all the atoms (molecules) of
the systems. Assuming that all the scattering units in the system
are equal, this expression can be further simplified by the
factorization of the form factor of these scattering units, and by
the introduction of the dynamic structure factor $S(Q,E)$ defined
as:

\begin{eqnarray}
S(Q,E)=\sum_{F_{n},I_{n}} P_{I_{n}} \delta(E-E_F+E_I)) \left |
\langle F_n | \sum_j e^{i \mathbf{Q} \cdot \mathbf{R}_j} |I_n
\rangle \right |^2 \nonumber
\end{eqnarray}

By representing the $\delta$-function above as a time integral,
indicating by $\langle ... \rangle$ the thermal average $\langle o
\rangle=\sum_{I} P_{I}\langle I |\hat o |I\rangle$ and using the
completeness operator $\sum_{F_n} |F_n \rangle \langle F_n| =1$
the dynamic structure factor can be also written in the more
familiar form:

\begin{eqnarray}
S(Q,E)=\frac{1}{2\pi N} \int dt e^{\frac{iEt}{\hbar}} \sum_{j,k}
\langle e^{i \mathbf{Q} \cdot \mathbf{R}_j(t)} e^{-i \mathbf{Q}
\cdot \mathbf{R}_j(0)} \rangle \nonumber
\end{eqnarray}

where $N$ is the number of particles in the system and the sum
over $(j,k)$ extend over these $N$ particles. The double
differential cross-section can then finally be re-written as:

\begin{equation}
\frac{\partial^2\sigma}{\partial E \partial \Omega}=
(\frac{e^2}{m_e c^2})^2 \frac{k_f}{k_i} (\mathbf{\epsilon}_i \cdot
\mathbf{\epsilon}_f)^2 \left | f(Q) \right |^2 S(Q,E) \label{csx}
\end{equation}

In the limit $Q\rightarrow 0$, the form factor $f(Q)$ is equal to
the number of electrons in the scattering atom, i.e. $f(Q)=Z$. For
increasing values of $Q$, the form factor decays almost
exponentially with a decay constants determined by the size of the
radial distributions of the electrons in the atomic shells of the
considered atom. At $Q$-values large with respect to the inverse
of these dimensions, therefore, the inelastic X-ray scattering
from density fluctuations is strongly reduced.

The cross-section derived so far is valid for a system composed of
a single atomic species, This derivation, however, can be easily
generalized to molecular or crystalline systems by substituting
the atomic form factor with either the molecular form factor, or
the elementary cell form factor respectively. The situation
becomes more involved if the system is multi-component and
disordered. In this case the factorization of the form factor is
still possible only assuming some specific distribution among the
different atoms. In the limiting case that such distribution is
completely random, an incoherent contribution appears in the
scattering cross-section.

\subsubsection{From cross section to count rate. \label{crate}}

The $Z$-dependence of the Thomson scattering cross-section seems
to imply a facilitation in studying systems with high $Z$. In
reality, this is no longer true when the effect of photoelectric
absorption is taken into consideration. Indeed, neglecting
multiple scattering events, the signal detected in an IXS
experiment from an infinitesimal slab of thickness $\delta x$
orthogonal to $k_i$ can be written as:

\begin{equation}
dN=N_0 (\frac{\partial^2\sigma}{\partial E \partial \Omega})
\Delta E \Delta \Omega \rho dx \label{scatt_inf}
\end{equation}

where $N_0$ is the flux of the incident photons, $N$ is the flux
of scattered photons in an energy interval $\Delta E$ and in a
solid angle $\Delta \Omega$, $\rho$ is the number of scattering
units per unit volume and $\mu$ is the total absorption
coefficient. Dealing with a macroscopic sample of length $L$, in
the relevant case of a nearly forward scattering geometry,
Eq.(\ref{scatt_inf}) becomes

\begin{equation}
dN(x)=N_0 e^{-\mu x}(\frac{\partial^2\sigma}{\partial E \partial
\Omega}) \Delta E \Delta \Omega \rho dx e^{-\mu (L-x)}
\label{scatt_1ph_diff}
\end{equation}

\noindent which, integrated over the whole sample length, yields:

\begin{equation}
N=N_0 (\frac{\partial^2\sigma}{\partial E \partial \Omega}) \Delta
E \Delta \Omega \rho L e^{-\mu L} \label{scatt_1ph}
\end{equation}

\begin{figure} [h]
\vspace{-2cm}
\includegraphics[width=.4\textwidth]{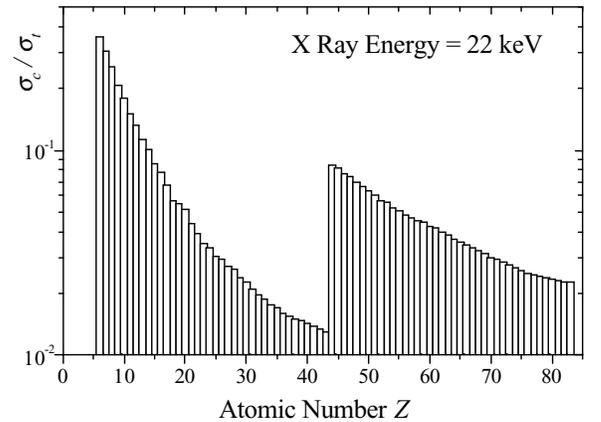}
\vspace{-2.2cm} \caption[cosanostra]{Ratio between the total
number of photons scattered by the Thomson process and those lost
through all the other processes, among which photoelectric
absorption, in a sample of length $\mu^{-1}$ calculated as a
function of the atomic number Z for photons of incident energy of
$\approx 22$ keV} \label{thom_cross}
\end{figure}

Let us discuss the L-dependence of this function. It is obvious
that N attains a maximum (the optimal sample length) when
$L=\mu^{-1}$, and that the value of N at this maximum point is
proportional to $\mu^{-1}$. Considering an X-ray energy of
approximately 20 keV and $Z>3$, $\mu$ is almost completely
determined by the photoelectric absorption process. This process
gives approximately $\mu \approx Z^4$ with modifications at
energies close to electron absorption thresholds. Consequently the
effective scattering volume is very much reduced in materials with
a high $Z$ (as $Z^4$), while the cross section increases as $Z^2$,
making the study of these materials by all means more difficult
than for those with low $Z$. The behavior of the optimal signal
intensity as a function of atomic number in monatomic systems with
sample length $L=\frac{1}{\mu}$ can be deduced from the data
reported in Fig. \ref{thom_cross}. There, we show the quantity
$\frac{\sigma_c \rho}{\mu}$, with $\sigma_c=(r_0 Z)^2$, which
gives directly a measure of the efficiency of the method at the
considered photon energy: in this example we took an incident
photon energy of 22 keV. The quantity $\frac{\sigma_c \rho}{\mu}$
is by definition the ratio between $\sigma_c$ and $\sigma_t$,
where $\sigma_t$ is the (measured) total X-ray cross-section of
the considered atom. This analysis is useful however only when it
is possible to study samples of optimal length. In cases where the
sample size is limited either by its availability or by the sample
environment (extreme pressure, high/low temperature, high magnetic
field...) it is obvious that one has great advantages in studying
high $Z$ materials.

Equation (\ref{scatt_1ph}) accounts for one scattering events
only. An estimate for the two scattering process intensity can be
obtained by invoking the forward scattering approximation again,
indeed, the ratio of the two over the one scattering rates
$N^{(2)}/N^{(1)}$ reads:

\begin{eqnarray}
\frac{N^{(2)}}{N^{(1)}} = \frac{\pi \rho h r_0^2 \int_0^\pi \left
[ f(\theta)S(\theta)\right ]^2 d\theta }{Z^2 S(0)} \nonumber
\end{eqnarray}

in which $h$ is sample transverse dimension traversed by the
incident beam. The integral accounts for all the possible two
scattering process leading to a final forward scattering. This
expression shows how to suppress multiple scattering one has to
reduce the transverse beam dimension. A similar estimate for
neutron scattering is prevented by the much more complicated
scattering paths, since, within a similar $Q$ range, the
scattering angle in INS is normally much lager. However, it can be
noted that in the case of neutrons the typical transverse beam
size is much larger ($\approx 10\div 100 $ mm) than the IXS ones
($\approx 100$ $\mu$m), thus resulting in a more important
contribution requiring accurate corrections.

\subsubsection{Kinematics of the scattering processes. \label{sec_klim}}

Another important difference between X-rays and neutrons
scattering lies in the kinematics of the scattering processes. The
momentum and energy conservation laws impose that:

\begin{eqnarray}
&&\mathbf{Q} = \mathbf{k}_i - \mathbf{k}_f \nonumber \\
&&E = E_f-E_i \nonumber \\
&&Q^2 = k^2_i + k^2_f -2 k_i k_f cos \theta \nonumber
\end{eqnarray}

where $\theta$ is the scattering angle between the incident and
scattered particles. The relation between momentum and energy in
the case of photons is given by:

\begin{eqnarray}
E(k)=hck \hspace{2.0cm} \nonumber
\end{eqnarray}

and therefore one obtains:

\begin{equation}
\left (\frac{Q}{k_i} \right )^2 = 1 + \left ( 1 - \frac{E}{E_i}
\right )^2 -2 \left (1 - \frac{E}{E_i} \right ) cos \theta
\hspace{1cm} \label{cin_x}
\end{equation}

Considering that the energy losses or gains associated to
phonon-like excitations are always much smaller than the energy of
the incident photon ($E << E_i$) this relation reduces to:

\begin{eqnarray}
\left (\frac{Q}{k_i} \right )= 2 sin \frac{\theta}{2} \hspace{1cm}
(E<<E_i) \nonumber
\end{eqnarray}

This last relation shows that, in the limit of small energy
transfers, the ratio between the exchanged momentum and the
incident photon momentum is completely determined by the
scattering angle. Therefore, in inelastic X-ray scattering, there
are basically no limitations in the energy transfer at a given
momentum transfer for phonon-like excitations.

\begin{figure} [h]
\vspace{-.3cm}
\includegraphics[width=.5\textwidth]{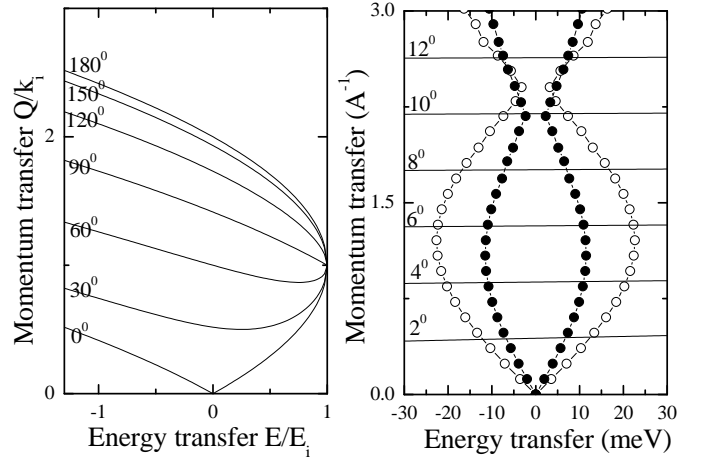}
\vspace{-6.2cm} \caption[cosanostra]{Kinematic region accessible
to IXS in reduced $E/E_i$ and $Q/k_i$ units (left panel). The
right panel shows the realistic cases of molten lithium (open
circles) and sodium (full circles) for an incident energy of 25
keV.} \label{kin_ixs}
\end{figure}

\subsubsection{X-rays vs. Neutrons.}

At variance with the previous equation, if the probe particles
have mass $m_p$

\begin{eqnarray}
E(k)=\frac{\hbar ^2 k^2}{2m_p} \hspace{1cm} \nonumber
\end{eqnarray}

and, therefore:

\begin{equation}
\left (\frac{Q}{k_i} \right )^2 = 1 + \left ( 1 - \frac{E}{E_i}
\right )-2 \sqrt{1 - \frac{E}{E_i}} cos \theta \hspace{1cm}
\label{cin_n}
\end{equation}

\begin{figure} [h]
\includegraphics[width=.7\textwidth]{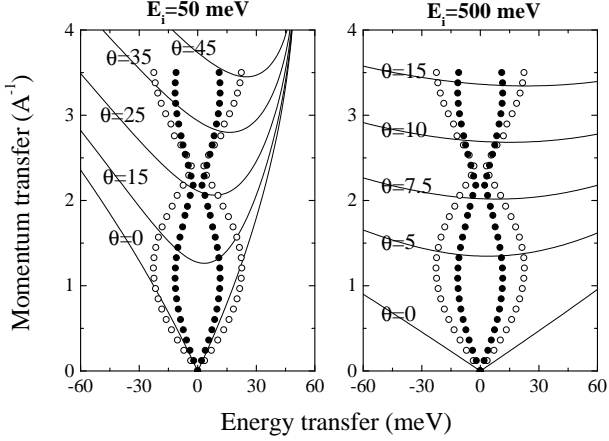}
\vspace{-2.8cm} \caption[cosanostra]{Kinematic region accessible
to neutron scattering experiments for incident energies $E_i=50$
meV (left panel) and $E_i=500$ meV (right panel), reported for
different scattering angles. Open and full circles are the
(approximated) sound dispersion of molten lithium and sodium,
respectively.} \label{kin_neu}
\end{figure}

In this case of thermal neutron scattering, the approximation
$E<<E_i$ no longer holds, and the kinematics of the scattering
experiments is determined by Eq. (\ref{cin_n}). As an example, in
Fig. \ref{kin_neu}, we report the accessible kinematics regions in
the $\frac{E}{E_i}$ vs $\frac{Q}{k_i}$ plane for two different
incident energies, indicating paths at constant scattering angles.
In the same figure, similarly to Fig. \ref{kin_ixs}, we also
report the approximate dispersion curves for liquid lithium and
sodium. In the best situation, i.e. in forward scattering where
the accessible region is as wide as possible, the limiting curve
is linear around $Q=0 (E=0)$, and its tangent is $\frac{E}{E_i} =
2 \frac{Q}{k_i}$. Recalling that $E_i = \frac{\hbar
^2k_i^2}{2m_p}$, one gets $E = v_N \hbar Q$, with $v_N$ the
velocity of the incoming neutron. As the dispersion relation for
acoustic phonon is linear, $E = v_s\hbar Q$, with $v_s$ the
velocity of sound, it is clear that whenever $v_s$ is larger than
$v_N$ the excitations peaks lie outside the accessible region and,
therefore, when $v_s
> v_N$ the neutron technique cannot be applied to study the
acoustic branch. This limitation does not apply to the case of
X-rays, as, according to Eq. (\ref{cin_x}), there are basically no
limits to the energy region accessible at a given scattering
angle.

As discussed before, the presence of relevant absorption phenomena
is the main effect that determine the scattering volume in an IXS
experiment. This implies, therefore that the probability that a
photon is scattered from the sample is small, and this strongly
suppresses the multiple scattering processes. In IXS experiments,
indeed, the multiple scattering can be disregarded, thus avoiding
the use correction procedures. This is therefore an important
advantage with respect to the neutrons case, where, on the
contrary, the sample length is determined by the scattering
(rather than absorption) length. In Fig. \ref{ixs-neu_cross} are
reported for comparison the (coherent) scattering lengths of the
elements for the X-rays and neutrons cases.

Finally, it is worth to compare the double differential scattering
cross section for X-rays obtained before (Eq. (\ref{csx})) with
the similar quantity derived for neutron scattering (in the
hypothesis of fully coherent scattering). The latter quantity is
not derived here for brevity and can be found in many textbooks
\cite{LOVESEY,MARSHALL,EGELSTAFF}. The two cross-sections read as:

\begin{equation}
\frac{\partial^2}{\partial E \partial \Omega}= \left \{
\begin{array}{l}
r_0^2 \frac{k_f}{k_i} \left ( \mathbf{\epsilon}_f \mathbf{\epsilon}_i \right )^2 |f(Q)|^2 S(Q,E) \hspace{.2cm} xray\\
\hspace{.4cm} b^2 \hspace{.6cm} \frac{k_f}{k_i} \hspace{2.25cm} S(Q,E) \hspace{.2cm} neutron \\
\end{array}
\right.
\end{equation}

Beside to the proportionality of both the cross sections to the
dynamics structure factor, it is worth to underline that:

\begin{enumerate}
\item  The two cross section are proportional to a characteristic
scattering length squared ($r_0$ in the case of X-ray and $b$ in
the case of neutrons) that are comparable in magnitude (see Fig.
\label{ixs-neu_cross}). The further factor $|f(Q)|^2$ ($\propto
Z^2$ for small $Q$'s) in the case of X-ray does not bring an
increase of the actual signal in the experiment because, as
discussed before, increasing $Z$ also limit the optimal scattering
volume due to the increase of the photoelectric absorption.

\item In both cases the phase space of the incident and final
plane waves gives rise to the factor  $\frac{k_f}{k_i}$, however
while in the X-rays case  $k_f\approx k_i$ , and this factor is
very close to 1, in the neutron case this term give rise to a $Q$
dependence of the scattered intensity. \item  No polarization
terms are present in the cross section for neutron, while in the
case of X-rays the term tells us that the Thomson scattering
arises from a scalar interaction and therefore the polarizations
of the incident and scattered photons must be parallel.

\item Finally, the X-ray scattering cross section contains the
form factor $f(Q)$, i.e. the Fourier transform of the charge
density spatial distribution. As the charge density is localized
around the nuclei in a space region of typical dimension of few
tenth to few hundredth of nm, the function $f(Q)$ decreases
appreciably on a $Q$ range of several inverse nm$^{-1}$, thus it
does not depresses  too much the scattering cross section in the
mesoscopic region of interest. In the case of neutrons this form
factor is not present (actually it is equal to 1) as the neutrons
interact with the nuclear matter, localized in typical dimension
of $10^{-6}$ nm. The neutron form factor is therefore constant in
the whole accessible $Q$ region.
\end{enumerate}

From the discussion made so far, it should be now quite easy to
understand how important is the development of the X-ray method,
which can access, in principle, an extremely large region of the
$E-Q$ plane. Particularly important is the small $Q$ region, where
the acoustic excitations have energies which are not of easy
access to the neutron spectroscopies.

\begin{figure} [h]
\vspace{-2cm}
\includegraphics[width=.4\textwidth]{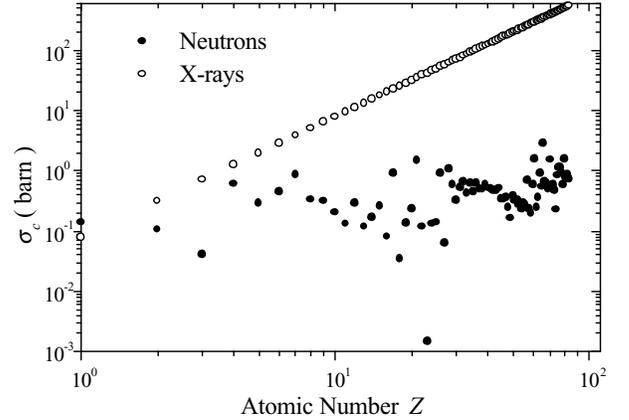}
\vspace{-2.2cm} \caption[cosanostra]{The coherent scattering cross
section of the elements for X-ray (open circles) and for neutrons
(full circles) reported as a function of the atomic number Z.}
\label{ixs-neu_cross}
\end{figure}

\subsection{From the experimental data to the dynamical
quantities \label{sec_fetdq}}

In order to extract quantitative information from the experimental
intensity, i.e. to perform measurements of $S_q(Q,\omega )$ on an
absolute scale, the most direct way is to use a reference
scatterer and this is customarily done in neutron experiments. In
IXS, for instance, such a procedure can be quite difficult because
of the $Q$-dependence of the form factor and of the analyzers
efficiencies. For these reasons indirect method are always
preferred. One possibility is to exploit the lowest order sum
rules of $S_q(Q,\omega )$ \cite{scop_jpc}: in particular for the
first two frequency moments one has the

\begin{eqnarray*}
\langle \omega  ^0 \rangle _{S_q}  &=&\int S_q(Q,\omega )d\omega =S(Q), \\
\langle \omega  ^1 \rangle _{S_q}  &=&\int \omega S_q(Q,\omega
)d\omega =\hbar Q^2/2m.
\end{eqnarray*}

where the second equality follows from Eq. (\ref{dispari}) applied for $n=1$%
. The measured raw intensity is related to the dynamic structure
factor through

\begin{equation}
I(Q,\omega )=A(Q)\int d\omega ^{\prime }S_q(Q,\omega ^{\prime
})R(\omega -\omega ^{\prime })  \label{convo}
\end{equation}

where $R(\omega )$ is the experimental resolution function and
$A(Q)$ is a factor taking into account the scattering geometries,
the experimental setup and the atomic form factor. The first
moments of the experimental data, $\langle \omega  ^0 \rangle _I$
and $\langle \omega  ^1 \rangle _I$, and those of the resolution
function, $\langle \omega  ^0 \rangle _R$ and $\langle \omega  ^1
\rangle _R$, are related to $\langle \omega ^0 \rangle _S $ and
$\langle \omega  ^1 \rangle _S $ by:

\begin{eqnarray*}
\langle \omega  ^0 \rangle _I &=&A(Q)\langle \omega  ^0 \rangle _{S_q}\langle \omega  ^0 \rangle _R, \\
\langle \omega  ^1 \rangle _I &=&A(Q)(\langle \omega  ^0 \rangle
_{S_q}\langle \omega  ^1 \rangle _R+\langle \omega  ^1 \rangle
_{S_q}\langle \omega ^0 \rangle _R).
\end{eqnarray*}

From the previous equation one derives that

\begin{equation}
S_{q}(Q)=\frac{\hbar Q^{2}}{2M}(\langle \omega  ^1 \rangle
_I/\langle \omega  ^0 \rangle _I-\langle \omega  ^1 \rangle
_R/\langle \omega ^0 \rangle _R)^{-1}. \label{norma}
\end{equation}

This procedure, therefore, can been adopted to establish an absolute scale for $%
S_q(Q,\omega )$ using the experimentally determined $I(Q,\omega )$ and $%
R(\omega )$.

\subsection{Handling liquid metals}

Working with liquid metals poses several practical problems. In
particular alkali metals are highly reactive and need to be kept
under a protective atmosphere. A relatively small impurity (less
than 100 ppm) in Ar or nitrogen will cause a film to form on the
surface of the liquid metal.

In addition, glass is often limited in its use as a container for
most liquid metals. Liquid metals are often strongly reducing.
Glass, composed chiefly of silicon dioxide, is penetrated by the
metal atoms, which can reduce the silicon by forming a metal
oxide. As a result, the glass becomes discolored and brittle. For
these reasons Pyrex can not be used above $600$ K and pure quartz
above $900$ K.

Preferred materials for working with liquid metals are the
refractory metals. This refers to the titanium group (Ti, Zr) as
well as the vanadium and chromium groups. These transition metals
are much less likely to undergo reduction and be solvated by
liquid metals. The disadvantage in addition to cost is that the
refractory metals have certain properties that make fabrication
difficult. A trade-off is to use iron plated with chromium. Other
less reactive transition metals, such as the noble metals, are
soluble in many liquid metal solutions. Austenitic stainless steel
can be suitable up to $1000$ K

Liquid metals also share a common chemistry. The increasing
electropositivity of the metals composing the liquid metal
solution will determine the liquid's reactivity. Mercury, which is
not very electropositive, is stable in air. Alkali metals, which
are the most electropositive group of elements, are air-sensitive.
Li reacts slowly with air, yet dissolves and reacts quite readily
with nitrogen. The other alkalis are insensitive to nitrogen but
react with other gasses. All the alkali metals violently react
with water on this basis: $Cs>Rb>K>Na>>Li$ \cite{OSE}. The
hydrogen generated in the water-alkali reaction can, in turn,
react explosively with oxygen.

Reactivities reflect those of the solid material. The more
reactive liquid metals, usually alkali ones, often develop a film
if exposed to air. The high surface tensions of many liquid metals
enable this film to remain in place. However, rates are greatly
increased if turbulence is introduced because the protective
coating is often dissolved into the liquid metal. Products of
reaction, such as hydrides and oxides, are often redissolved into
the liquid metal solution.

By virtue of the previously mentioned difficulties, experiments
with alkali metals are often very hard to perform. In the case of
Inelastic X ray Scattering, in particular, such difficulties are
enhanced by sample dimension requirement. As we have seen in sec.
\ref{crate}, indeed, in an optimal IXS experiment the sample
length has to be comparable with the absorption length. With a few
exceptions (Li and Na), this typically means sub-millimeter sample
thickness. Common choices are, therefore, sample cells made of
compatible metals provided with sealed sapphire or diamond
windows.

\begin{figure} [h]
\centering
\includegraphics[width=.5\textwidth]{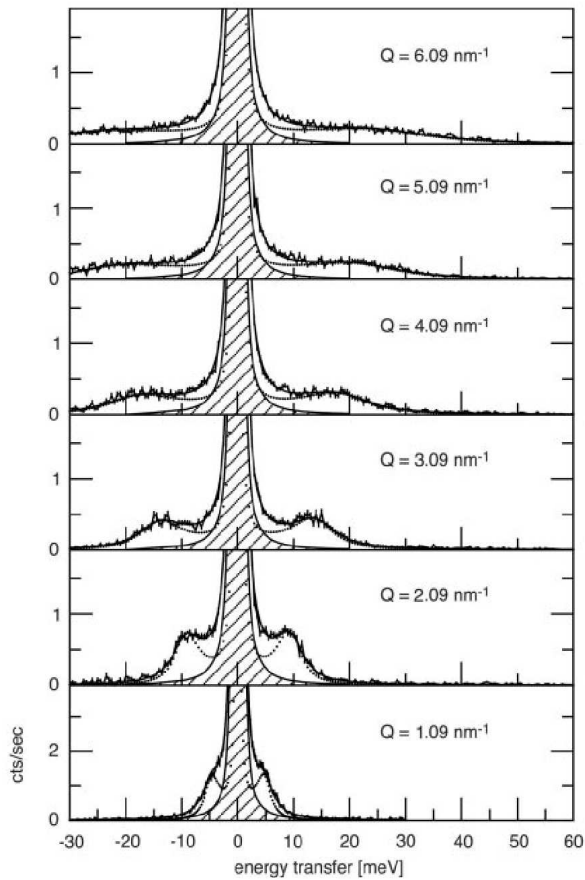}
\caption[H. Sinn - sinn@aps.anl.gov - Fig.1 - Science 299, 2047
(2003)]{Coupling IXS with levitation techniques: Constant $Q$
spectra of liquid alumina. From \cite{sin_al2o3}} \label{al2o3}
\end{figure}

Large efforts have been done recently to overcome the difficulty
of performing X-ray experiments on liquid metals, the most
remarkable example being the so called Tamura-type cells made with
a single crystal sapphire with Be windows pressurized under He
\cite{tam_cell1}, performing up to $1900$ K and $2$ Kbars. More
recently, a new sample environment especially tailored for alkali
metals has been proposed \cite{tam_cell2}. In this case the cell
is entirely made of molybdenum, with the windows made by single-
crystal disks of controlled orientation electrolytically thinned
at $\approx 40$ $\mu m$.

A totally different approach is the one of contact-less
techniques. In this case the sample is levitated either
electrostatically or by means of a controlled gas jet. The main
difficulty in this technique is related to the sample stability in
the x-ray beam, but recent impressive advancement have been done
in this field. A new beamline for electrostatic levitation (BESL)
has been developed at the Advanced Photon Source (APS), and the
relevance of icosahedral ordering in the supercooling capabilities
of liquid metals has been investigated \cite{kel_lev}. Another
example is a recent X-ray scattering experiment performed on
liquid Al$_2$O$_3$ in which alumina droplet of 3-4 mm diameter
have been levitated by gas jet flow on the inelastic scattering
beamline 3ID-C \cite{sin_al2o3}.

\section{Experimental results \label{sec_res}}

In this section we review, to the best of our knowledge, the
experimental results reported so far, ordered according to the
sample group in the periodic table. No results are available so
far for elements belonging to group II.

\subsection{Alkali metals}

Alkali metals do not occur freely in nature, they are very
reactive and can explode if exposed to water. These metals have
only one electron in their outer shell and, as with all metals,
they are malleable, ductile, and good conductors of heat and
electricity. Alkali metals are softer than most other metals.
Among the metallic elements they share the simplest pairwise
interaction potential, which is also the closest to the Lennard
Jones one. As a consequence, their structural properties are also
particulary simple, with a structure factor resembling the one of
hard spheres. Also the dynamics, therefore, is expected to mimic
the theoretical and numerical results achieved for Lennard Jones
and hard sphere systems.

\subsubsection{Lithium}

Liquid lithium is probably the system which better reveals the
complementarity of neutrons and X-rays as far as inelastic
scattering is concerned. Due to the high absorption cross section
of the $^6$Li isotope ($\sigma_a=940$ b) neutron scattering
experiments must necessarily be performed on $^7$Li enriched
samples, which is the dominant specie in the natural abundance.
The high sound velocity ($c_t \approx 4500$ m/s), and the almost
equivalent neutron scattering cross sections ($\sigma_i=0.68$ b;
$\sigma_c=0.62$ b for $^7$Li), pose severe limitations to the use
of INS aiming at the determination of collective properties, while
this technique turns out to be extremely useful for the
investigation of the single particle motion. The first INS studies
on this system can be traced back to the work of De Jong and
Verkerk \cite{dej_phd,ver_li,dej_li}, who showed the presence of
collective modes with a series of accurate experiments. Though
they had to face the above mentioned drawbacks, indeed, they were
able to point out some significant issues: by modelling the
coherent contribution with the extended hydrodynamic model (see
Eq. (\ref{ehm}) of section \ref{sec_kin}) they measured the
dispersion curve above $Q_m/2$ (see Fig. \ref{bur2}) and they
reported deviations from the Landau-Plazek ratio, which is
expected to hold in the hydrodynamic regime (see Eq. (\ref{lp}) of
section \ref{sec_collhydro}). On the other side they shed light on
the single particle motion, accurately determining the incoherent
contribution to the dynamic structure factor within the framework
of section \ref{sec_kin} (Eqs (\ref{mct1}-\ref{mct4})). They
corroborated the Mode Coupling predictions \cite{desh_mc},
extracting values of the diffusion coefficient and determining its
temperature dependence.

\begin{figure} [h]
\centering
\includegraphics[width=.4\textwidth]{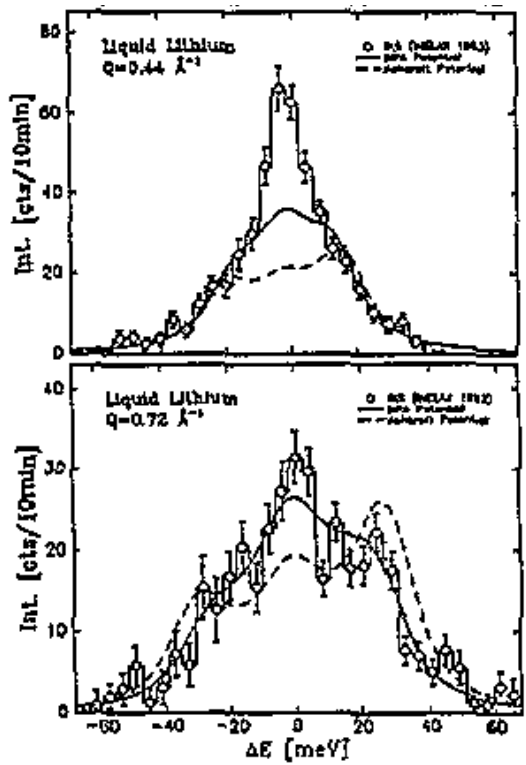}
\caption[E. Burkel, burkel@physik1.uni-rostock.de - Fig. 4 - J.
Phys. Cmatt 6 A225 (1984)]{Pioneering (1991) low resolution IXS
determination of the dynamic structure factor in liquid lithium
with INELAX. From \cite{bur}.} \label{bur1}
\end{figure}

\begin{figure} [h]
\centering
\includegraphics[width=.3\textwidth,angle=270]{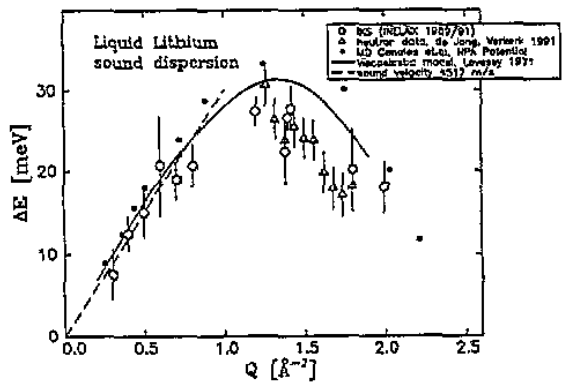}
\caption[E. Burkel, burkel@physik1.uni-rostock.de - Fig. 3 - J.
Phys. Cmatt 6 A225 (1984)]{Dispersion curve of liquid lithium
achieved with IXS at INELAX. Theoretical predictions and INS
results at higher $Q$'s are also reported. From \cite{bur}.}
\label{bur2}
\end{figure}

An exhaustive characterization of the coherent dynamics was
provided by the advent of Inelastic X ray Scattering developed in
the early nineties, and liquid lithium has been the benchmark of
such development. Being the lightest of the liquid metals, indeed,
lithium played a privileged role in IXS, for the favorable signal
to noise ratio and for the high sound velocity which allowed to
resolve the inelastic spectral component minimizing the initial
difficulty of achieving energy resolutions comparable to neutrons.

Since the pioneering work of Burkel \cite{BURKEL} with the INELAX
instrument (see fig. \ref{bur1} and \ref{bur2}), a decisive step
forward achieved with the advent of the third generation sources
which, combined to a brilliant technique for manufacturing silicon
crystal analyzers, allowed to exploit IXS to gather insight into
the microscopic dynamics of disordered systems.

\begin{figure} [h]
\centering
\includegraphics[width=.5\textwidth]{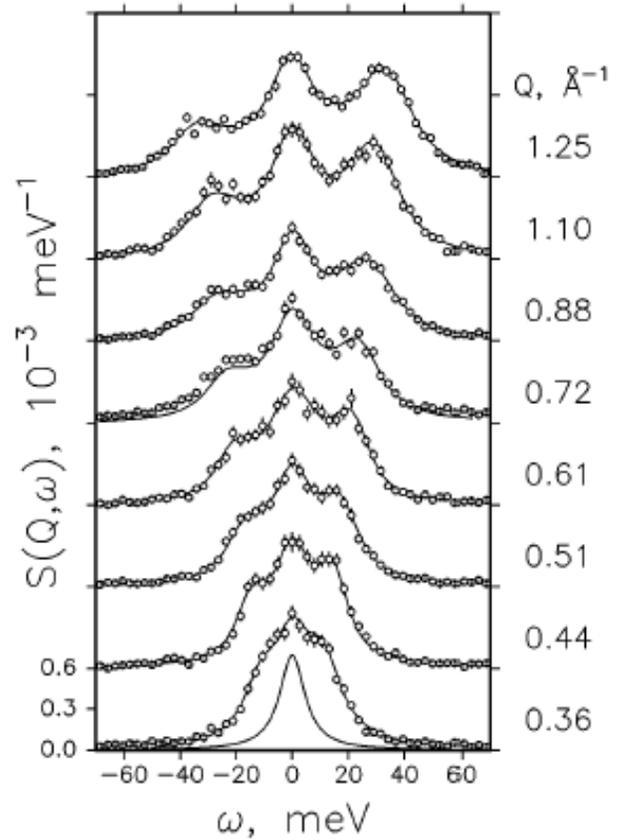}
\caption[H. Sinn - Fig. 1 - PRL 78, 1715 (1997)]{First IXS
measurements on liquid lithium performed on a third generation
source (ESRF) \cite{sinn}. Energy resolution is here $\delta E=11$
meV. The continuous line is the best fit according to the extended
hydrodynamic model of Eq. (\ref{ehm}). From \cite{sinn}}
\label{li_sinn}
\end{figure}

The first remarkable result on a third generation facility (ESRF)
was provided by Sinn et al. \cite{sinn} who, measuring energy
spectra at fixed wavevectors, reported clear evidence of
collective modes, being able to give significant hints for the
choice of the most appropriate pseudopotential to describe liquid
metals in numerical simulations \cite{can_lit}. In the same work,
following the extended hydrodynamic model outlined in section
\ref{sec_kin} \cite{desh_hyd0,desh_hyd}, it was also reported
evidence for positive dispersion, i.e. for a sound velocity value
exceeding the hydrodynamic one. This phenomenon was ascribed to a
transition from a liquid to a solid-like response.

\begin{figure} [h]
\centering
\includegraphics[width=.5\textwidth]{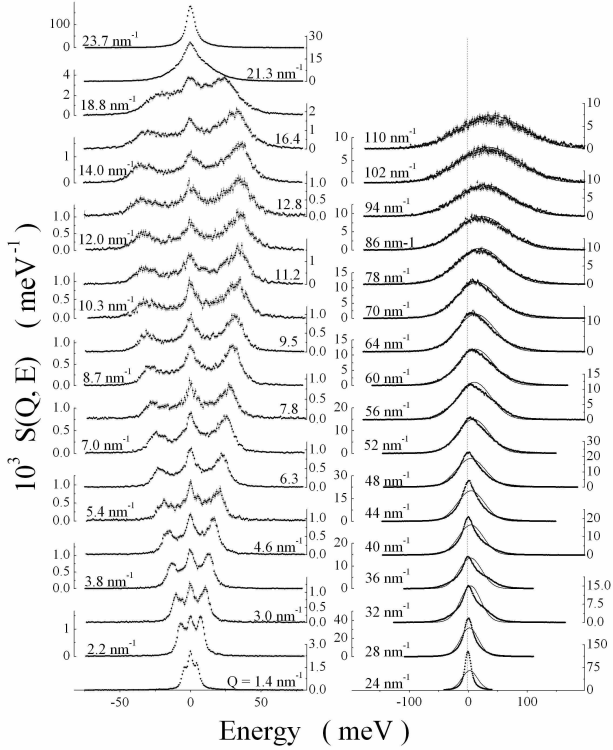}
\caption[cosanostra EPL 50, 189 (2000)]{IXS measurement of liquid
lithium in a wide energy momentum-region \cite{scop_epl}. The
transition from hydrodynamic to gaussian-like response (continuous
line in the right panel) can be clearly noticed. Energy
resolutions are here $1.5$, $3.0$ and $7.0$ meV, increasing with
the exchanged momentum} \label{all_li}
\end{figure}

Following the development of the IXS technique, new experiments
have been more recently performed on liquid Li in the extended
region $1.4$ to $110$ nm$ ^{-1} $, corresponding to
$Q/Q_{m}\approx 5\cdot 10^{-2}\div 5$, which are reported in Fig.
\ref{all_li}.

\begin{figure} [h]
\centering
\includegraphics[width=.5\textwidth]{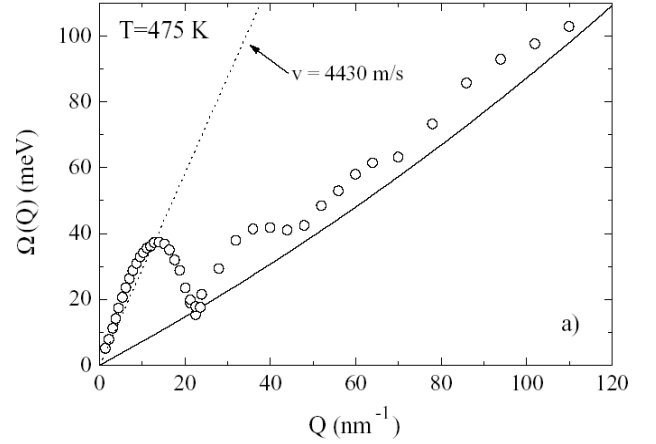}
\vspace{-7cm} \caption[cosanostra, adapted, forse va bene
cosi]{Sound velocity as deduced by the maxima of the current
correlation spectra, from the best fit with quantum corrected and
resolution convoluted models.} \label{li_fulldisp}
\end{figure}

In Fig. \ref{li_fulldisp} is reported the dispersion relation
determined in the same energy-wavevector region, and the
transition between the two distinct dynamical regime is here
evidenced by the sound velocity behavior. Beyond the first
quasi-hydrodynamic region (an initial nearly linear dispersion),
structural effects take place suppressing the sound propagation
around $Q_{m}/2$ due to strong negative interference. With
increasing $Q$ values, the points in Fig. \ref{li_fulldisp} show a
second pseudo-BZ, followed by a series of oscillations that damp
out with increasing $Q$ - here, $\omega _{l}(Q)$ is approaching
the single particle behavior. These oscillations are in anti-phase
with those of $S(Q)$ and are therefore associated with the local
order in the liquid.

\begin{figure} [h]
\centering
\includegraphics[width=.4\textwidth]{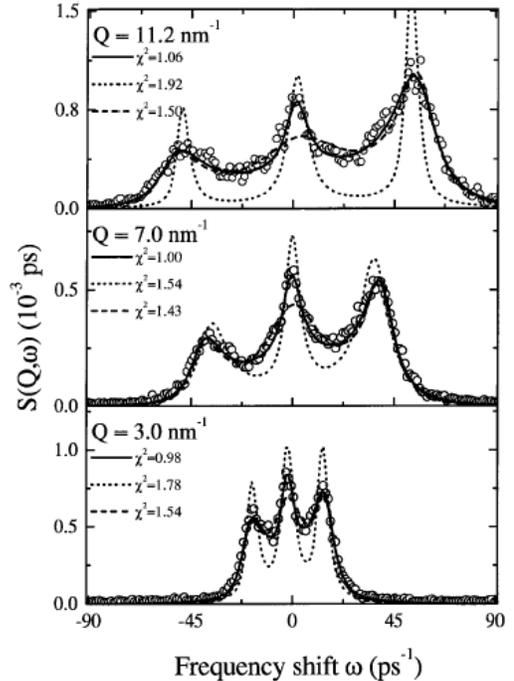}
\vspace{-1cm} \caption[cosanostra PRL \textbf{OK}]{Memory function
at work: refined lineshape analysis of high resolution IXS
spectra. Both thermal and viscous channel are taken into account,
mimicking this latter with one (Eq. (\ref{memoryL1}), dotted line)
or two (Eq. (\ref{x3tempi}), continuous line) exponential
processes.} \label{li_2times}
\end{figure}

While at low $Q$'s the dynamic structure factor is qualitatively
described by an extended hydrodynamic treatment (Eq. (\ref{ehm} of
section \ref{sec_kin}), at wavevectors distinctly larger than
$Q_m$ the single particle response is attained through the
mechanism described in section \ref{sec_swl}, ultimately leading
to the expressions well accounted by a combination of two Eq.
(\ref{sqw_free_quant}), accounting for each of the two isotopes
$^6$Li and $^7$Li.

Thanks to the improvement in the energy resolution, which is
nowadays comparable to the one of INS spectrometers in the same
energy-wavevectors domain (1.5 meV at present), an approach based
on the generalized hydrodynamics has been developed, which allowed
to point out the presence and the role of relaxation processes
driving the collective dynamic at the microscopic probed
wavelenghts \cite{scop_jpc,scop_prlli}. Within a memory function
framework \cite{mori_mf}, it has been ascertained the presence of
two distinct viscous relaxation channel (see fig. \ref{li_2times})
beyond the thermal relaxation, clarifying the origin and the
nature of sound dispersion and attenuation properties in simple
fluids. Of the two processes, active over well separated
timescales, one is related to the well known transition between a
low frequency, liquid-like response to an high frequency,
solid-like response. The second mechanism is instead a general
relaxation process peculiar of the vibrational dynamics which is
present regardless the thermodynamic state of the system. In this
context, the positive dispersion has been shown to be strongly
related to this latter process, being the solid like response
already attained over the wavevectors range probed in IXS (or INS)
experiments.

\begin{figure} [h]
\centering
\includegraphics[width=.5\textwidth]{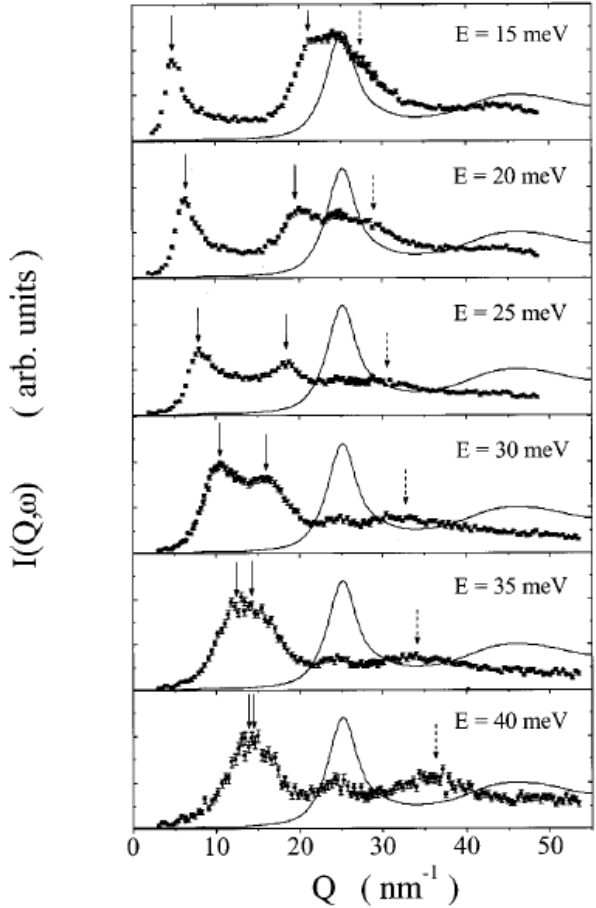}
\caption[cosanostra PRB \textbf{OK}]{Constant energy slices of the
dynamic structure factor determined by IXS. Umklapp modes are
visible on the sides of the main structure factor peak}
\label{li_umk}
\end{figure}

An alternative route to the investigation of collective dynamics,
which is dual to the one followed in the above mentioned
experiments and which easily achievable through IXS, is the
determination of the dynamic structure factor performing Q-scan
for fixed values of the energy transfer, reported in fig.
\ref{li_umk} for the case of lithium \cite{scop_prbumk}. In this
way, one is able to have a direct sight over the so called umklapp
modes, i.e. excitations characterized by wavevectors which differ
by multiple of the reciprocal lattice spacing, which have been
early reported by means of INS in liquid lead
\cite{rand_umk,coc_umk,dor_umk}.

\subsubsection{Sodium}

Pioneering experimental determinations of the scattering law in
liquid sodium can be traced back to the time of the IAEA symposium
held in Chalk River \cite{coc_na,rand_na}. In this system the
ratio between the incoherent to coherent cross section is very
close to one (see table \ref{table}), therefore the separation
between the two contributions is of crucial importance. Soon after
Randolph's experiment, his data were analyzed in terms of mean
square displacement of an atom \cite{des_napb}. This framework
(described in section \ref{sec_gaussapp}) poses on the Gaussian
assumption for the incoherent cross section, while the coherent
contribution was evaluated according to the effective mass
approximation \cite{dej_ema}:

\begin{eqnarray}
S_s(Q,\omega)&=&\frac{1}{\pi} \int_0^\infty dt cos(\omega t) exp
\left
[-\frac{Q^2 \langle r^2(t)\rangle}{6} \right ] \nonumber \\
S(Q,\omega)&=&\frac{1}{\pi} \int_0^\infty dt cos(\omega t) exp
\left [-\frac{Q^2 \langle r^2(t)\rangle}{6S(Q)} \right ] \nonumber
\end{eqnarray}

The mean square displacement was then determined describing the
atomic motion in terms of independent harmonic oscillators of
frequency $\omega_0$ and lifetime $\tau_0$, which, in turns, are
related to the spectral density of the velocity autocorrelation
function $f(\omega)$:

\begin{eqnarray}
\omega_0^2&=&\int_0^\infty d\omega \omega^2
f(\omega)=\frac{\langle
(\overrightarrow{\nabla} U)^2 \rangle}{3m} \nonumber \\
\tau_0&=&\frac{T}{mD\omega_0^2} \nonumber \\
f(\omega)&=&\frac{2}{\pi}\frac{\omega_0^2 /
\tau_0}{(\omega^2-\omega_0^2)^2+(\omega^2 / \tau_0)^2} \nonumber
\end{eqnarray}

\begin{figure} [h] \centering
\vspace{-2cm}
\includegraphics[width=.53\textwidth]{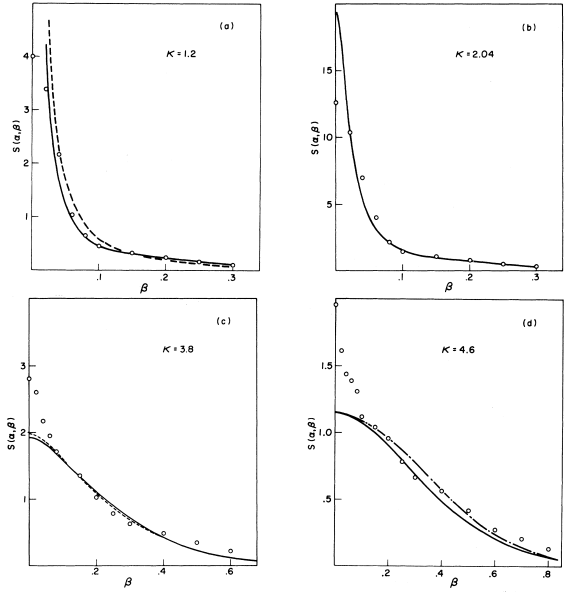}
\vspace{-2.4cm} \caption[R.C. Desai rashmi.desai@utoronto.ca -
Fig.3 - Phys. Rev. 180, 299 (1969).]{Randolph's measurement
\cite{rand_na} of the dynamic structure factor of liquid sodium in
reduced units: $\alpha=\hbar^2Q^2/2mT$ and $\beta=\hbar \omega /
T$ for four different values of momentum transfer. Lineshape
analysis according to different models \cite{des_napb} Continuous
line: EMA+calculated MSD. Short dashed line: EMA+computer
simulation computation of the MSD. Long dashed line: hydrodynamic
prediction. Dotted-dashed line: free streaming limit} \label{nady}
\end{figure}

The basic ingredients of this approach are, therefore, the
knowledge of the static structure factor, of the macroscopic
diffusion coefficient and of the mean squared force $\langle
(\overrightarrow{\nabla} U)^2 \rangle$. The results of this
description are tested against the experimental data in fig.
\ref{nady}. From the same figure, it emerges how the single
particle regime is already attained at the lowest reported $Q$
value, i.e. $Q=12$ nm$^{-1}$, while the low frequency discrepancy
has been tentatively ascribed to finite instrumental resolution
and to multiple scattering effect.

\begin{figure} [h]
\centering
\includegraphics[width=.42\textwidth]{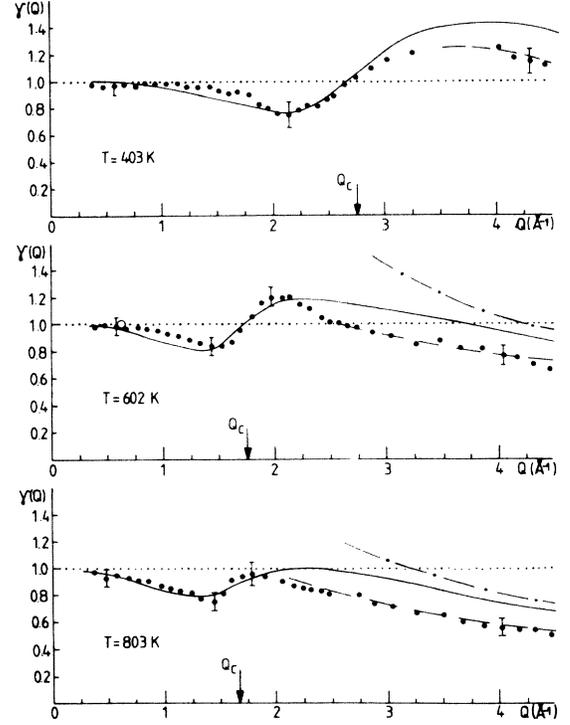}
\caption[C. Morkel chiedere a Pilgrim - Fig. 6 - PRA 33, 3383
(1986)]{Reduced quasielastic linewidth $\omega_{1/2}/DQ^2$ in
liquid sodium at three different temperatures \cite{mor_na}. The
dotted line is the Fickian limit while dash dotted line is the
perfect gas behavior ($\propto 1/Q$). The dashed line is the hard
sphere prediction \cite{coh_hs}, while the continuous line is the
result obtained within mode coupling theory \cite{got_na}}
\label{gammana}
\end{figure}

Twenty years later, new INS data were reported
\cite{mor_na,sod_na}, addressing in more detail the incoherent
scattering contribution and showing how the diffusion process is
actually more complex. Morkel and Gl{\"a}ser, following for the
coherent contribution the Lovesey's prescription \cite{lov_visco},
and adopting for the incoherent part the Nelkin-Gatak model
\cite{nelkin_inco} described in section \ref{sec_ng}, extracted
the reduced halfwidth $\omega_{1/2}$ of the incoherent
contribution finding a crossover between the hydrodynamic
(lorenzian) and single particle (gaussian) regimes. In fig.
\ref{gammana} the linewidth $\omega_{1/2}/DQ^2$ is reported, and
it clearly emerges how the single particle limit
($\omega_{1/2}/DQ^2\propto 1/Q$) is not yet attained even at
$Q\approx 40$ nm$^{-1}$, which contrast the earlier assumptions of
Desai and Yip. After a low $Q$ diffusion retardation the mobility
increases in the transition region and finally tends to the free
gas limit. The whole $Q$ dependency is well described within the
Enskog's hard sphere gas \cite{sears_inco}, in terms of the
expression (\ref{sears}). An alternative description of the single
particle dynamics can be recovered within the memory function
approach, though it fails in the high $Q$ region \cite{got_na}.

The first IXS determination of the collective dynamics in liquid
sodium is due to Pilgrim and collaborators \cite{pil_na}. In this
work, the coherent dynamic structure factor was measured at
several temperatures, and analyzed according to the extended
hydrodynamic model previously applied in liquid lithium
\cite{sinn}. The take-home message is the presence of a positive
dispersion effect which does not show significant temperature
dependence. This result seems to rule out an interpretation of the
positive dispersion in terms of an activated process, as is the
case in hydrogen bonding systems \cite{monaco_water}.

\begin{figure} [h]
\centering
\includegraphics[width=.5\textwidth]{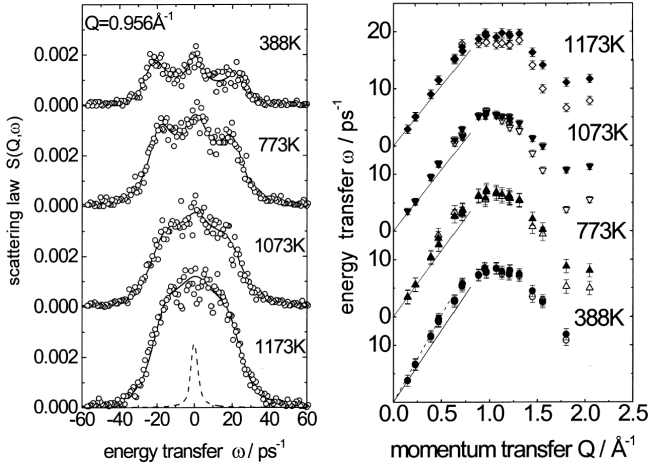}
\vspace{-6cm} \caption[Pilgrim - Fig1 and 2
pilgrim@mailer.uni-marburg.de JNCS 250-252, 96 (1999)]{Left panel:
IXS determination of the $S(Q,\omega)$ in liquid sodium for
selected temperatures. Right panel: dispersion curves at different
temperature.} \label{napilgrim}
\end{figure}

IXS experiments on liquid lithium were then repeated with
increased energy resolution \cite{scop_prena}, and analyzed within
the same two viscous relaxation processes proposed for liquid
lithium \cite{scop_prlli}. The same data have been also
interpreted within the framework of the scale invariance of
relaxation processes \cite{yul_na}, a theory originally developed
for liquid cesium \cite{yulm_cs}, which has been recently shown to
be equivalent to the memory function approach in the sense that
one solves the chain of equations \ref{memory} with some
\textit{ad hoc} closure relation.

\subsubsection{Potassium}

The first experimental data on liquid potassium appeared
surprisingly late relatively to the other liquid metals
\cite{nov_k,nov_k1,nov_k2}. Moreover, the kinematic region $Q-E$
spanned in this experiment was quite narrow ($10<Q<13$ nm$^{-1}$)
and only partial information on the microscopic dynamics could be
obtained.

\begin{figure} [h]
\centering \vspace{-2cm}
\includegraphics[width=.4\textwidth]{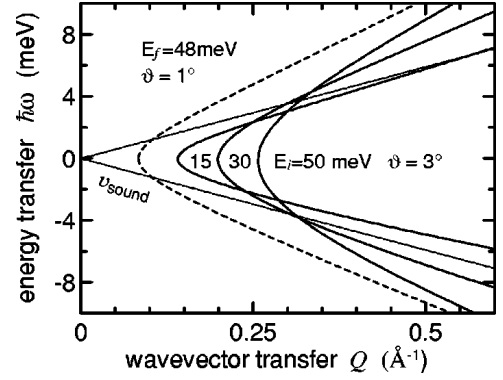}
\vspace{-2cm}\caption[OK]{Sketch of the kinematic regions accessed
in the experiments of Bove et al. (dashed line) and Cabrillo et
al. (continuous lines). The linear sound dispersion is also
reported.} \label{kin_k}
\end{figure}

Very recently, two sets of INS experiments have been reported in
molten K just above the melting temperature, one at the ISIS
source \cite{cab_k}, and the other at the ILL \cite{bov_k}. In the
first case, two time of flight spectrometers were utilized (IRIS
and MARI) aiming at a combined study of the dynamic structure
factor with different energy resolutions for the narrow
quasielastic and the broader inelastic component. The experiment
of Bove et al. has been instead performed on the triple axis
spectrometer IN1 optimized to access a broader kinematic region,
as shown in fig.\ref{kin_k} where a detail of the energy-momentum
region accessed in the two experiments is shown.

\begin{figure} [h]
\centering
\includegraphics[width=.4\textwidth]{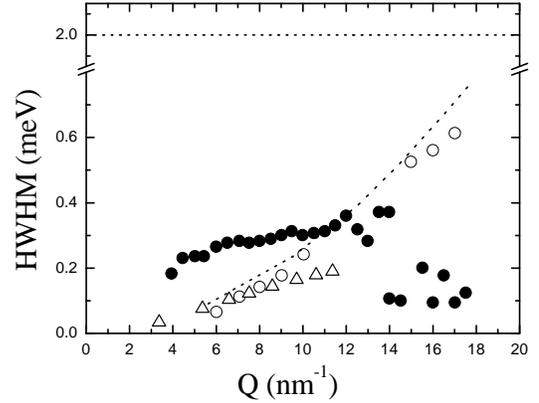}
\vspace{-4.5cm}\caption[cosanostra]{Quasielastic linewidth
according to recent INS measurements. Full and open circles are
the coherent and incoherent contributions, respectively,
determined with TOF \cite{cab_k}. Open triangles are the
incoherent linewidth as measured with TAS \cite{bov_k}, the
observed $2$ meV coherent contribution is also indicated. The
lower dotted line indicates the fickian approximation.}
\label{diff_k}
\end{figure}

The data taken on IRIS allowed for an accurate, high resolution,
determination of the diffusive processes underlying the incoherent
dynamics. The results support the hydrodynamics predictions
corrected by the mode coupling terms (see eqs.
(\ref{mct1}-\ref{mct4}) of section \ref{sec_mct}), as reported in
fig. \ref{diff_k}. Beyond the diffusive mode, Cabrillo et al.
identify a second contribution, coherent in nature according to
the authors, which is weaker, broader and almost $Q$ independent
up to $Q\approx Q_m$, while becomes narrower above $Q_m$. The $Q$
dependence of this coherent mode is rationalized in terms of
extended heat mode (Eq. (\ref{zh_hs}) of section \ref{sec_kin})
but its ultimate origin is rather ambiguous, especially in view of
the INS measurements taken at IN1 and MARI.

\begin{figure} [h]
\centering
\includegraphics[width=.4\textwidth]{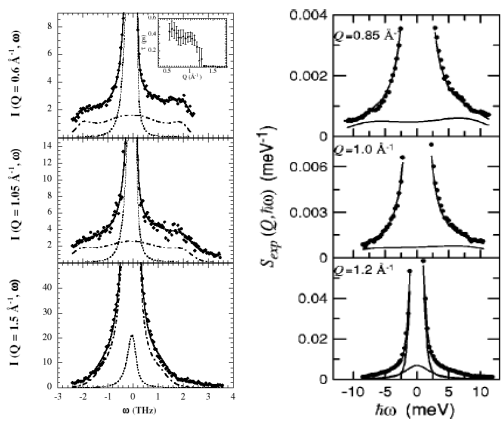}
\vspace{-4.5cm} \caption[Cabrillo - Fig. 2 - PRL 89, 075508
(2002)]{Left panel: TOF determination of the DSF in liquid
potassium (circles) \cite{cab_k}. The dash-dotted line depicts the
coherent contribution. Right panel: TAS measurements in a similar
momentum transfer region (circles) \cite{bov_k}. The continuous
line is the inelastic contribution to the collective dynamics}
\label{kins}
\end{figure}

The results of these two experiments are reported in fig.
\ref{kins} for similar fixed Q values \cite{cab_k,bov_k}. As can
be easily noticed the possibility (offered by IN1) of extending
the INS measurements at low Q is paid in terms of resolution. In
both cases, however, evidence for inelastic coherent scattering is
reported, though the incoherent scattering largely dominates in
the region where collective modes are more visible. The two sets
of data have been analyzed according to different approaches by
the respective authors. Cabrillo et al utilized a memory function
approach truncating the continued fraction at $n=2$, motivating
this assumption as necessary to account for the nearly $Q$
independence of the coherent quasielastic contribution reported in
fig. \ref{diff_k}. Odd enough, as evinced from Fig. \ref{kins}
(left panel), neither the inverse relaxation time, nor the raw
quasielastic width that they extract with this model favourably
compare with the coherent linewidth reported in fig. \ref{diff_k}.
Bove et al., on the other side, utilize a Damped Harmonic
Oscillator for the \textit{purely} inelastic term and two
lorentian for the quasielastic coherent and incoherent
contributions, respectively. The results for the incoherent part,
achieved within the jump diffusion model described in section
\ref{sec_jd}, are consistent with the high resolution measurements
(IRIS) of Cabrillo et al. (see fig.\ref{diff_k}). On the other
side, the coherent contribution turns out to be much broader
(FWHM$\approx 4$ meV, i.e. a relaxation time $\tau\approx 0.32$
ps), in contrast with the Cabrillo's data reported in
fig.\ref{diff_k} (full circles), but in qualitative agreement with
the quasielastic linewidth and the relaxation time of the same
author's measurements reported in fig.\ref{kins}.

Both the experiments extract the dispersion curves, in one case
following the exact hydrodynamic prescription as the maxima of the
current correlation function \cite{cab_k} and, in the other
\cite{bov_k} as the DHO frequency, which coincides with the
current correlation maximum if the presence of the quasielastic
coherent term is neglected. The two independent determinations are
indeed in good agreement, except at large wavevectors where the
data of Bove et al. are systematically higher though with some
scattering. The sound velocity values exhibit the usual excess in
respect to the hydrodynamic value. This high frequency sound is
ascribed by both the studies as a reminiscence of solid like
behavior, i.e. as the upper edge of a transition occurring from
the low $Q$, hydrodynamic domain to the high frequency regime.
This claim stems on the basis of the similarity of the sound
velocity value of molten potassium with the value for the
crystalline acoustic phonons along the [1 0 0] direction.

\begin{figure} [h]
\centering
\includegraphics[width=.4\textwidth]{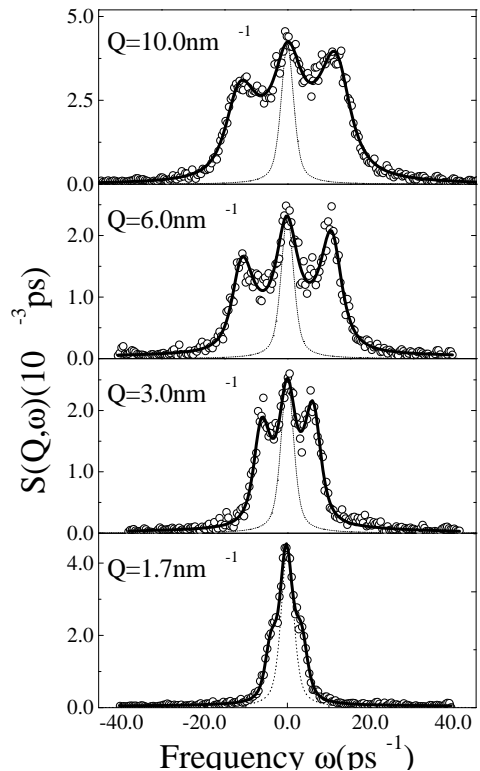}
\caption[Monaco et al. chiedere J.Chem.Phys.]{High resolution IXS
measurements in liquid potassium (open circles) \cite{mon_k}. The
continuous line is the lineshape description according to a
multiple relaxation memory function model (see text).}
\label{kixs}
\end{figure}

\begin{figure} [h]
\centering
\includegraphics[width=.4\textwidth]{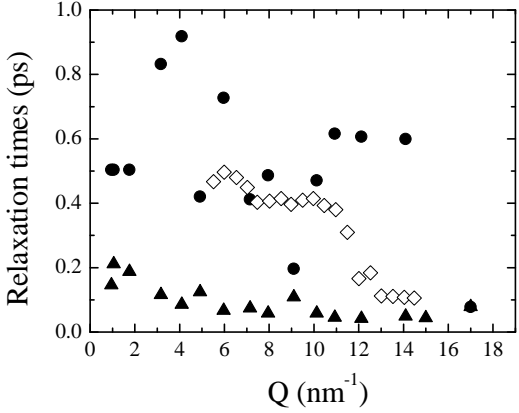}
\vspace{-5cm}\caption[Monaco et al, OK]{Viscous relaxation times
as measured by means of IXS \cite{mon_k}. Circles: structural
relaxation time. Triangles: microscopic relaxation time. The
relaxation time obtained by means of INS is also reported
\cite{cab_k}, showing how it averages between the two mechanisms
reported by IXS.} \label{tau_k}
\end{figure}

\begin{figure} [h]
\centering
\includegraphics[width=.4\textwidth]{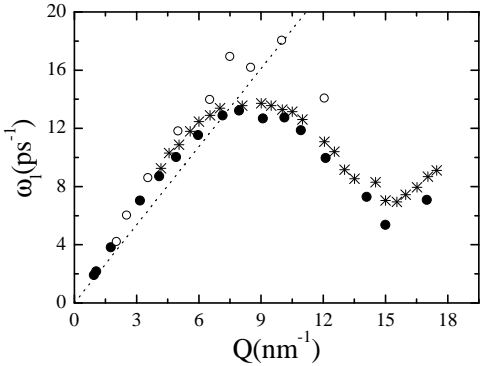}
\vspace{-5.5cm} \caption[Cosanostra OK]{Dispersion curves (maxima
of the current correlation function) measured by INS (open circles
\cite{bov_k}, stars \cite{cab_k}) and IXS (full circles
\cite{mon_k})} \label{kdisp}
\end{figure}

A recent IXS experiment on molten K \cite{mon_k} contributed to
shed some light on some aspects of the collective dynamics, giving
a coherent picture in terms of relaxation processes which is
common to several other simple fluids and, more generally, to
glass forming materials and molecular liquids (see fig
\ref{kixs}).

\begin{figure} [h]
\centering
\includegraphics[width=.4\textwidth]{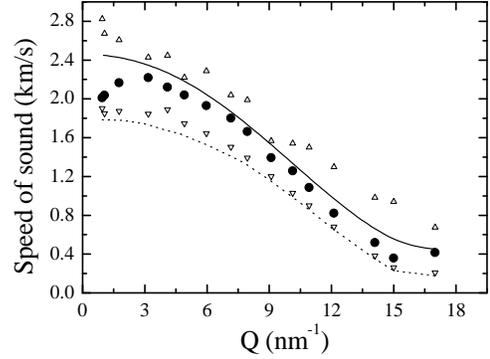}
\vspace{-5.5 cm} \caption[Cosanostra OK]{Sound velocities as
determined by IXS: apparent (circles, from the maxima of
$C_L(Q,\omega)$, isothermal (dotted line) and infinite frequency
limit (continuous line) determined from Eqs.(\ref{a2}) and
(\ref{winf}), respectively. The unrelaxed sound velocity values,
due to the structural relaxation only (down triangle) and to the
whole relaxation process (uptriangle) are also reported as
estimated by the fitting.} \label{kvelo}
\end{figure}

First, it has been shown how the coherent dynamics is driven by
thermal and viscous processes. These latter, which are dominant,
proceeds over two different timescales. Consequently, the FWHM of
the quasielastic (coherent) contribution is {\textit per se} not
directly associated to any relevant timescale. The thermal
process, indeed, is characterized by a timescale largely exceeding
the Brillouin frequency, while both the viscous processes controls
the quasielastic width. The corresponding relaxation times can be
instead determined within the memory function formalism of Eq.
(\ref{x3tempi}), obtaining the results reported in fig.
\ref{tau_k}. In the same plot, the relaxation time obtained by
Cabrillo et al. (consistent with the determination of Bove et al.)
is also reported. As one might expect, this value is somehow
averaging between the two distinct viscous processes. More
specifically, at low Q the one time approximation of Cabrillo et
al. seem to mimic the slower process, while at higher Q is
describing the faster process. This hypothesis is consistent with
the observation, reported in other alkali metals, of a decreasing
weight of the slow relaxation process on increasing the wavevector
\cite{scop_jpc}. As far as the sound propagation properties are
concerned, the IXS experiment analyzed in terms of generalized
hydrodynamics suggest a minor role of the structural process,
which accounts for approximatively $10\%$ of the whole positive
dispersion effect, which is dominated by the faster process (see
fig. \ref{kvelo}). This observation, already reported for many
other simple liquids, poses against the commonly invoked
explanation of the positive dispersion in terms of transition from
liquid to solid like regime (structural relaxation).

\subsubsection{Rubidium}

Liquid Rubidium has been the first of the alkali metals to be
addressed by a very famous neutron scattering experiment
\cite{cop_rb}, immediately substantiated by molecular dynamic
simulations \cite{rahman_sim}. The reason for such interest lies
in the possibility of extracting information on the collective
dynamics, given the almost negligible incoherent cross section
($\sigma_i / \sigma_c \approx 10^{-4}$) and the relatively low
sound velocity value. The result of this experiment contributed to
open the route to the understanding of the collective dynamics in
simple liquids, showing that the presence of an high frequency
inelastic mode is an intrinsic property of the alkali metals not
related to quantum properties or critical thermal population
effect as suggested by earlier works on liquid hydrogen
\cite{carn_h2}. Though the experiment was affected by an
elaborated multiple scattering subtraction (due to the lack of an
absolute normalization), which lead to a possibly unreliable
quasielastic spectral component, some other chords of interest
where hit. At variance with earlier results on liquid lead
\cite{dor_pb}, no evidence of secondary modes of transverse nature
was reported in this system. Finally, a mild positive dispersion
effect was observed (though Copley and Rowe looked at the maximum
of $S(Q,\omega)$ rather than to the maximum of $J(Q,\omega)$)
which was tentatively ascribed to the distinction of zero sound
and first sound as discussed by Egelstaff \cite{EGELSTAFF}.

\begin{figure} [h]
\centering
\includegraphics[width=.4\textwidth]{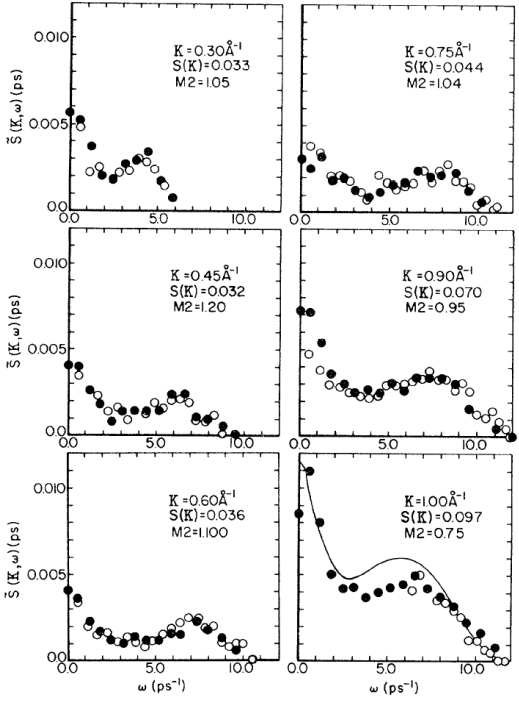}
\caption[Copley and Rowe, PRL 32, 49 (1974) Fig.1]{The
experimental determination of constant $Q$ slices of the DSF by
means of INS scattering in liquid Rubidium. A new era for the
study of collective properties in simple liquids. From
\cite{cop_rb}} \label{rbins}
\end{figure}

More recently, an inelastic scattering experiment was performed
with cold neutrons \cite{chi_rb} aiming at the determination of
the dynamic scattering law in an extended temperature region
beyond the one explored by Copley and Rowe. The kinematic accessed
region is above $Q=9$ nm$^{-1}$, and therefore the observed
excitations lies beyond the linear dispersion region. This work
has the merit to stress the importance of the choice of the
appropriate dynamical variable and of the fitting model to
determine the dispersion curve.

The old data of Copley and Rowe have been more recently reanalyzed
in terms of generalized hydrodynamics \cite{mor_rbcs}, comparing
the results to the ones obtained in molten Cesium, which are
discussed in section \ref{sec_Cs}.

To our knowledge no IXS measurements have been reported on liquid
Rubidium. The main difficulties for such an experiment would be
the very small absorption length (about $200$ $\mu$m) and the
quite low sound velocity value ($\approx 1400$ m/s) which would
confine the elastic modes on the tail of the resolution function.

\subsubsection{Cesium \label{sec_Cs}}

The experimental determination of the dynamic scattering law in
liquid cesium is dated to the early nineties \cite{gla_cs,bod_cs}.
Due to the relatively small incoherent cross section and to the
low sound velocity value, after Rubidium liquid Cs is the most
favorable alkali metal aiming at the study of collective dynamics
by means of INS. Despite its late outlet, the work of Gl{\"a}ser
and Bodensteiner reports an impressive state-of-the-art triple
axis experiment, and a robust data reduction performed with
innovative algorithms \cite{bod_phd}.

\begin{figure} [h]
\centering
\includegraphics[width=.4\textwidth]{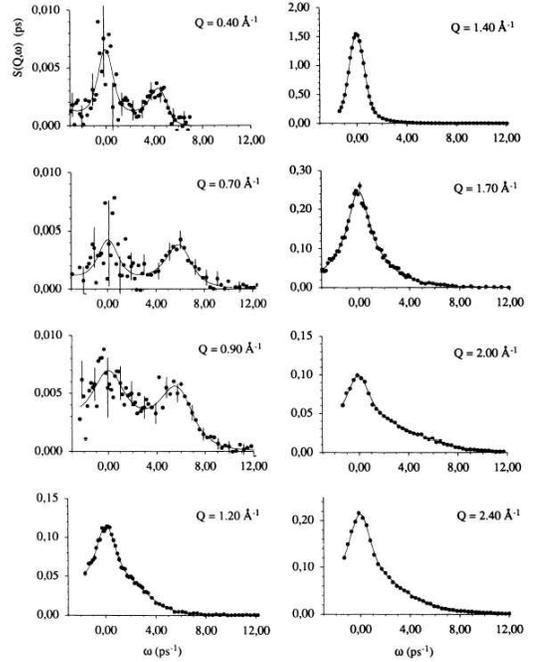}
\caption[Bodensteiner Glaser Morkel Fig. 5 PRE 45, 45
(1992)]{Dynamic structure factor of Cesium at the melting point
(circles). The continuous line is the viscoelastic approximation.
From \cite{bod_cs}} \label{csins}
\end{figure}

\begin{figure} [h]
\centering
\includegraphics[width=.28\textwidth,angle=270]{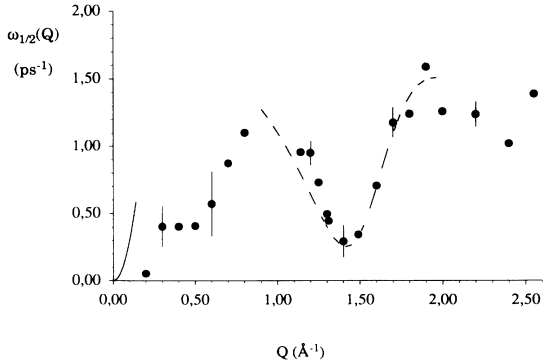}
\vspace{-.1cm} \caption[Bodensteiner Glaser Morkel Fig. 8 PRE 45,
45 (1992)]{Linewidth of the coherent quasielastic spectral
component in molten Cesium (circles). The continuous line is the
thermal value $D_TQ^2$, while the dashed line is the hard sphere
prediction from Eq.(\ref{zh_hs}). From \cite{bod_cs}}
\label{cswidth}
\end{figure}

The experiment is focused on the determination of the collective
properties and, though a careful subtraction of the incoherent
contribution is performed, once more it emerges the intrinsic
difficulty of determining the quasielastic part of the coherent
spectrum. This notwithstanding, the data clearly show the
departure of the collective dynamics from the strict hydrodynamic
region and the evolution toward the free streaming limit. This
effect is quantified by the behavior of the FWHM of the
quasielastic line reported in Fig.\ref{cswidth}: while the
hydrodynamic prediction, based on purely adiabatic thermal
fluctuations, predicts $\omega_{1/2}=D_TQ^2$, the actual linewidth
is always below this limit in the whole explored region,
indicating the dominant presence of viscous processes in the
quasielastic spectrum, as recently pointed out in other alkali
metals and liquid aluminium \cite{scop_comm} against an opposite
interpretation in terms of linearized hydrodynamic models
\cite{sing_pre,sing_rep}. The De Gennes narrowing was also
observed \cite{dej_ema}, and the $Q$ dependence of the linewidth
was described within the hard sphere extended mode approximation
(Eq.\ref{zh_hs}, section \ref{sec_kin}). The free streaming limit
is not yet attained at wavevectors as large as twice the position
of the main peak of the static structure factor \cite{bod_cs}. The
lineshape analysis was performed with several approaches, within
the extended hydrodynamic model \cite{desh_hyd0}, with the
viscoelastic model \cite{lov_visco} and with two relaxation times
accounting for both viscous and thermal processes. In this latter
case, it was found a negligible role of the thermal process on
approaching the first maximum of the structure factor, though the
fitted values of the thermal relaxation time were in significant
disagreement with the expected values $1/D_T Q^2$ (fig 9 of ref.
\cite{gla_cs}). It was then pointed out the impossibility of
discriminating the different models, given the s/n ratio of the
available data. The $Q-$ dependence of the longitudinal viscosity,
extracted from $S(Q,\omega=0)$ value (according to the
prescription of generalized hydrodynamics), showed a decreasing
behavior previously observed in Lennard-Jones systems
\cite{aila_visco,tank_visco} and more recently in several other
liquid metals \cite{scop_prlga,mon_k}. Finally, the maxima of
current correlation spectra showed the usual positive dispersion
effect, which was interpreted as precursor of the solidification
according to the usual idea of a transition between a liquid and
solid-like regime.

Neutron scattering data on liquid cesium have also been used as a
benchmark to develop an approach based on the idea of timescale
invariance of the relaxation processes \cite{yulm_cs}. Within the
Zwanzig Mori projectors formalism, one can construct an infinite,
non Markovian, set of interconnected kinetic equations relating
each memory function with the one of higher order \cite{mori_mf}:

\begin{eqnarray}
\frac{dF(Q,t)}{dt}&=&-\Omega_{1}^{2}\int_{0}^{t}d\tau
M_{1}(Q,\tau)F(Q,t-\tau) \nonumber \\
\frac{dM_{1}(Q,t)}{dt}&=&-\Omega_{2}^{2}\int_{0}^{t}d\tau
M_{2}(Q,\tau)M_{1}(Q,t-\tau) \nonumber \\
..............&.&................................................ \nonumber \\
\frac{dM_{i}(Q,t)}{dt}&=&-\Omega_{i+1}^{2}\int_{0}^{t}d\tau
M_{i+1}(Q,\tau) M_{i}(Q,t-\tau). \label{chain} \end{eqnarray}

\noindent where $F(Q,t)$ is the normalized density correlation
function and $\Omega_i$ are characteristic frequencies of the
process. Following the Bogoliugov approach of the reduced
description, one hypotizes the time scale invariance of the
relaxation processes beyond a certain level, defining a closure
level $M_{i+1}(t) \approx M_{i}(t)$ and thus getting an explicit
expression for the DSF in terms of the spectral moments.

To our knowledge no IXS experiments on liquid Cs have been
reported, for the same kind of difficulties as in liquid Rb.

\subsection{Alkaline earth elements}

\subsubsection{Magnesium}

Magnesium is one of the simplest divalent elements. The coherent
dynamic structure factor has been recently determined at the
SPring8 IXS beamline \cite{kaw_mg}. The dispersion curve shows a
$8 \%$ deviation from the adiabatic sound velocity, with a maximum
value lying halfway the hydrodynamic and the $c_\infty$ value. An
average relaxation time was determined ($\tau=0.094$ ps), which is
about one third of the one of the neighboring alkali element
liquid Na \cite{pil_na}. The analysis of the quasielastic line
revealed a $Q^2$ broadening in the quasi-hydrodynamic regime,
while around the De Gennes narrowing region the linewidth was
successfully reproduced by the de Schepper and Cohen model
\cite{desh_hs}, i.e. through Eq. (\ref{zh_hs}). Molecular dynamics
simulations have been recently performed in this system by both
classical molecular dynamics and orbital free \textit{ab initio}
simulations \cite{alem_mg,gonz_potalc}. The two approaches give
very similar results as far as the phonon dispersion is concerned,
while the quasielastic contribution is less pronounced in the
\textit{ab initio} calculation. In this respect, a comparison with
the experimental data \cite{kaw_mg} would be extremely
interesting, though one should first convolute the calculated
$S(Q,\omega)$ with the instrumental IXS resolution and take into
account for the quantum correction arising from detailed balance
condition.

\subsection{Group III elements}

\subsubsection{Aluminium}

One of the most puzzling results of early neutron spectroscopy is
the striking similarity between the spectra of polycrystalline and
liquid Aluminium observed in Stockholm in 1959 and published in
the final form a few years later \cite{lar_al}. In this experiment
TOF data were collected on a cold neutron spectrometer, but at
those times multiple scattering corrections were almost impossible
and therefore the spectra did not obtain a detailed explanation. A
second TOF experiment was performed in the same period, but again
the results were laking a detailed interpretation \cite{coc_al}.
Fifteen years later the original experiment of Larsson was
revisited, in an effort to re-analyze the results in the light of
up to date theoretical developments. More specifically, the
experimental results were used to test the old convolution
approximation \cite{vine_convo}, mean field approaches
\cite{sing_meanf} and kinetic theory \cite{sjog_kin,sjogr_al}.
Nevertheless, these data have been collected at very large energy
and wavevectors value and always presented as time of flight scan,
and therefore they are not very helpful to establish the existence
of collective modes in this system. Aluminium, indeed, is an
almost purely coherent scatterer, but the high sound velocity
value prevents the study of acoustic modes by means of INS (see
table \ref{table}). This is testified by a more recent INS
experiment performed at IN4 (ILL) \cite{eder_al}. Although
multiple scattering correction and constant -$Q$ cuts reduction
have been performed in this case, no evidence of collective modes
could be reported due to the restricted kinematic region
corresponding to the slow neutrons utilized ($\approx 55$ meV).

\begin{figure} [h]
\centering
\includegraphics[width=.45\textwidth]{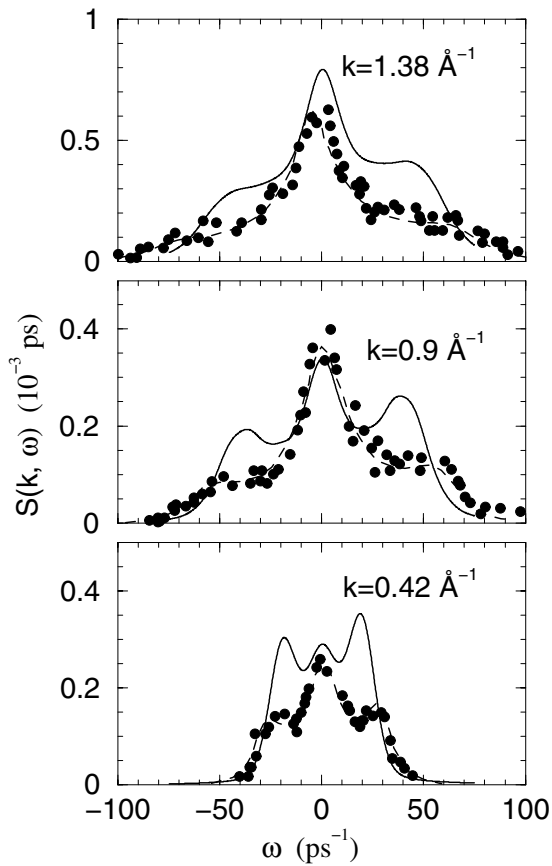}
\caption[Gonzalez Fig. 6 JPC 13, 7801 (2001)]{Dynamic structure
factor of molten aluminium. Comparison between the OF-AIMD
calculation (continuous line) \cite{gonz_al} and the experimental
IXS values with their fitting based on generalized hydrodynamics
\cite{scop_preal}. From \cite{gonz_potalc}} \label{al}
\end{figure}

Much of the knowledge about the microscopic dynamics in liquid Al
relies, in fact, on the numerical work of Ebbsj{\"o}
\cite{ebb_al}, who calculated the dynamic structure factor
utilizing two different local pseudopotentials and the local
pseudopotential originally developed by Ashcroft
\cite{ashc_potalk}. The dynamic structure factor has been shown to
be somehow reminiscent of the viscoelastic model with the addition
of a gaussian term, accounting for the approach to the high $Q$,
free streaming limit. He was then able to predict the existence of
collective modes for $Q<10$ nm$^{-1}$, though he reported sound
velocity values distinctly larger then the adiabatic value over
the whole explored range ($Q>3$ nm$^{-1}$). Triggered by this
observation, a modified version of the viscoelastic approach was
developed \cite{gas_al} and tested against the data of Ebbsj{\"o}.
At the same time, it has been proposed a single relaxation process
scenario based on a $sech$ memory function shape within the
Mori-Zwanzig scheme \cite{tank_al}.

\begin{figure} [h]
\centering \vspace{-2cm}
\includegraphics[width=.45\textwidth]{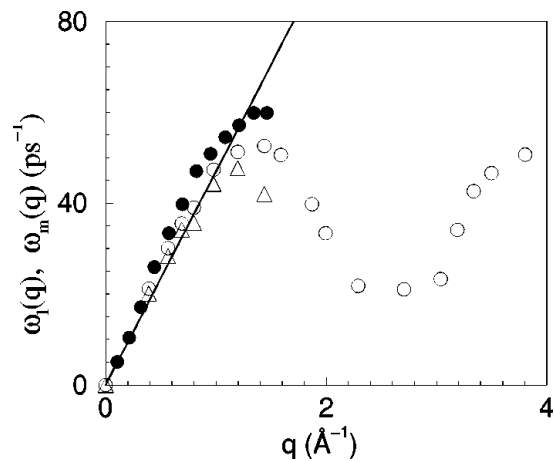} \vspace{-3cm}
\caption[Gonzalez Fig. 10 PRB 65, 184201 (2001)]{Sound dispersion
of liquid aluminium from the maxima of the current correlation
function: open circles, OF-AIMD calculation \cite{gonz_al}, full
circles IXS experimental values \cite{scop_preal}. The dispersion
from the maxima of the dynamic structure factor numerically
evaluated is also reported (open triangles), as well as the
hydrodynamic value (continuous line). From \cite{gonz_al}}
\label{aldisp}
\end{figure}

\begin{figure} [h]
\centering
\includegraphics[width=.45\textwidth]{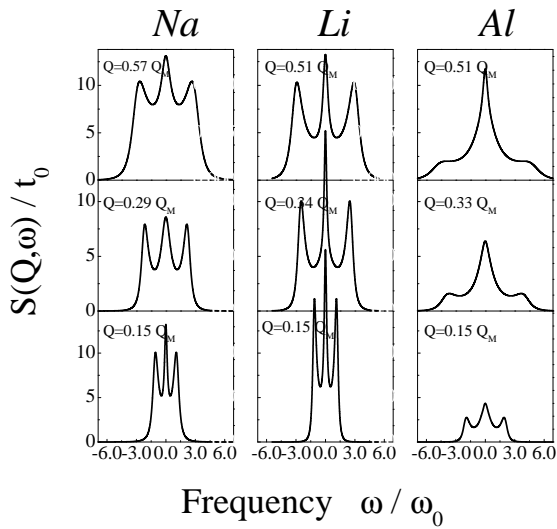} \vspace{-4cm}
\caption[Cosanostra OK]{Resolution deconvoluted, classical
lineshape utilized to described the IXS spectra of molten Li
\cite{scop_prlli,scop_prena,scop_preal}, Na and Al, reported on
relative momentum and energy scale (see text)} \label{linaal}
\end{figure}

The first experimental observation of collective modes in liquid
Al has been reported much more recently by means of IXS
\cite{scop_preal}. An high frequency regime has been observed for
$Q>5$ nm$^{-1}$, while below this value the dynamics approaches
the hydrodynamic limit, though the transition is not fully
accomplished at the lowest investigated wavevector, $Q=1$
nm$^{-1}$. The scenario arising from the IXS study is much similar
to the one characterizing alkali metals, though with significant
quantitative differences such as a more intense quasielasic
component testifying a more important role of the structural
relaxation in this system.

The IXS data have been recently used to test orbital free
\textit{ab initio} calculations (OF-AIMD)
\cite{gonz_al,gonz_potalc}. The overall agreement is quite
satisfactory though the numerical calculations show somehow lower
sound velocity value and tend to overemphasize the inelastic
components. From Fig.\ref{aldisp} one can argue the importance of
the dynamical variable representing the sound velocity. The
presence of positive dispersion, in particular, is strongly
affected by the choice of the maxima of $C_L(Q,\omega)$ rather
than the ones of $S(Q,\omega)$.

In Fig. \ref{linaal}, finally, we report a comparison of the
lineshape obtained from the resolution-deconvoluted, classical
lineshape utilized to fit the IXS spectra of liquid lithium,
sodium and aluminium at the same reduced values of momentum
($Q/Q_M$) and energy ($\omega t_0=\omega / \omega_0$ with
$t_0=\sqrt{m/k_BT_m}/Q_M$) transfer. As can be clearly evinced,
the attitude of alkali metals to sustain density fluctuation is
much more pronounced than in other simple liquid metals.

\subsubsection{Gallium}

Among simple liquid metals, Ga is endowed with peculiar structural
and electronic properties. In addition to the low melting point
($T_m=303$ K), it shows an extended polymorphism in the solid
phase with complex crystal structures where a competition between
metallic and covalent bindings takes place. Despite the nearly
free electron electronic DOS anomalies associated with some
covalency residue are present. Moreover, the first peak of the
$S(q)$ presents a hump characteristics of non close-packed liquid
structures \cite{bf}.

Early inelastic scattering on liquid Ga were performed at the
beginning of the seventies with neutrons
\cite{loff_phd,pag_ga,bos_ga,gla_ga}. Due to the quite large sound
velocity value, compared to the available kinematic range, these
studies were mainly addressed to the investigation of the
quasielastic part of the dynamic structure factor.

\begin{figure} [h]
\centering
\includegraphics[width=.45\textwidth]{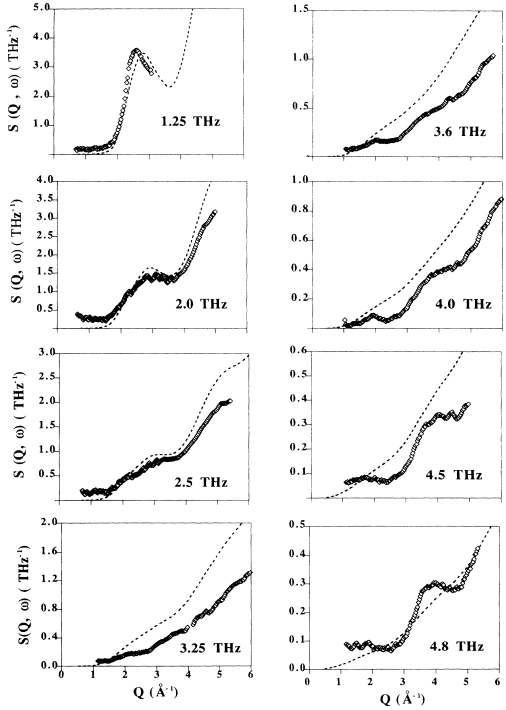}
\caption[F.J. Bermejo et al Fig. 5 PRE 49, 3133 (1994)]{INS
constant energy scans (circles) compared to the model of Buchenau
for plane wave excitations in solids \cite{buch_Q}. From
\cite{ber_ga1}.} \label{gaberqscan}
\end{figure}

Twenty years later, another series of INS experiments were
performed on a triple axis instrument \cite{ber_ga1} with the aim
of ascertaining the presence of low $Q$ collective modes just
above the melting temperature. Indeed, although by virtue of the
above mentioned anomalies liquid Ga seems to elude the picture of
the high frequency dynamics emerging in all the monatomic liquids,
on the basis of the shape of the interaction potential evidence
for collective modes should be expected below $Q_m/2$. Quite
surprisingly, no evidence of inelastic signal was reported in the
region were longitudinal modes were expected on the basis of the
hydrodynamic sound velocity. This result was interpreted as an
overdamping effect traced back to the high value of longitudinal
viscosity ($\approx 10$ cP). Additionally, by comparing constant
energy scan (see Fig.\ref{gaberqscan}) to the expression of
Buchenau for acoustic-like plane wave excitations in amorphous
solids \cite{buch_Q}, an excess of scattering was reported for
frequency distinctly larger then the maximum frequency of the
acoustic branch. This result was interpreted as the fingerprint of
the presence of high energy optic-like modes.

A few years later, then, a new set of experiments were performed
at higher temperatures by the same authors \cite{ber_ga2}, and a
contrasting behavior with the previous findings was reported. More
specifically, the appearance of non-overdamped sound modes was
reported, accompanied by a second, higher frequency mode of
presumably optical origin. The discrepancy between the low and
high temperature experiment was ascribed to a viscosity drop of a
factor $\approx 7$ and therefore to a narrowing of the acoustic
mode according to the hydrodynamic expression \ref{gidro}.

The presence of an higher frequency mode appearing at in the
constant $Q$ scan for wavevectors larger then $Q_m$ seems to
corroborate the presence of an optic-like excitation suggested by
the previously mentioned constant energy scans \cite{ber_ga1}.

\begin{figure} [h]
\centering
\includegraphics[width=.45\textwidth,height=11 cm]{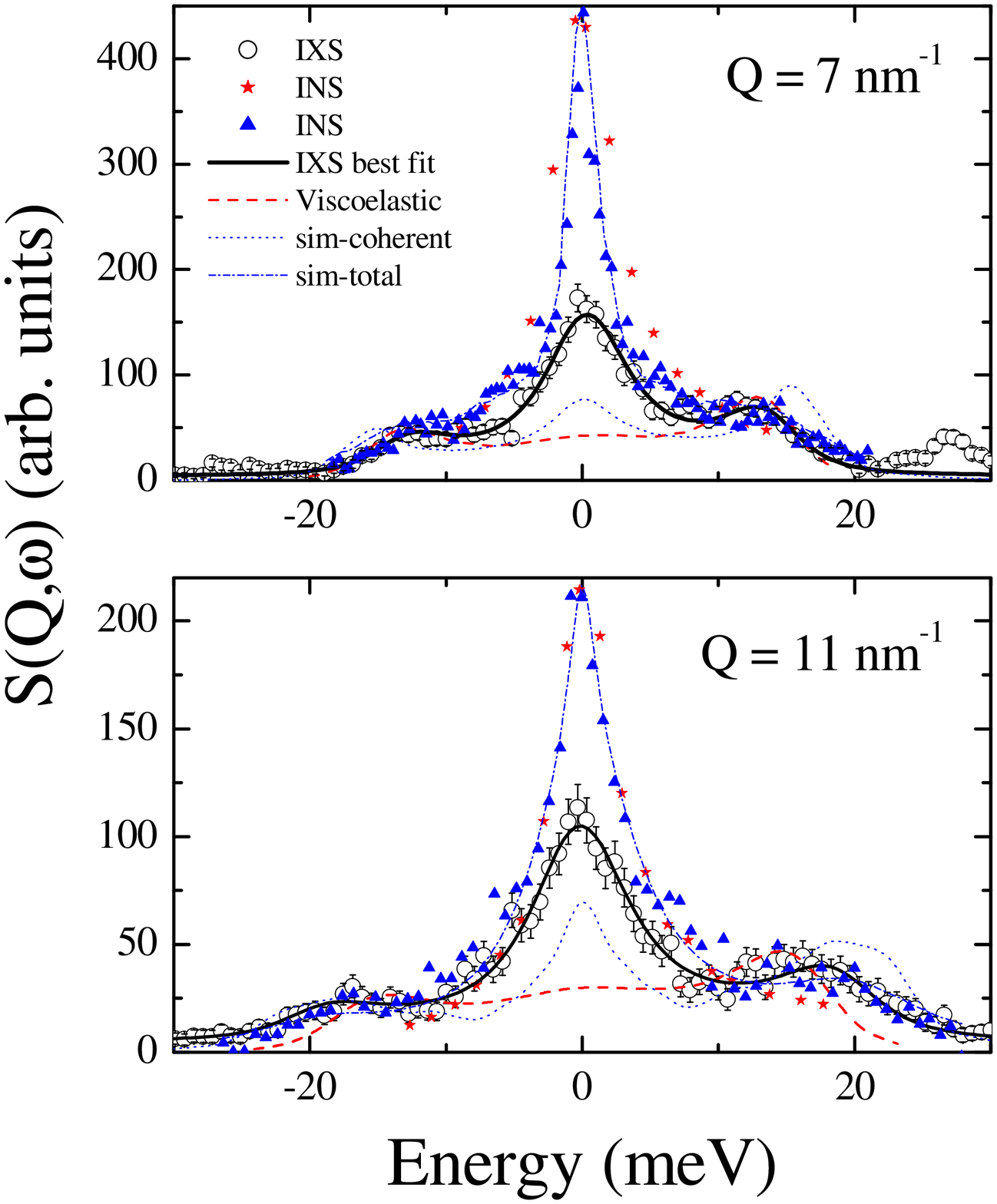}
\caption[Cosanostra OK]{Comparison between the INS (stars
\cite{ber_ga1} and triangles \cite{bov_ga}) and IXS (open circles
with error bars \cite{scop_prlga}) determinations of $S(Q,\omega)$
in gallium at the melting temperature, for two different values of
the (constant) momentum transfer. The dotted line is the
viscoelastic prediction \cite{ber_ga1} (convoluted with the INS
instrumental resolution and accounting for the detailed balance
condition), while the continuous line is the best fit to the IXS
data utilizing a memory function accounting for the thermal
relaxation and two viscous processes (see text). Molecular
dynamics symulations for the coherent (dotted line) and total
(dash-dotted line) $S(Q,\omega)$, convoluted to the INS
resolution, are also reported \cite{bov_ga}.} \label{gaixsins}
\end{figure}

A recent IXS experiment performed on liquid Gallium just above the
melting point \cite{scop_prlga} portraits the collective dynamics
in a much similar fashion to the one of Alkali metals and of
liquid Al, thus removing the anomaly suggested by the neutron
experiments. Collective modes, in fact, have been unambiguously
observed in the low temperature region where neutrons suggested an
overdamped regime. This result suggests the inadequacy of the
expression \ref{gidro} to estimate Brillouin linewidths, which can
be easily understood in terms of the generalized hydrodynamics
results reported for alkali metals \cite{scop_jpc}: outside the
truly hydrodynamic region, the viscous relaxation dynamics
proceeds over two distinct physical mechanisms, the structural
relaxation and the short-lived rattling dynamics. On the high
frequency region of interest, the structural relaxation is frozen
(the system is responding as a solid) and therefore the viscosity
associated to this process does not contribute to the sound
damping. Moreover, the thermal contribution in Eq. (\ref{gidro})
might not be correct at wavevectors as large as a few nm$^{-1}$,
since the adiabatic regime could be replaced by as isothermal one,
as already pointed out \cite{scop_comm}. Consequently,
Eq.\ref{gidro} is an overestimate of the actual linewidth (which
in the case of liquid lithium has been quantified as a factor two
\cite{scop_jpc}). In Fig. \ref{gaixsins} we report the two
experiments for similar values of the momentum transfer. The
viscoelastic prediction \cite{lov_visco,ber_ga1} is also reported,
showing how it clearly underestimates the quasielastic
contribution, though it provides a reasonable estimate of the
genuine inelastic mode. The IXS findings have been recently
corroborated by a new accurate INS experiment, used to test the
reliability of a model interaction potential by comparing the
dynamic structure factors \cite{bov_ga}.

Summing up, the lack of low temperature collective excitations
reported in this system with neutrons is probably due to the
difficulty of a reliable extraction of the coherent part of the
scattering function. On the other side, the interesting
observation of possibly optic-like high frequency modes certainly
calls for further investigation although, in our opinion, does not
justify \textit{per se} a description of the acoustic dynamics in
terms of crystalline reminiscent dynamics. The analogy with
several IXS results on different systems suggests indeed a
prominent role of the topological disorder in characterizing the
acoustic branch.

\begin{figure} [h]
\centering
\includegraphics[width=.45\textwidth]{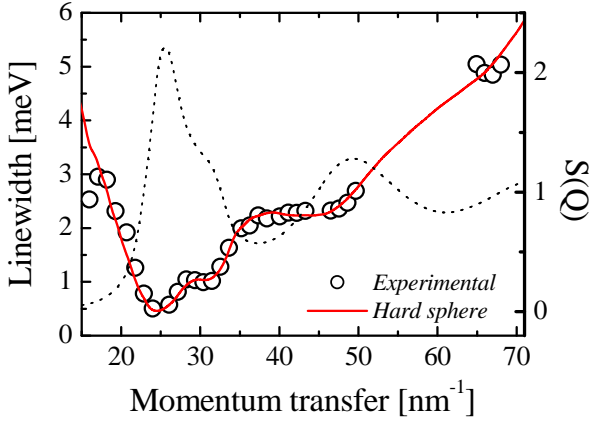}
\vspace{-4. cm} \caption[Cosanostra OK]{Full width at half maximum
of the dynamic structure factor as determined by IXS in the
kinetic regime (open circles). The prediction according to
Enskog's theory is also shown (continuous line)
\cite{scop_prlga2}. } \label{gahs}
\end{figure}

Very recently, an IXS experiment explored the high $Q$ region
($20<Q<70$ nm$^{-1}$, i.e. lengthscales smaller than the size of
the first coordination shell \cite{scop_prlga2}. While generalized
hydrodynamics provides a coherent picture of the dynamics in the
lower $Q$ region, not much is known about collective dynamics at
such short lenghtscales. For hard spheres, Enskog's kinetic theory
predicts in this region the dominant effect of a generalized heat
mode. Liquid Gallium, however, by no means can be modelled as an
hard sphere fluid, for the above mentioned structural and
electronic properties. Surprisingly, it turned out that a
description in terms of heat mode (Eq. (\ref{zh_hs})) still
applies, at the price of introducing an \textit{effective} hard
sphere diameter (larger than the one associated to the first
$S(Q)$ maximum), which probably accounts for the associative
tendency of this liquid (dimer-like structures).

\subsection{Group IV elements}

\subsubsection{Silicon}

Due to its several unusual properties, liquid Si is always
classified as a non simple liquid. While in the crystalline phase
Si is a diamond structure semiconductor, it undergoes a
semiconductor-metal transition upon melting, which is accompanied
by a density increases of about $10\%$, and by significant
structural changes (the coordination number grows from four in the
solid state to about seven in the liquid). Similarly to Gallium,
the static structure factor, S(Q), exhibits a shoulder on the
high-Q side of the first peak \cite{waseda_sn}, a feature that
cannot be reproduced using a simple hard-sphere model, appropriate
for alkali metals.

No neutron scattering data exist to the best of our knowledge,
while very recently two Inelastic X ray Scattering experiments
have been performed both at the ESRF \cite{hos_si1} and SPring8
\cite{hos_si2}.

\begin{figure} [h]
\centering \vspace{-1.8 cm}
\includegraphics[width=.45\textwidth]{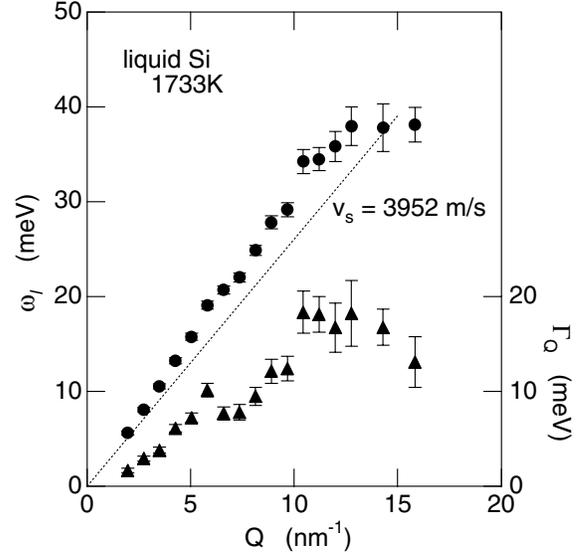}
\vspace{-1.8 cm} \caption[hosokawa spring8]{Sound velocity and
attenuation in molten silicon \cite{hos_si2}} \label{si}
\end{figure}

In the first of the two above referenced experiments, a positive
dispersion of $15 \%$ has been found. In the second experiment, an
higher resolution and more accurate study has been carried out,
allowing to follow the transition from the high frequency to the
low frequency regimes. The data were analyzed in terms of DHO
model for the inelastic component and Lorenzian for the
quasielastic, and no significant quantitative differences were
detected utilizing the same memory function scheme proposed for
other IXS studies on liquid metals \cite{scop_jpc}.

In the vicinity of the main peak of the static structure factor,
the lorenzian shape for the quasielastic component turned out to
be inadequate, and has been replaced by a combination of Lorenzian
and gaussian contributions (pseudo-Voigt).

\subsubsection{Germanium}

Liquid Ge shares the same peculiarities as liquid Silicon, though
with some slight quantitative differences.

IXS data on liquid Ge have been recently obtained \cite{hos_ge},
and they show evidence for collective propagating modes. An
analysis based on a Lorenzian shape for the quasi-elastic and a
DHO for the inelastic modes, revealed the absence of positive
dispersion effects in the investigated Q range ($2\div 28$
nm$^{-1}$). On our opinion, this result calls for further
investigations, as this is an almost unique feature in respect to
the other monoatomic liquid metals investigated sofar (especially
Silicon, which has very similar structural properties). The De
Gennes narrowing has been analyzed in terms of extended
hydrodynamic heat mode, utilizing the analytical expression
obtained within hard-sphere approximation \cite{coh_hs}, but the
quite large error bars and the limited spanned $Q$ range prevents
to draw a final conclusion.

\subsubsection{Tin}

Tin is the heaviest of 4B element. Its structural properties are
quite similar to the one of alkali metals, with a coordination
number close to 12 but with the typical shoulder on the high Q
side of the main $S(Q)$ peak \cite{waseda_sn} which is typical of
Si, Ge and Ga.

The first INS experiments in liquid Tin have to be traced back to
the early sixties. Similarly to other early neutron scattering
experiment no clear picture of the microscopic dynamics could be
outlined. In some case \cite{pal_tin}the vineyard approximation
\cite{vine_convo} was used to analyze the data, while in another
study experimental strategy for suppressing multiple scattering
were tested \cite{broc_tin}. Although a wavevector-energy plot was
obtained from raw TOF data \cite{coc_tin}, according to Copley and
Lovesey no side peaks should be observed trasforming TOF data to
$S(Q,\omega)$ on constant $Q$ slices \cite{coplov_rev}.

Constant $Q$ IXS spectra of liquid tin have been very recently
reported for low ($T=593 K$) and high ($T=1273 K$) temperatures at
the ESRF \cite{hos_sn}. The sound velocity seems to exceed the
hydrodynamic value at both temperatures of $6\%$ and$12\%$,
respectively. This notwithstanding, these quantitative estimates
must be taken with care, due to the quite large error bars.
Moreover, the dispersion curves have been determined from the DHO
frequency, neglecting the effect of the quasielastic component.
For $Q$ values close to the main peak of the structure factor, as
in the case of liquid Si \cite{hos_si2}, the lineshape of
$S(Q,\omega)$ turned out to be empirically described by a
combination of Lorentian and Gaussian contributions or,
equivalently, by a memory function analysis similar to the one
reported for alkali metals \cite{scop_prl}. As a matter of fact,
at such large $Q$ values, as previously observed for liquid
lithium \cite{scop_epl,scop_jpc}, the microscopic dynamics
undergoes a transition from the (generalized) hydrodynamic
behavior to the free streaming limit and a detailed description of
such transition is still missing.

\subsubsection{Lead}

Molten lead has been one of the first metals to be investigated by
Inelastic Neutron Scattering, as the first experiment can be
traced back to the fifties \cite{egel_pb,brock_pb}. Details of the
experiments performed up to 1975
\cite{dor_umk,rand_umk,dor_pb,coc_pb} have been exhaustively
reviewed by Copley and Rowe \cite{coplov_rev}. One of the most
interesting results has been the evidence of both a longitudinal
and a transverse branch in the dynamic structure factor, though
this result was presented with some caution as the transverse mode
could be an artifact arising from multiple scattering effects.

In the early eighties new INS studies performed with both TAS and
TOF spectrometers appeared \cite{sod_pb,sod_pb1}, aiming to
validate the presence of a dispersion relation and of a transverse
branch. A longitudinal mode, compatible with the higher frequency
excitation previously reported by Dorner \cite{dor_pb} was
reported, whose sound velocity is consistent with hydrodynamic
value. The accessible kinematic range is too limited to asses any
evidence of positive dispersion effect. No evidence for a lower
frequency excitation was instead reported, corroborating the
hypothesis that such feature is an artifact stemming from multiple
scattering.

Liquid lead has been recently in focus of molecular dynamic
simulations aiming to describe collective dynamics in terms of
generalized kinetic modes \cite{bry_pb1,bry_pb2}. Beyond the
hydrodynamic region, different branches corresponding to sound and
heat waves have been identified, and their nature has been
extensively discussed.

\subsection{Group V elements}

\subsubsection{Bismuth}

The first inelastic scattering investigation of molten bismuth was
originally reported by Cocking \cite{coc_bi}, who reported a
dispersion curve extracted by TOF neutron spectra. At the IAEA
symposium of 1968 two experiments Bi were presented: in one case
two dispersion branches were obtained from TOF data \cite{tun_bi},
though the low frequency excitation was probably due to an
artefact of a missing multiple scattering correction. In the
second study data were converted to constant $Q$ spectra
\cite{mat_bi}.

\begin{figure} [h]
\centering
\includegraphics[width=.4\textwidth]{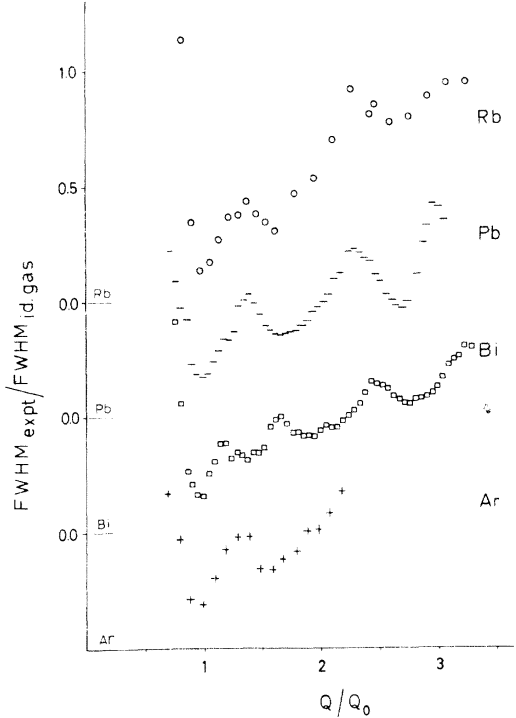}
\vspace{-.5cm} \caption[U. Dahlborg and Olsonn  Fig. 11 PRA 25,
2712 (1982)]{FWHM of molten Bi compared to other simple liquids.
It can be noted the presence of an intermediate minimum between
the first two maxima which is not present in the other simple
fluids. From \cite{dal_bi}.} \label{bi_ins}
\end{figure}

Liquid Bi has been recently the subject of new INS investigations
\cite{dal_bi}. Measurements were performed just above melting at
$T=578$ K, in a wavevector region spanning from just below the
first maximium in $S(Q)$ up to wavevector as high as $70$
nm$^{-1}$. This kinematic region lies above the region were
collective modes could be expected, so the study mainly deals with
the quasielastic spectral component. Generalized hydrodynamics
models based upon a single relaxation time were tested against the
experimental data utilizing different approximation for the memory
function shape \cite{lov_visco,aila_visco}. Experimental data were
then unfolded by instrumental resolution modelling the
quasielastic shape as Lorenzian and the resolution as gaussian.
The main point of the paper is the determination of the $Q$
dependence of the spectral FWHM, which is also compared to other
systems. In particular, it is pointed out how, beside the expected
De Gennes narrowing occurring at $Q_m$, the FWHM shows a minimum
rather then the expected maximum at $Q \approx 1.5 Q_m$ (which
characterize the dynamics in several simple liquids such as Pb, Rb
and Ar). This anomaly is related by the authors to the shoulder
observed in the static structure factor just above $Q_m$, and
therefore identified as a non-simple nature of liquid Bi.

The kinetic region has been investigated in liquid Bi within the
generalized collective mode approach \cite{bry_bi1}. The presence
of high frequency kinetic branches has been ascertained, and it
has been pointed out that their weight is too small to make them
visible in the dynamic structure factor. This result seems to be
in agreement with recent IXS findings on a very similar system,
namely liquid Ga, in which only acoustic modes were reported
\cite{scop_prlga}.

\subsection{Transition metals}

\subsubsection{Mercury}

Experimental studies of microscopic dynamics in liquid mercury are
very recent compared to the systems reviewed so far, and they were
presented at the LAM XI conference (Yokohama, Japan). The first
investigation was obtained by means of INS at the IN1 facility of
the ILL \cite{bov_hg,bov_hg_lam}. In this work, an detailed
investigation of the dynamic structure factor is undertaken at
room temperature, and presented as constant $Q$ cuts in the range
$2.5 \div 12$ nm$^{-1}$ with a high energy resolution of $\delta E
\approx 1$ meV. The data are analyzed with an empirical model
consisting of a DHO for the purely inelastic part and the sum of
two lorentian functions accounting for the quasielastic
contribution. While the inelastic component is no doubt of
coherent nature, the narrower of the two lorentians is ascribed to
incoherent scattering, and modelled as a simple diffusive term of
linewidth $DQ^2$. Given the values of $D$, this results in a quasi
elastic contribution which could not be resolved by the much
broader resolution function. The linewidth of the broader
lorentian is almost $Q$ independent, and its origin is ascribed by
the authors to an incoherent process, on the basis of the
coincidence of the DHO area with independent (static) structure
factor determinations. Turning our attention to the collective
dynamics, the extrapolated high frequency value of the sound
velocity ($c_\infty(Q\rightarrow 0)=2100 \pm 80$ m/s) , obtained
by the DHO frequency parameter (therefore approximately equal to
the maximum of the current correlation function, the difference
being due to potential quasielastic coherent contribution), turns
out to largely exceed the hydrodynamic value ($c_s=1451$ m/s),
suggesting a huge positive dispersion effect close to $50\%$
largely exceeding similar effects reported in other metals. This
result is rationalized in terms of Bohm-Staver model, which
provides the expression of Eq.(\ref{BS}) for the sound velocity
yielding, for molten Hg, $c_l(Q\rightarrow 0)\approx2090$.

\begin{figure} [h]
\centering
\includegraphics[width=.4\textwidth]{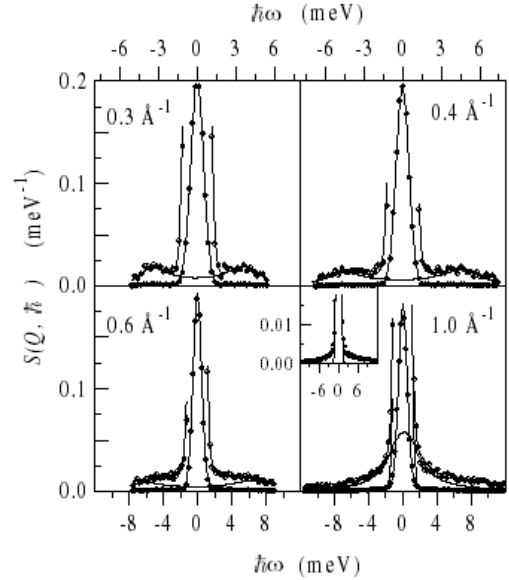}
\vspace{-2.5cm} \caption[Bove Fig. 1 PRL 87, 215504 (2001)]{DSF of
molten Hg measured by means of INS at the indicated $Q$ values.
The inelastic and the quasielastic componens, modeled with two
lorenzians and a DHO, respectively, are also shown. From
\cite{bov_hg}} \label{hg_ins}
\end{figure}

Nearly in coincidence with the neutron experiment of Bove
\textit{et al.}, an IXS study of the coherent dynamics in liquid
Hg appeared \cite{hos_hg}, in which the dynamic structure factor
is examined in a wavevector region extended up to $Q_m \approx 25$
nm$^{-1}$, with a factor 2 worse resolution, but in a
significantly larger energy region. As in the work of Bove
\textit{et al.}, the genuine inelastic features of the data are
modelled with a DHO function, but it appears clearly from the raw
data that a coherent quasielastic term dominates the $\omega
\approx 0$ region. This latter contribution is modelled with a
lorenzian shape. Although the authors neglect the presence of this
quasielastic in the calculation of the sound velocity value,
taking the DHO frequency as the relevant parameter they obtain a
value of $c_\infty(Q\rightarrow 0)=1840$ m/s, an estimate which is
directly comparable with the corresponding INS determination. This
discrepancy, which may be due to the limited energy range at the
low $Q$'s of the INS experiment, as well as to the non negligible
resolution effect on the lowest $Q$'s of the IXS experiment, calls
in our opinion for further investigations and suggests to take
with some care any interpretations in terms of Bohm Staver model
of the positive dispersion at least by the quantitative point ov
view. Hosokawa \textit{et al.}, on the other hand, cast the
anomalous dispersion in the usual framework of the shear
relaxation. The IXS work also confirms the non negligible presence
of quasielastic signal in the coherent dynamic structure factor,
suggesting that it should be taken into account also in INS data
treatment. We expect that the broader lorenzian contribution
reported by Bove \textit{et al.}, for example, could be at least
partially coherent in nature.

\begin{figure} [h]
\centering
\includegraphics[width=.4\textwidth]{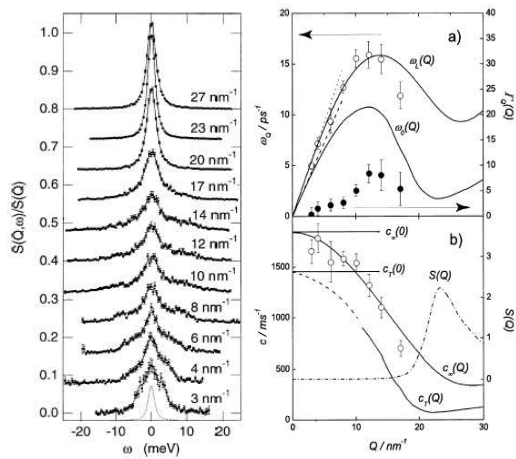}
\vspace{-4cm} \caption[S. Hosokawa Fig. 1 and 2 JNCS 312-314, 163
(2002)]{Left panel: IXS determination of the DSF in Hg near the
melting temperature. Right upper panel: dispersion relation and
sound attenuation properties as deduced by a DHO + one Lorenzian
fit. The low and high frequency limits are also reported. Right
lower panel: Corresponding sound velocities. From \cite{hos_hg}.}
\label{hg_ixs}
\end{figure}

A more recent INS study contributed to shed some light on the
possible origin of the quasielastic spectral components
\cite{bad_hg}. The experiment has been performed on the TOF
spectrometer MARI, optimized to access a restricted kinematic
region ($-6<\omega<6$ meV at the lowest accessed momentum transfer
$Q\approx 4$ nm$^{-1}$) with an energy resolution sufficient to
study the diffusive dynamics ($\delta E=0.4$ and $\delta=0.8$ meV
at the two incident energy utilized). The narrower incoherent
contribution (self diffusion), resolution limited in the
experiment of Bove \textit{et al.}, was now detected with a
procedure similar to the one applied in liquid potassium
\cite{cab_k}, i.e. fitting the data with two lorenzian in a
restricted wavevector region where the two quasielastic features
are well separated, determining the analytical $Q$ dependence of
the FWHM of the diffusive term, and finally focusing on the
broader component over the whole momentum transfer region keeping
all the parameters of the diffusive term fixed. With this
procedure, the narrower lorenzian is confirmed to be incoherent in
nature, and well described by Fick's law properly modified
according to the revised Enskog's theory \cite{kag_hs}. As far as
the broader component is concerned, the authors point out that
more than one lorenzian is needed to describe it at increasing
wavevector, then they discuss its possible origin. First, they
point out that the thermal origin of this broad component is ruled
out by its low $Q$ intensity, which is by far larger than the one
expected by the Landau Plazeck ratio. Moreover, the linewidth
reaches a constant value on decreasing wavevector, instead of
following the $Q^2$ dependency of the heat mode. Second, they
examine the possibility of a cage diffusion mechanism, as proposed
in MD simulations \cite{bov_hgsim}. In this respect, they point
out how the experimentally observed mode intensity is too large
than expected, but they propose an enhancement mechanism based on
valence fluctuations which could be active at low wavevectors
amplifying the expected intensity.

Very recently, new state-of-the-art IXS experiments have been
reported in expanded mercury near the critical point ($T_c=1751$
K, $P_c=1673$ bars and $\rho _c=5.8$ g cm$^{-3}$), aiming at the
investigation of collective dynamics at the metal-non metal
transition \cite{inui_sn}. Despite extremely difficult
experimental conditions, the speed of sound has been accurately
measured and no significant changes have been observed in the
transition from the metallic ($m\rho _c=13.6$ g cm$^{-3}$) to the
insulating ($m\rho _c=9.0$ g cm$^{-3}$) phase, while the
ultrasonic sound velocity exhibits a significant drop across the
same thermodynamic point \cite{yao_sn}. Only upon further
expansion in the insulating phase the high frequency sound
velocity ultimately drops reaching the adiabatic value. This
result seems to indicate the presence of very large positive
dispersion as peculiar of the metal-non metal transition, opening
a new experimental route to the investigation of the interplay
between acoustic and transport properties.

Summing up, though the microscopic dynamics in molten Hg has been
the subject of deep investigations in the last few years, some
aspects still remain controversial. The sound velocity as
determined by INS and IXS are not fully consistent with each
other, though both techniques show positive dispersion effect the
INS result show a very large effect never observed sofar. On the
one side it has been emphasized the role of electronic properties
\cite{bov_hg}, while, on the other side, the collective dynamics
as determined by IXS closely resemble the one of several other
simple fluids \cite{hos_hg}. The most intriguing aspect concerns,
however, the interpretation of the quasielastic component of
$S(Q,\omega)$. Neutron scattering data suggest a negligible effect
of thermal fluctuations \cite{bad_hg}, adding a piece of
information to a recently debated issue
\cite{scop_comm,sing_pre,sing_rep}. On the other side, the IXS
data unambiguously show the presence of a coherent quasielastic
dynamics, which no doubt has to show up in neutron experiment as
well. The broad quasielastic component as observed with two
different experiments show however opposite $Q$ trends,
monotonically increasing \cite{bov_hg} and decreasing
\cite{bad_hg}, respectively. Badyal \textit{et al.} suggest a cage
diffusion process, enabled by valence fluctuations mechanism. We
believe that the broad mode observed in INS could be closely
related to the coherent quasielastic scattering reported in the
IXS data. In this case, the similarity with several other
investigated systems would suggest an interpretation in terms of a
high frequency structural relaxation process
\cite{scop_prlli,scop_preal,scop_prena,scop_prlga,mon_k}. Very
recent IXS investigations, however, suggest an enhancement of
positive dispersion at the metal non metal transition, pointing
out how changes in electronic transport properties dramatically
affect acoustic properties \cite{inui_sn}.

\subsubsection{Nickel}

Early investigations in liquid Nickel have been reported in 1977
with TOF technique \cite{joh_ni}. Two different isotopic
concentrations, one with the natural abundance ratio and the other
a wholly incoherent mixture, were chosen in order to study
separately the coherent and incoherent scattering contributions.
These latter, in turns, have been related each other through the
Vineyard approximation \cite{vine_convo}. The spectral density
$g(\omega)$ was then extracted from the low $Q$ limit of the self
dynamic law. The coherent $S(Q,\omega=0)$ were reported for
wavevectors above $Q=22$ nm$^{-1}$, therefore beyond the first
brillouin pseudozone. Consequently, this study was not able to
ascertain the existence of low $Q$ collective modes.

A neutron scattering experiment has been recently performed at the
IN1 facility of the ILL \cite{ber_ni}. Constant $Q$ spectra have
been collected from $Q=8$ nm$^{-1}$ all the way up to wavevectors
as high as twice the main peak in the static structure factor. The
data have been analyzed utilizing two lorentian terms for the
quasielastic (coherent and incoherent) contribution and one DHO
function for the purely inelastic spectral features. Evidence of
collective modes has been reported in the whole investigated Q
domain, despite the relatively high viscosity value ($\eta_s
\approx 5.7$ mPa s, i.e. one order of magnitude larger than the
typical values for alkali metals). The reason for this apparent
oddness can be traced back to arguments similar to the one
applying to the case of gallium, i.e. to the freezing of the
diffusional motion over the probed high frequency regime, which
reduces the effective Brillouin damping in respect to the
hydrodynamic prediction of Eq. (\ref{gidro}).

\begin{figure} [h]
\centering
\includegraphics[width=.7\textwidth]{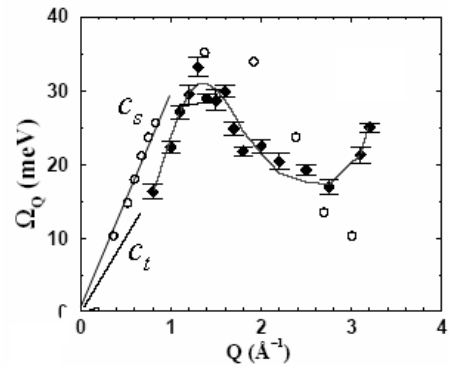}
\vspace{-12.cm} \caption[F.J. Bermejo half of Fig. 2 PRL 85, 106
(2000)]{Sound velocity in liquid Nickel determined by INS (full
circles) and MD (open circles). The isothermal ($c_t$) and
adiabatic ($c_s$) values are also indicated.} \label{ni_disp}
\end{figure}

More interestingly, the low $Q$ limit of the sound velocity seems
to approach the isothermal and not the adiabatic value, as shown
in Fig. \ref{ni_disp}. It is worth to stress how, given the large
value of $\gamma$ (and therefore the large differences between
$c_t$ and $c_s$), this observation strikingly holds beyond the
reported experimental error. If confirmed, this result would
substantiate the hypothesis of an intermediate isothermal domain
bridging the hydrodynamic limit and the high frequency regime,
which has been recently matter of debate
\cite{sing_pre,scop_comm,sing_rep}. On the other side, previous
molecular dynamics simulations indicate higher values of the sound
velocity which agree quite well with the adiabatic response
\cite{alem_sim}. Summing up, liquid Ni seems to be an ideal system
to test the evolution of the thermal relaxation once the
hydrodynamic limit is abandoned. An IXS investigation would be
helpful to clarify this issue, though the small absorption length,
the high melting temperature and the reactivity of Ni poses severe
experimental challenges.

\subsubsection{Copper}

Time of flight neutron scattering data have been reported for this
system long time ago in solid and liquid phase. The accessed
kinematic range was quite broad ($Q>10nm^{-1}$ and $E<30$ meV) and
the data have been analyzed within pioneering models
\cite{egel_model}. More recently, an IXS experiment has been
performed at the ESRF, though this work is still in progress a
very preliminary estimate of the sound velocity gives a value of
$4230 \pm 70$m/s, well above the hydrodynamic value.

\subsection{Solutions of metals}

Alkali metals easily dissolve in water, molten alkali halides and
ammonia, resulting in a free electron and a positively charged
ion. Given the relatively low electronic concentration of the
saturated solutions, these are ideal systems to challenge the
validity extents of plasma-based theories introduced in section
\ref{plasma}, and the relative approximations for the dielectric
function. Models like the RPA, indeed, are expected to hold in
systems with low $r_s$ (or, equivalently, high electronic density)
such as bulk metals, while they reach their limits on increasing
the $r_s$ value.

A second interesting aspect concerns the presence in liquid metals
of the so called Kohn anomaly, i.e. a kink in the dispersion curve
occurring at $Q=2k_F$ and reflecting a singularity of the
dielectric function which is observed in metallic crystals, but
which is not yet established in the disordered phase. The high
frequency dynamics of metal-ammonia systems, in particular, have
been recently investigated by means of X-rays and neutrons. High
resolution IXS performed in lithium-ammonia solutions allowed to
detect high frequency excitations, softening at twice the Fermi
wavevector $k_f$ \cite{burns_prl1}. Unfortunately, at the
investigated concentrations $2k_F$ is close to $Q=Q_m$, i.e. the
main peak of the static structure factor. This coincidence
generates an ambiguity in the interpretation of the observed dip
in the sound dispersion, as it is not clear whether this feature
is related to the structural periodicity as in non ordinary fluids
or has to do with a purely electronic effect. The reported sound
velocity values are intermediate between the one of the bare ions
and the one appropriate for the Li(NH$_3$)$_4$ vibrating network,
though close to this latter. In a later study \cite{burns_prb},
collective excitations are rationalized in terms of ionic plasma
oscillations, and the sound velocity values measured at several
metal concentrations are compared with the prediction of the
Bohm-Staver expression of Eq.(\ref{BS}), taking as relevant mass
either the bare ionic value or the metal-ammonia unit. In both
cases the predictions do not agree with the measured values, and
this seems to be a signature of the well known failure of the RPA
approximation in the low-density regime, where the
electron-electron interactions are relevant. A contemporary INS
study on deuterated ammonia \cite{sacc_ammonia} also addresses the
deviations from the BS model. In this case an improvement is
achieved accounting for both the finite ionic size and the
electron-electron interactions. As far as the first aspect is
concerned, an alternative renormalization of the free ionic plasma
frequency is undertaken, while an expression for the dielectric
screening function, going beyond the RPA approximation of
Eq.(\ref{TF}) and accounting for local fields effects, is
proposed. As far as the Kohn anomaly is concerned, this study
suggests that the value of $2k_f$ ($\approx 10$ nm$^{-1}$ in
saturated lithium-ammonia solutions) has to be compared with
$Q\approx 20$nm$^{-1}$, which is actually the second peak of
$S(Q)$, related to the $N-N$ periodicity, rather than with the
first peak ($Q_m \approx 20$nm$^{-1}$) related to the
Li(NH$_3$)$_4$ periodicity. This might arise from the different X
ray and neutrons cross sections: while in the first case the two
peaks have similar intensities, the neutron diffraction strongly
enhance the $N-N$ peak. A later study has shown how an unambiguous
separation between $2K_F$ and $Q_m$ occurs in low density
solutions, although in this case the kink observed in the
dispersion curve is almost within the error bars
\cite{giura_liamm}. Further studies, therefore, seem to be
necessary to draw a conclusive picture about the presence of the
Kohn anomaly in metallic fluids.

\begin{figure} [h]
\centering
\includegraphics[width=.5\textwidth]{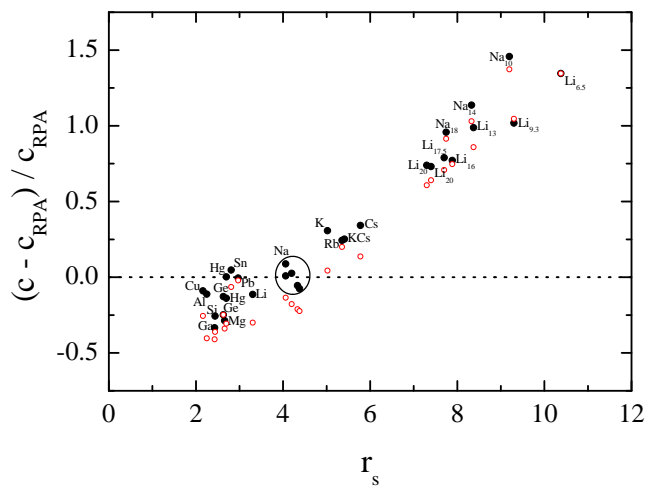}
\vspace{-6.cm} \caption[mia]{Random phase approximation at work:
data taken from Table \ref{table} and \cite{bov_ga} for pure
elements and from \cite{giura_liamm,burns_prb} for alkali-ammonia
solutions, with the BS values estimated through Eq. (\ref{BS}).
Full circles: the measured sound velocity, $c$, is the maximum
high frequency value determined over the $Q$ range probed by
inelastic scattering techniques, i.e. includes the positive
dispersion effect. Open circles: $c$ is here the adiabatic value.
The subscript of Li and Na indicates the concentrations in ammonia
solutions. The Na group is relative to several different
temperatures.} \label{rpa}
\end{figure}

The validity extent of the RPA approximation for the determination
of the sound velocity via the Bohm-Staver expression (\ref{BS}) is
depicted in fig. \ref{rpa}, in which we report the relative
deviations of the BS calculated values ($c_{RPA}$) from the
experimental ones, for systems of different $r_s$, ranging from
pure metals to alkali-ammonia solutions. In the latter case
($r_s>6$) the BS predictions underestimate more than $50 \%$ the
calculated values. The deviations, however, show a trend which
monotonically decreases towards low $r_s$ elements and finally get
negative for $r_s$ values close to 2.

A final remark concerns the choice of the dynamical variable to
calculate the sound velocity when one is looking at subtle effects
as in the present case. First, according to the hydrodynamic
definition of sound velocity in liquids, one should look at the
maxima of the current correlation spectra. While the difference
with the maxima of $S(Q,\omega)$ is usually negligible in
crystals, indeed, there can be a significant discrepancy in
strongly overdamped cases such as the one metallic solutions. In
this respect, the choice of the DHO to describe $S(Q,\omega)$
implicitly overcame this problem, as the characteristic frequency
of this model coincides indeed with the maxima of $C_L(Q,\omega)$.
Second, in all the reported studies, the "phonon" velocity is
extracted through ad hoc models (DHO, extended hydrodynamic model
etc.) looking only at the genuine inelastic component. Again, in
liquids, the full $C_L(Q,\omega)$ spectrum should be considered,
as when $S(Q,\omega)$ has a significant quasielastic contribution
this latter can affect the peak positions of $C_L(Q,\omega)$ (see,
for instance, Fig. (\ref{aldisp}) for the aluminum case).

Inelastic X-ray Scattering, with lower energy resolution (a few
hundreds of $meV$) and in a broader energy range (up to a few eV),
allows to study electronic excitations (plasmons). In this case
the plasma oscillation is brought about by free electrons, while
the background is constituted by the ionic network. Some recent
studies \cite{burns_prl2,burns_prl3} point out a decrease of the
plasmon dispersion at low metal concentrations, which, in turn, is
ascribed to the failure of the RPA approximation.

\section{Summary and perspectives \label{sec_sum}}

In this section we will try to summarize the scenario arising from
the measurements reported so far. In respect to the collective
properties, it seems useful to discuss the different dynamical
regimes probed at different wavevectors and frequencies. Although
the two domains are in principle independent (as testified, for
instance, by the two separate generalization of the classical
hydrodynamics introduced in sec. \ref{sec_collnonhydro}), the
existence of a dispersion relation ultimately allows one to think
in terms of wavelenghts only. In our point of view, though precise
boundaries can not be traced, one can identify in liquid metals
the following dynamical regimes:

\begin{figure} [t]
\centering
\includegraphics[width=.5\textwidth]{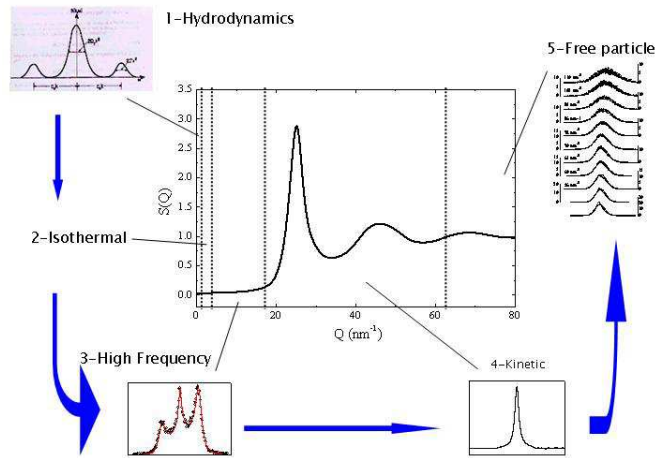}
\vspace{-5cm} \caption[cosanostra]{Sketch of the different
dynamical regimes on decreasing the wavelength.} \label{forse}
\end{figure}

\begin{itemize}
\item The hydrodynamic, $Q\rightarrow 0$ limit, that, in liquid
metals, basically means $Q\lesssim 0.1$ nm$^{-1}$. In this region
simple hydrodynamic treatment based on three microscopic dynamical
variables (density, current, energy) provides an exhaustive
description of the main features. Although not accessible by
neutron and X-ray spectroscopic techniques, the hydrodynamic
predictions are in very good agreement with acoustic measurements
of sound velocity and attenuation properties. Moreover, the strict
analogy of long wavelength fluctuations in conductive and ordinary
liquids (accessible via light scattering) corroborates the
validity of the simple hydrodynamic theory. This region is
characterized by \textit{adiabatic} sound propagation, and the
whole dynamical features are ruled by macroscopic transport
parameters (viscosity, thermal diffusivity, specific heats). In
this regime, other approaches tailored for conductive fluids such
as plasma oscillation theories provides alternative descriptions,
which turns out to be increasingly accurate for low $r_s$ systems.

\item An (hypothetic) isothermal region, which should be
observable in the $0.2 \lesssim Q \lesssim 3$ nm$^{-1}$ momentum
range. Upon increasing the wavevector outside the strict
hydrodynamic limit, indeed, the lifetime of the entropy
fluctuations becomes increasingly shorter ($\tau=(\gamma
D_TQ^2)^{-1}$). Since the frequency of the corresponding density
fluctuation increases almost linearly ($\omega\approx cQ$), one
should expect a transition at $Q^*\approx c/D_T$. Given the
typical thermal diffusivity values of liquid metals this crossover
should be located at a few fractions of nm$^{-1}$. Since with both
INS and IXS one normally access momentum transfers above $1$
nm$^{-1}$, this region can be hardly explored, and no direct and
convincing indication for its existence are available at present.
Some old INS data on liquid lead \cite{FABER} as well as a more
recent experiment on liquid nickel \cite{ber_ni}, however, seems
to suggest indication for such existence. The case of Nickel, in
particular, is an interesting one and would deserve new
investigations since in this system the specific heat ratio is
particularly high and therefore the difference in sound velocities
between an ordinary adiabatic regime and an isothermal one would
be of the order of $40\%$. It is worth to stress that the
prediction for the existence of an isothermal regime poses on: i)
a negligible $Q$ dependence of the thermodynamic quantities and
ii) the validity of a one component effective description in which
the effective thermal diffusivity is well described by the sum of
the electronic and ionic contributions. Both this points have been
recently discussed in the analysis of alkali metals and liquid
aluminium IXS spectra \cite{scop_comm,sing_pre,sing_rep}.

\item A generalized hydrodynamic region, probed above $Q\approx 3$
nm$^{-1}$, in which the frequency-wavevector dependence of the
transport properties heavily affects the sound mode. The upper
limit of validity of this region is rather system dependent,
normally higher for alkali metals (say up to $0.7 Q_m$). The
natural framework is here the memory function formalism, which can
be developed at different levels of accuracy, ranging from the
celebrated Lovesey's model \cite{lov_visco} (accounting for a
single average relaxation time for the second order memory
function) up to more refined memory function models based on
multiple relaxation phenomena which, firstly introduced for
Lennard-Jones systems \cite{lev_2t}, have been more recently
adapted and tested against IXS investigations of liquid metals
\cite{scop_prlga,scop_prlli,mon_k,scop_jpc} and numerical
simulations of model undercooled and glassy alkali
\cite{scop_presim}.

In this respect, it is worth to point out how the high points of a
memory function approach are not solely in a better agreement with
experimental data, which, in general, heavily depends on the
quality of the experimental data and, of course, on the number of
the model parameters \cite{scop_prlli}. In most cases, indeed,
simplified phenomenological models such as the damped harmonic
oscillator and a lorenzian function for the inelastic and
quasielastic feature, respectively, provide satisfactory
agreements.

On the contrary, the memory function framework allows to grasp an
insight behind the mechanisms ruling the relaxation dynamics,
extracting relevant information about quantities which are not
directly related to any spectral features, such as the $Q$
dependencies of the relaxation time(s) and of the longitudinal
viscosity.

\begin{figure} [h]
\centering
\includegraphics[width=.45\textwidth]{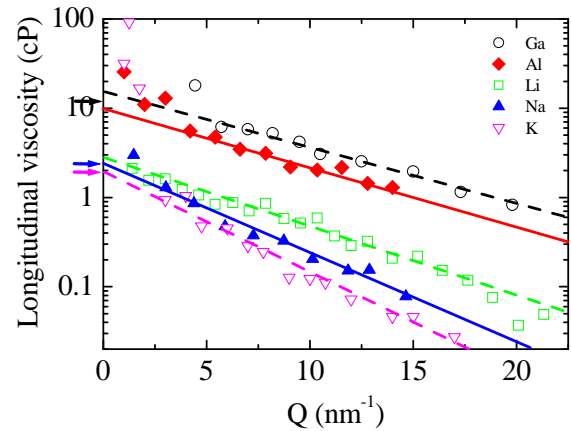}
\vspace{-3cm} \caption[cosanostra]{Generalized longitudinal
viscosities for a collection of liquid metals, extracted by the
experimental data throughout a memory function approach. The long
wavelenght limit, determined by ultrasonic measurements, is also
reported for K, Na, Ga (bottom to top).} \label{allvisco}
\end{figure}

Following the prescriptions illustrated in section \ref{sec_mf},
indeed, the longitudinal viscosity is related to the total area
under the viscous contribution to the memory function. In
Fig.(\ref{allvisco}), for instance, we report the (generalized)
longitudinal viscosities extracted in this way for several
investigated systems. As can be seen, the low $Q$ extrapolation of
the experimental values compares well with independent acoustic
determinations (when available), but also allows to determined the
$Q$ generalization of this important transport properties, which
can be directly estimated only by numerical simulations
approaches.

In this $Q$ region, hard sphere-based theories provide alternative
descriptions in terms of extended hydrodynamic models
\cite{kag_hs} but they still miss a convincing explanation of one
of the most important points: the physical interpretation beyond
the relaxation of the sound mode, which is now a firmly
established evidence supported by uncountable experimental
investigations.

\begin{figure} [h]
\centering
\includegraphics[width=.5\textwidth]{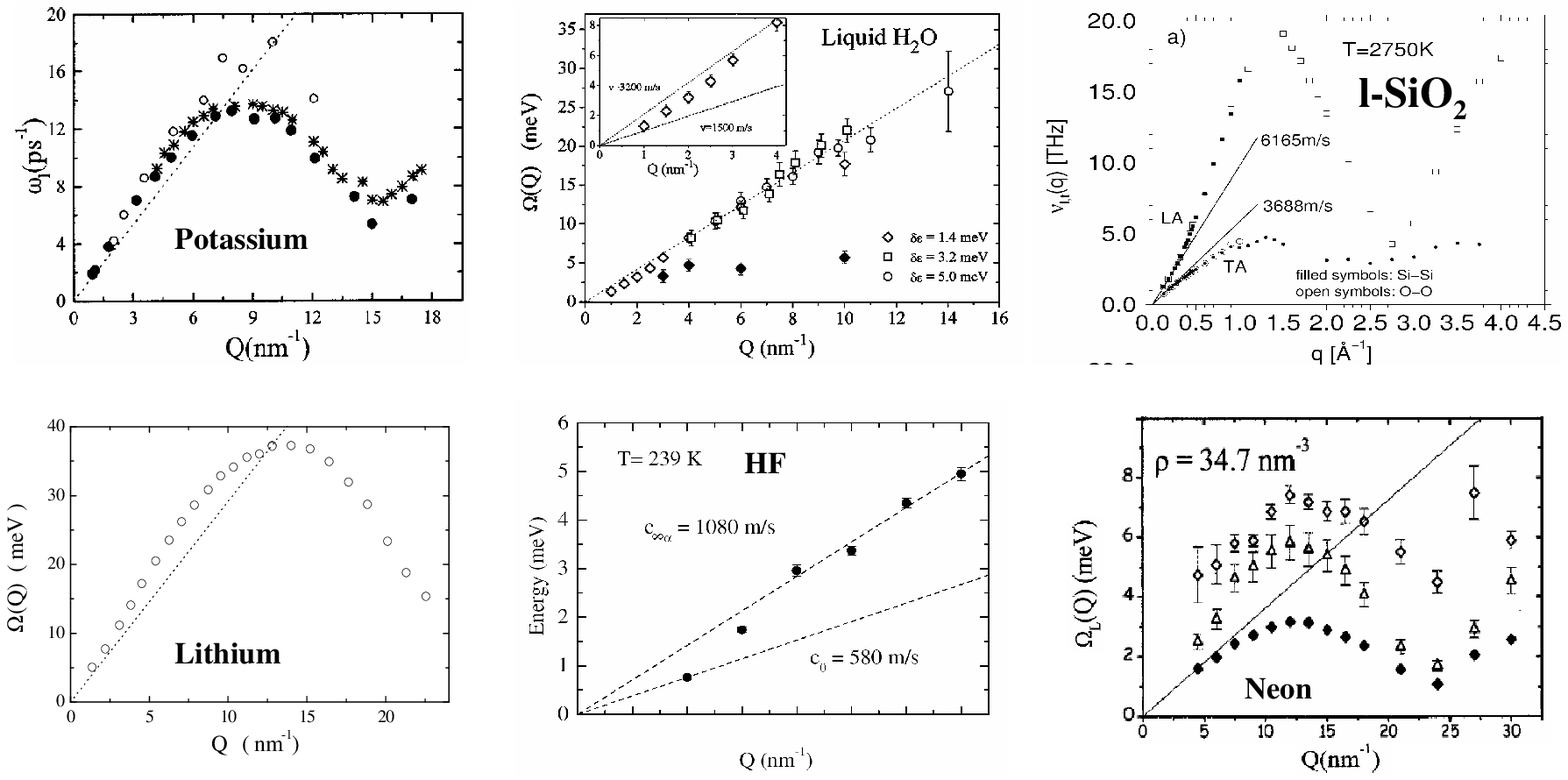}
\vspace{-8cm} \caption[cosanostra]{Sound dispersion in several
kind of liquids. Alkali metals (Li \cite{scop_epl} and K
\cite{mon_k}, dashed line is the isothermal sound velocity);
Hydrogen bonding systems (Water \cite{monaco_water} and hydrogen
Fluoride \cite{ange_HF}) low and high frequency sound velocities
are also indicated by the dotted lines); liquid neon
\cite{cun_ar}: adiabatic (circles and line) apparent (open
triangles) and high frequency (open diamonds) sound velocities;
liquid silica, molecular dynamics simulations
\cite{kob_si02posdisp}, in this case also the transverse branch is
reported.} \label{pdisp1}
\end{figure}

This latter aspect, which usually emerges in terms of a speed up
of the sound velocity taking place in the $1\div 10$ nm$^{-1}$
region, is one of the most interesting aspects which is still
lacking an explanation. Broadly speaking, it is always referred to
as a shear relaxation, but the ultimate nature of the involved
physical processes still have to be clarified.

Actually, Mode Coupling Theory \cite{ern_pdis} provides a
description of the acoustic dispersion curve in terms of even
powers of $Q$ \cite{ern_pdis}, but its interpretation is
restricted to monoatomic systems in the liquid phase, while the
observed phenomenology seems to be a more general feature of the
disordered systems.

Interestingly, indeed, a qualitatively similar phenomenon (see
Fig. \ref{pdisp1}) has also been reported numerically in fused
silica \cite{kob_si02posdisp}) and experimentally in Lennard jones
fluids \cite{cun_ar} and hydrogen bonding systems
(\cite{monaco_water,ange_HF}). While in this latter case the
positive dispersion has been shown to be an activated process,
related to the structural relaxation, a different scenario seems
to characterize the other systems. Interestingly, indeed, (see
Fig. \ref{pdisp2}) the same behavior of the sound velocity also
appear in \textit{glasses} either in experimental IXS measurements
in g-Se and vitreous silica \cite{scop_se,ruz_sio2}, in numerical
works on model glasses of Lennard Jones systems \cite{gcr_prlsim},
vitreous silica \cite{kob_si02posdisp} and alkali metals
\cite{scop_presim} and, finally, in theoretical calculations for
an hard sphere glass \cite{got_hs}. In the case of alkali metals,
in particular, it has been shown how the speed up of the sound
velocity persists upon cooling well below the glass transition,
thus ruling out the possible role of the structural relaxation in
this effect. The presence of positive dispersion at THz
frequencies in glasses, quantitatively comparable to the one
observed in liquids, therefore, challenges the interpretation of
the positive dispersion in terms of a transition from a liquid
like to a solid like behavior, an effect which seems to be
quantitatively negligible (with the remarkable exception of
hydrogen bonding systems). Accepting a description of the
collective dynamics proceedings over two distinct viscous
relaxations, therefore, the ultimate responsible for the bending
up of the dispersion curve seems to be the faster of the two
observed processes, which turns out to have a mild temperature
dependence. In this respect, the physical nature of this faster
process calls for deeper understanding. It is worth to point out
however, that in Lennard Jones systems the positive dispersion is
recovered within an \textit{harmonic} description of the dynamic
structure factor in terms of eigenvalues and eigenvectors, a
result which relate the positive dispersion to the properties of
the dynamical matrix and, ultimately, to the topological disorder
of the inherent equilibrium position, being the interaction
potentials comparable in glasses and crystals.

\begin{figure} [h]
\centering
\includegraphics[width=.5\textwidth]{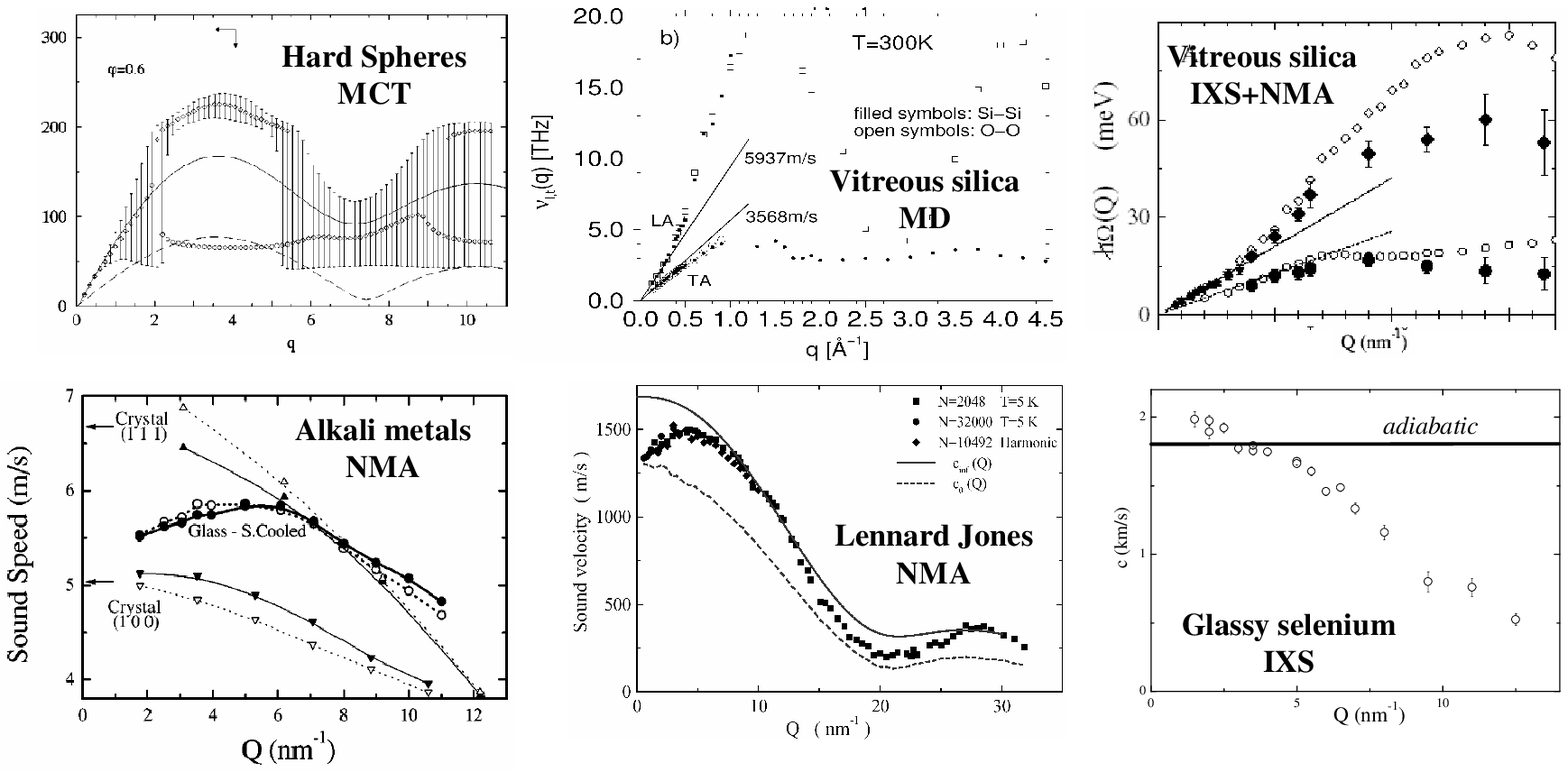}
\vspace{-8cm} \caption[cosanostra]{Sound velocities in several
glassy systems. Hard spheres glass, exact solution within MCT
\cite{got_hs}; Metallic glass obtained quenching a model system
interacting via Price-Tosi pseudopotential \cite{scop_presim}, the
crystalline counterpart is also reported; Lennard Jones glass
obtained in a similar way \cite{gcr_prlsim}, reported with the low
and high frequency sound velocities; SiO2, MD as in
Fig.\ref{pdisp1} but in the glassy state \cite{kob_si02posdisp}
and experimentally determined by means of IXS scattering
\cite{ruz_sio2}. In both cases the transverse branch is also
reported; glassy selenium \cite{scop_se}, again the sound velocity
exceeds the adiabatic value.} \label{pdisp2}
\end{figure}

\item A kinetic regime, valid from around $Q_m$, up to a few
oscillations of the structure factor. Here the hard sphere
description provides remarkably accurate predictions, in terms of
a quasielastic "extended heat mode" whose linewidth is described
in terms of the Enskog's diffusion coefficient and of an
equivalent hard sphere diameter. The extent of validity of kinetic
theory in this momentum range has been widely tested
\cite{kag_hs,coh_hs,desh_hs,coh_hs1} in several hard sphere-like
systems (like alkali metals and lennard-jones fluids), and it also
apply in colloidal systems. It would be interesting to challenge
such theory in less simple liquid metals. Despite the success of
hard sphere model, there is still an obscure point concerning the
real origin of such extended heat mode: while in the low $Q$ limit
it tends to the entropy fluctuation mode, indeed, at finite $Q$'s
it certainly involves mass diffusion processes. Looking at things
from the constant energy point of view, umklapp modes resembling
crystalline phonons in Brillouin zones higher than the first seems
to be still poorly understood
\cite{scop_prbumk,rand_umk,coc_umk,dor_umk}.

\item An high $Q$ region, probed as soon as the Van Hove distinct
function vanishes. In this limit, both IXS and INS experiments
provide the same information, about the atomic motion over
timescales shorter than the interparticle collision time. An
almost exact description for this regime is available, which can
also account for quantum aspects such as recoil energy an quantum
corrections to the spectral moments.

\end{itemize}

The single particle dynamics seems to be better understood if
compared to the collective motion, at least by a coarse grained
point of view of an hydrodynamic diffusive mode with finite $Q$
corrections, evolving towards a ballistic regime. This is probably
due to the intrinsic difficulty of isolating a (wide enough)
constant $Q$ \textit{coherent} energy spectrum from an INS
experiment. Since the study of the strictly coherent spectrum
became possible only in the last decade, indeed, a lot of efforts
have been devoted in the past to the single particle case. Though
all the approaches described in this review describe, on average,
equally well the available experimental data, a memory function
formalism paralleling the one for the collective case could
provide the route to relate the single particle motion and the
collective dynamics in the microscopic regime.

Raising the level of detail of the description of the single
particle dynamics, however, a major experimental challenge seems
to be the identification of the different processes giving rise to
the quasielastic incoherent scattering. Recent INS results,
indeed, suggest the presence of two distinct physical mechanisms,
active over different timescales, underlying the diffusive motion
\cite{bov_hg,bov_k,bad_hg}. The combined presence of coherent and
incoherent scattering, however, makes such identification still
unclear although, in principle, the IXS signal might be used to
subtract the coherent contribution from the INS spectra, thus
extracting the purely incoherent dynamics. In this respect, the
synergy of combined IXS and INS studies on a same sample seems us
imperative and could help to shed light on this point. The IXS
signal indeed, might be used to subtract the coherent contribution
from the INS spectra, thus extracting the purely incoherent
dynamics.

\section{Acknowledgements}

J.-P. Hansen made extensive comments on this manuscript for which
we are most grateful. We thank L.E. Bove and T. Bryk for several
discussions and interesting comments on the preprint, and S.
Cazzato for his help in compiling the data reported in table I.
T.S. gratefully acknowledges his debt to U. Balucani for the
vivifying influence exerted on his outset in the field of simple
liquids. Most of our IXS activity greatly benefited from the
support of the staff of the beamlines ID16-ID28 at the ESRF and
BL35XU at SPring8. The assistance of the technical staffs of the
ESRF (D. Gambetti, B. Gorges and C. Henriquet), of the University
of L'Aquila (O. Consorte) and of the University of Rome "La
Sapienza" (I. Deen, M. Pallagrossi, C. Piacenti and A. Salvati) is
also acknowledged. Last but not least, thanks are due to all
authors, editors, and publishers who granted us permission to
include in this review previously published illustrations, images,
and figures.

\newpage
\begin{table*}[h]
\centering
\renewcommand{\arraystretch}{0.5}
\begin{tabular}{|c|c|c|c|c|c|c|c|} \hline\hline Sample&$T [K]$&$\gamma$&$c_s [m/s]$&$c_t [m/s]$&$max\{c_{l}\} [m/s]$&$\sigma_{inc}/\sigma_{coh}$&$D_T$ $[nm^2/ps]$\\ \hline
$\mathrm{Li}$&453&1.08\footnotemark[3]$^,$\footnotemark[4]1.065\footnotemark[15]&4554\footnotemark[15]&4466\footnotemark[31]&5762\footnotemark[31]&&19.1\footnotemark[5]\\
&488&&&&5423\footnotemark[1]$^,$\footnotemark[2]&&\\
&500&&&&&&20.3\footnotemark[5]$^,$\footnotemark[6]\\
&600&1.092\footnotemark[15]&&4356\footnotemark[31]&5560\footnotemark[31]&&\\
&&&&&&0.99\footnotemark[6]1.1\footnotemark[2]& \\
\hline
$\mathrm{Na}$&371&1.12\footnotemark[11]1.091\footnotemark[15]&2531\footnotemark[15]&&&&68.8\footnotemark[5]\\
&388&1.11\footnotemark[3]$^,$\footnotemark[4]&2514\footnotemark[7]&&3160\footnotemark[7]&&\\
&390&&&&2930\footnotemark[8]&&\\
&500&&&&&&68.4\footnotemark[5]\\
&773&&2310\footnotemark[7]&&2881\footnotemark[7]&&\\
&1073&&2150\footnotemark[7]&&2577\footnotemark[7]&&\\
&1173&&2093\footnotemark[7]&&2492\footnotemark[7]&&\\
&&&&&&0.84\footnotemark[9]1.006\footnotemark[10]0.976\footnotemark[7]&\\
\hline
$\mathrm{Mg}$&923&1.29\footnotemark[3]&4070\footnotemark[3]&&&&37\footnotemark[5]\\
&973&&4038\footnotemark[32]&&4380\footnotemark[32]&&\\
&1000&&&&&&39.8\footnotemark[5]\\
&&&&&&0.06\footnotemark[6]&\\
\hline
$\mathrm{Al}$&933&1.4\footnotemark[3]&4750\footnotemark[3]&&&&35.2\footnotemark[5]\\
&1000&&&4670\footnotemark[12]&7075\footnotemark[12]&&36.4\footnotemark[5]\\
\hline
$\mathrm{Si}$&1683&1.57\footnotemark[3]&3977\footnotemark[3]&&&&9.4\footnotemark[5]$^,$\footnotemark[14]\\
&1753&&3952\footnotemark[13]&&4597\footnotemark[13]&&\\
&&&&&&0.05\footnotemark[6]&\\
\hline
$\mathrm{K}$&336.7&1.11\footnotemark[11]1.102\footnotemark[15]&1880\footnotemark[15]&&&&81.4\footnotemark[5]\\
&343&1.105\footnotemark[15]&1877\footnotemark[15]&1605\footnotemark[17]1710\footnotemark[16]&2352\footnotemark[16]2260\footnotemark[17]&&\\
&350&&&&2360\footnotemark[18]&&\\
&&&&&&0.20\footnotemark[6]0.16\footnotemark[17]&\\
\hline
$\mathrm{Fe}$&1808&1.8\footnotemark[3]&4000$\div$4400\footnotemark[3]&&&&7.3\footnotemark[5]\\
\hline
$\mathrm{Co}$&1700&&&&&&7.8\footnotemark[5]$^,$\footnotemark[14]\\
&1765&1.8\footnotemark[3]&4033$\div$4090\footnotemark[3]&&&&\\
\hline
$\mathrm{Ni}$&1500&&&&&&16\footnotemark[5]$^,$\footnotemark[14]\\
&1728&1.98\footnotemark[3]&4036$\div 4045$\footnotemark[3]&&&&9.6\footnotemark[3]\\
&1763&1.88\footnotemark[19]&4280\footnotemark[19]&3121\footnotemark[19]&3855\footnotemark[19]&&\\
&&&&&&0.35\footnotemark[6]0.30\footnotemark[19]&\\
\hline
$\mathrm{Cu}$&1356&1.33\footnotemark[3]&3440$\div$3485\footnotemark[3]&&4230\footnotemark[33]&&42.1\footnotemark[5]\\
&&&&&&0.06\footnotemark[6]&\\
\hline
$\mathrm{Zn}$&693&1.25\footnotemark[11]1.26\footnotemark[3]&2835$\div$2850\footnotemark[3]&&&&15.7\footnotemark[5]\\
\hline
$\mathrm{Ga}$&303&1.08\footnotemark[11]&&&&&11.6\footnotemark[5]\\
&315&&2930\footnotemark[20]&2600\footnotemark[22]&3050\footnotemark[22]&&\\
&326&1.08\footnotemark[21]&&&3240\footnotemark[21]&&\\
&350&&&&&&13.6\footnotemark[5]\\
&&&&&&0.07\footnotemark[21]0.02\footnotemark[6]&\\
\hline
$\mathrm{Ge}$&1253&1.18\footnotemark[3]$^,$\footnotemark[23]&2682\footnotemark[23]&&2682\footnotemark[23]&&\\
&1063&&&&&&8$\div$9\footnotemark[5]\\
&&&&&&0.006\footnotemark[6]&\\
\hline
$\mathrm{Rb}$&312&1.15\footnotemark[11]1.097\footnotemark[15]&1260\footnotemark[15]&&&&61.5\footnotemark[5]\\
&320&&1370\footnotemark[24]&&1420\footnotemark[24]$^,$\footnotemark[1]&&\\
&&&&&&0.00055\footnotemark[6]&\\
\hline
$\mathrm{Ag}$&1233&1.32\footnotemark[3]&2710$\div$2770\footnotemark[3]&&&&66.5\footnotemark[5]\\
&&&&&&0.125\footnotemark[6]&\\
\hline
$\mathrm{Cd}$&594&1.25\footnotemark[3]1.25\footnotemark[11]&2235$\div$2255\footnotemark[3]&&&&39.8\footnotemark[5]$^,$\footnotemark[14]\\
&&&&&&2.3\footnotemark[6]&\\
\hline
$\mathrm{Sn}$&505&1.11\footnotemark[11]&&&&&17.3\footnotemark[5]\\
&593&1.09\footnotemark[3]$^,$\footnotemark[25]$^,$\footnotemark[4]&2443\footnotemark[25]&&2736\footnotemark[25]&&\\
&1273&&2228\footnotemark[25]&&2362\footnotemark[25]&&\\
&&&&&&0.01\footnotemark[25]0.007\footnotemark[6]&\\
\hline
$\mathrm{Sb}$&904&1.21\footnotemark[3]&1893$\div$1900\footnotemark[3]&&&&15.5\footnotemark[5]\\
&&&&&&0.046\footnotemark[6]& \\
\hline
$\mathrm{Te}$&723&1.033\footnotemark[3]&889\footnotemark[3]&&&&0.8$\div$1.3\footnotemark[5]\\
&&&&&&0.05\footnotemark[6]&\\
\hline
\end{tabular}
\end{table*}
\newpage
\begin{table*}[h]
\centering
\renewcommand{\arraystretch}{0.5}
\begin{tabular}{|c|c|c|c|c|c|c|c|} \hline\hline Sample&$T [K]$&$\gamma$&$c_s [m/s]$&$c_t [m/s]$&$max\{c_{l}\} [m/s]$&$\sigma_{inc}/\sigma_{coh}$&$D_T nm^2/ps$\\ \hline
$\mathrm{Cs}$&302&&967\footnotemark[15]&&&&44.6\footnotemark[5]\\
&308&1.102\footnotemark[26]1.099\footnotemark[15]&965\footnotemark[26]&&1140\footnotemark[26]&&\\
&&&&&&0.0596\footnotemark[26]&\\
\hline
$\mathrm{Au}$&1336&1.28\footnotemark[3]&2560\footnotemark[3]&&&&40.4\\
&&&&&&0.06\footnotemark[6]&\\
\hline
$\mathrm{Hg}$&234&&&&&&3.62\footnotemark[5]\\
&293&1.14\footnotemark[27]&1451\footnotemark[28]&&2100\footnotemark[15]1800\footnotemark[29]&&4.41\footnotemark[5]\\
&300&&&&&&4.41\footnotemark[5]\\
&&&&&&0.324\footnotemark[27]0.31\footnotemark[6]&\\
\hline
$\mathrm{Tl}$&576&1.143\footnotemark[3]&1665\footnotemark[3]&&&&25.2\footnotemark[5]\\
&&&&&&0.025\footnotemark[6]&\\
\hline
$\mathrm{Pb}$&600&&&&&&9.89\footnotemark[5]\\
&623&1.19\footnotemark[3]$^,$\footnotemark[4]&1770\footnotemark[30]&&&&9.89\footnotemark[5]\\
&700&&&&&&11.4\footnotemark[5]\\
&&&&&&0.000088\footnotemark[6]&\\
\hline
$\mathrm{Bi}$&544&1.15\footnotemark[11]&&&&&8.09\footnotemark[5]\\
 \hline\hline
\end{tabular}
\footnotetext[31]{\onlinecite{scop_jpc}} \footnotetext[1]{from the
max of S} \footnotetext[2]{\onlinecite{sinn}}
\footnotetext[3]{\onlinecite{ida}}
\footnotetext[4]{\onlinecite{hult}}
\footnotetext[5]{\onlinecite{tpm_DT}}
\footnotetext[32]{\onlinecite{kaw_mg}}
\footnotetext[6]{\onlinecite{mug_ncs}}
\footnotetext[7]{\onlinecite{pil_na}}
\footnotetext[8]{\onlinecite{scop_prena}}
\footnotetext[9]{\onlinecite{des_napb}}
\footnotetext[10]{\onlinecite{mor_na}}
\footnotetext[11]{\onlinecite{klep}}
\footnotetext[12]{\onlinecite{scop_preal}}
\footnotetext[13]{\onlinecite{hos_si1,hos_si2}}
\footnotetext[14]{Solid} \footnotetext[15]{\onlinecite{OSE}}
\footnotetext[16]{\onlinecite{mon_k}}
\footnotetext[17]{\onlinecite{cab_k}}
\footnotetext[18]{\onlinecite{bov_k}}
\footnotetext[33]{\onlinecite{caz_cu}}
\footnotetext[19]{\onlinecite{ber_ni}}
\footnotetext[20]{\onlinecite{inui_ga}}
\footnotetext[21]{\onlinecite{ber_ga2}}
\footnotetext[22]{\onlinecite{scop_prlga}}
\footnotetext[23]{\onlinecite{hos_ge}}
\footnotetext[24]{\onlinecite{cop_rb}}
\footnotetext[25]{\onlinecite{hos_sn}}
\footnotetext[26]{\onlinecite{bod_cs}}
\footnotetext[27]{\onlinecite{bad_hg}}
\footnotetext[28]{\onlinecite{bov_hg}}
\footnotetext[29]{\onlinecite{hos_hg}}
\footnotetext[30]{\onlinecite{sod_pb}} \caption{Summary of some
physical properties of liquid metals relevant for the dynamics.}
\label{table}
\end{table*}


\end{document}